\documentclass[a4paper,10pt]{article}

\usepackage{ae}
\usepackage[ansinew]{inputenc}

\usepackage{amssymb,amsmath,amsthm,mathtools}

\usepackage{setspace}
\usepackage{epsf}
\usepackage{psfrag}
\usepackage{mathrsfs}
\usepackage{graphicx}
\usepackage{epsfig}

\usepackage{multirow}

\def\bmath#1{\mbox{\boldmath $#1$}}

\usepackage{color}
\theoremstyle{plain}

\newtheorem{lem}{Lemma}
\newtheorem{cor}{Corolary}

\newtheorem{prop}{Proposition}[section]

\title{Covariant and gauge-invariant linear scalar perturbations in multiple scalar field cosmologies}

\author{
Artur Alho$^{1,2}$\thanks{E-mail: aalho@math.ist.utl.pt}~~and
Filipe C. Mena$^1$\thanks{E-mail: fmena@math.uminho.pt}\\\\
{\small $^1$Centro de Matem\'atica, Universidade do Minho, Campus de Gualtar, 4710-057 Braga,
Portugal} \\{\small $^2$Centro de An\'alise Matem\'atica, Geometria e Sistemas Din\^amicos, Instituto Superior T\'ecnico, 1049-001 Lisboa, Portugal}}

\begin{document}

\maketitle

\begin{abstract}
We derive a set of equations monitoring the evolution of covariant and gauge-invariant linear scalar perturbations 
of Friedman-Lema\^itre-Robertson-Walker models with multiple interacting nonlinear scalar fields. We use a dynamical systems' 
approach in order to perform a stability analysis for some classes of scalar field potentials. In particular, using a recent 
approximation for the inflationary dynamics of the background solution, we derive conditions under which homogenization 
occurs for chaotic (quadratic and quartic potentials) and new inflation. We also prove a cosmic no-hair result for power-law 
inflation and its generalisation for two scalar fields with independent exponential potentials (assisted power-law inflation).
\end{abstract}

\section{Introduction}
Nonlinear scalar fields $\phi$ have been important to model the presently observed accelerated cosmological expansion 
\cite{CST06} as well as the inflationary phase of the early universe \cite{Lin08}. 

The simplest of these models has potentials ${\cal V(\phi)}$ with a strictly positive lower bound, which is a 
straightforward generalisation of the positive cosmological constant \cite{Wal83} and  mimics it at late times. 
In that case, the Bianchi models of types I-VIII were studied in detail by Rendall \cite{Ren04} and their nonlinear 
stability by Ringstr\"om in \cite{Rin08}. In subsequent works, Rendall considered potentials with zero lower bound when 
$\phi$ is either infinite \cite{Ren05} or finite \cite{Ren07}. For the former class of solutions, it was shown that if 
$\frac{d\mathcal{V}}{d\phi}/\mathcal{V}$ satisfies an upper bound which rules out too rapid exponential decay, 
accelerated expansion is expected to exist indefinitely and has a dynamical behaviour between the power-law \cite{Hal87,BB88} 
and exponential types, commonly termed as intermediate inflation \cite{Bar90,BS90,PB95,BP95}.
Later, in \cite{Ren07}, Rendall considered potentials which are positive and tend to zero, but which do not experience 
accelerated expansion indefinitely, as the Klein-Gordon field. These potentials are very useful for studying the early 
inflationary stage of the universe in models of chaotic inflation (see \cite{Lin08}), since they allow the physical 
process of reheating \cite{KLS94,KLS97}.

Dynamical systems's techniques can be applied when the Einstein field equations (EFE) reduce to a system of ordinary 
differential equations (ODEs). For an exponential potential, the flat and isotropic power-law inflationary solution was found by Halliwell using phase-plane 
methods \cite{Hal87} and by Burd and Barrow for Bianchi types I and III, as well as Kantowski-Sachs models \cite{BB88}. 
Polynomial type potential were also studied using dynamical systems' techniques by a number of authors: Isotropic models were 
first studied by Belinskii et al. \cite{BGZK85,BGKZ85,BGZK86} 
whereas spatially homogeneous and anisotropic models by \cite{MS86,KB90,Heu91}. In \cite{Ren02},  Rendall revisited some 
results of  \cite{BGKZ85} by giving a rigorous asymptotic analysis of the inflationary dynamics using center manifold theory.

A procedure which has proved to be very useful when using techniques from dynamical systems' theory applied to Cosmology is 
the reduction of the original system of equations, by using {\it Hubble-normalized-variables} \cite{WH89,WE97,WL05}. 
For scalar field cosmologies, such variables were defined by Coley et al. \cite{CIvdH97} in the context of a single scalar 
field with an exponential potential. There, the Bianchi type models I-VIII were studied in detail and previous results in 
the literature \cite{Lid92,AFI93,FI93} were treated in an unified way, in particular, it was possible to test whether a given 
model inflates and/or isotropizes at late times and thus test the validity of the cosmic no-hair conjecture  in those settings. 
More precisely, it was shown that the flat isotropic power-law inflationary solution is an attractor for all ever expanding Bianchi models 
with an exponential potential \cite{vdHO99}. The flat Friedman-Lema\^itre-Robertson-Walker (FLRW) model with exponential 
potential coupled to matter was studied in \cite{BC00}, whereas the case of several independent exponential potentials was considered in \cite{CvdH00}. 
For more details see Coley \cite{Col03} and references therein. 

Recently, Hubble-normalized variables have also been used in the study of the Klein-Gordon field 
by Ure\~na-L\'opez and Reyes-Ibarra in \cite{ULRI09,RIUL10}, see also \cite{KT08,KT10}. Contrary to the exponential potential 
situation, in this case the ODEs system does not decouple from the Raychaudhuri equation and it is necessary to introduce new 
expansion-normalized variables.
By treating this new variable as a potential parameter, 
they found an analytical approximation for the inflationary dynamics, which for small values of the parameter, works as 
a first order correction to the usual slow-roll approximation.
                                                                                                                                                                                                                                                                                                                                                                                                                                                                  
The above results concern spatially homogeneous backgrounds. Here, instead, we shall be interested in the evolution of 
inhomogeneous spacetimes resulting from linear perturbations of FLRW backgrounds and in the application of 
dynamical systems techniques to these settings.

In this paper, we use the approach to linear perturbation theory developed by Ellis and Bruni \cite{EB89,EBH90,BDE92}. 
This consists in starting from exact non-linear equations using the {\it 1+3 covariant formalism} 
which, in view of the fundamental lemma of Stewart and Walker \cite{SW74}, are then linearized about exact 
FLRW models.
The advantage of this approach with respect to other metric formalisms \cite{Bar80,Ste90}, relies on the fact that the perturbations variables are covariant and 
gauge-invariant by construction, having a clear geometrical and well defined physical interpretation \cite{BDE92}, see also \cite{UW12}.
Exact evolution equations for linear perturbations of FLRW with a perfect-fluid as matter source were given in 
\cite{EB89,EBH90} and the extension for an imperfect-fluid can be found in Hwang and Vishniac 
\cite{HV90}. The imperfect-fluid case was also applied to describe perturbations in a multi-component fluid by Dunsby et al. 
\cite{Dun91,DBE92} 
using the methods of King and Ellis \cite{KE73} and, to minimally coupled scalar-fields 
by Bruni et al \cite{BED92} using the results by Madsen \cite{Mad88} and the field-fluid relation.
More recently, this has also been applied to charged multifluids \cite{MDBST03} and magnetized cosmologies 
\cite{TB97,TB98}. 

To study the evolution of inhomogeneities, we shall then employ a dynamical systems` approach following the methods of 
Woszczyna \cite{Wos92II,Wos92,Bru93,Wos95}. These were also used to study stability problems in a universe with dust and 
radiation \cite{BP94}, magnetized cosmologies \cite{HD00} and locally rotational symmetric (LRS) Bianchi I models \cite{Dun93}.

The plan of the paper is the following: in Section 1 we revise the background dynamics and relevant results 
for the upcoming sections. We start by introducing the Hubble normalized variables and the resulting reduced dynamical system. 
This shall be done in a particular way so that case of exponential potentials and the approximations for polynomial potentials can be 
treated in a unified way, simplifying the analysis of last section. In sections 2 and 3, we construct the system of equations governing the evolution of linear scalar perturbations of a 
{\it Friedman-Lema\^itre-Robertson-Walker-scalar field} (FLRWsf) background with multiple interacting scalar fields generalising 
the works of \cite{DBE92,BED92}.
In Section 5, we shall then apply the dynamical systems' approach of Woszczyna and show how can it be generalised to the case of multiple scalar fields. 
In particular, we consider in detail the examples of one and two scalar fields with exponential potentials as well as the approximations for models with polynomial potentials.

\section{The background spacetime}
In this section, we revise the background setting as well as some results that will be used in subsequent sections. 
This will be done by presenting the formalism in an unified way for several classes of scalar field potentials.

We will consider $N$ minimally coupled scalar fields $\phi_{A},~A=1,...,N$, with arbitrary self-interaction potentials 
$\mathcal{V}_{A}=\mathcal{V}\left(\phi_{A}\right)$ and a general interaction potential between the 
fields $\mathcal{W}=\mathcal{W}\left(\phi_{1},...,\phi_{N}\right)$. The action associated to this scenario is given by
\begin{equation}
 S[\phi_A,\bmath{g}]=\int_{\mathcal{M}}d^{4}x\sqrt{-g}\left[\frac{1}{2\chi}R-\frac{1}{2}\sum^{N}_{A=1}\left(\nabla_{\lambda}\phi_{A}\right)\left(\nabla^{\lambda}\phi_{A}\right)-\sum^{N}_{A=1}\mathcal{V}_{A}-\mathcal{W}\right],
\end{equation}
where $g$ is the determinant of the metric ${\bf g}$, $R$ is the Ricci scalar and the Einstein summation convention is understood on the greek indices. We will also use units such that 
$\chi=\frac{8\pi G}{c^{4}}=1$. The energy-momentum (EM) tensor is then
\begin{equation}
 T_{\mu\nu}=\sum^{N}_{A=1}\left[\left(\nabla_{\mu}\phi_{A}\right)\left(\nabla_{\nu}\phi_{A}\right)-g_{\mu\nu}\left\{\frac{1}{2}\left(\nabla_{\lambda}\phi_{A}\right)\left(\nabla^{\lambda}\phi_{A}\right)+\mathcal{V}_{A}\right\}\right]-g_{\mu\nu}\mathcal{W},
\end{equation}
and the generalized Euler-Lagrange equations give a system of $N$ evolution equations for the scalar fields
\begin{equation}
 \Box_{\mathbf{g}}\phi_{A}-\frac{d\mathcal{V}_{A}}{d\phi_{A}}=\frac{\partial \mathcal{W}}{\partial\phi_{A}}\quad A=1,..,N,
\label{K-G}
\end{equation}
where  $\Box_{\mathbf{g}}$ is the D'Alembertian for the metric $\mathbf{g}$.
On a FLRW background and using comoving coordinates, the line element reads 
\begin{equation}
 ds^{2}=-dt^{2}+a^{2}(t)\left[\frac{1}{1-kr^{2}}dr^{2}+r^{2}d\Omega^{2}\right],
\end{equation}
where $a(t)$ is the scale factor, $t$ the proper time, $d\Omega^2$ the spherical 2-metric and $k=1,0,-1$ the curvature of spatial hypersurfaces. 
On such a background, the scalar fields $\phi_{A}$ are functions of time $t$ only. Making use of the momentum density variable defined by
\begin{equation}
 \psi_{A}:=\frac{\partial\mathcal{L}}{\partial\phi_{A}}=\dot{\phi}_{A},
\end{equation}
 where the dot denotes differentiation with respect to proper time, the EFEs together 
with the scalar field evolution equations give an autonomous system of first order ODEs
\begin{equation}
 \dot{\psi}_{A}=-3H\psi_{A}-\frac{d\mathcal{V}_{A}}{d\phi_{A}}-\frac{\partial\mathcal{W}}{\partial\phi_{A}}\quad,\quad A=1,...,N
\label{K-G3-1background}
\end{equation}
\begin{equation}
 \dot{H}=-H^{2}-\frac{1}{3}\sum^{N}_{A=1}\psi^{2}_{A}+\frac{1}{3}\sum^{N}_{A=1}\mathcal{V}_{A}+\frac{1}{3}\mathcal{W}
\label{raychauduri}
\end{equation}
and the Friedman constraint
\begin{equation}
 {^3}R=-6H^{2}+\sum^{N}_{A=1}\psi^{2}_{A}+2\sum^{N}_{A=1}\mathcal{V}_{A}+2\mathcal{W},
\label{Friedmann}
\end{equation}
where $H:=\dot{a}/a$ is the Hubble function and ${^3}R=6k/a^{2}$ the Ricci scalar of the spatial metric. 
An important quantity in cosmology is the deceleration parameter given by
\begin{equation}
 q:=-\frac{\ddot{a}a}{\dot{a}^2}=-\left[1+\frac{\dot{H}}{H^{2}}\right]
\label{decpar}
\end{equation}
and a model has accelerated expansion, $\ddot{a}>0$, if and only if, $q<0$. 
We will use the {\em Hubble normalized variables} for scalar field cosmologies defined by
\begin{equation}\label{ENV}
 \Psi_{A}:=\frac{\psi_{A}}{\sqrt{6}H}\quad,\quad\Phi_{A}:=\left(\frac{\mathcal{V}_{A}}{3H^2}\right)^{\frac{1}{2n}}\quad,\quad\Theta:=\left(\frac{\mathcal{W}}{3H^2}\right)^{\frac{1}{2n}}\quad,\quad K:=-\frac{{^3}R}{6H^{2}}\;,
\end{equation}
where $n\in\mathbb{N}$ takes values for specific potentials. For instance, in the case of an exponential potential, $n=1$, the 
variables coincide with those of \cite{CIvdH97} and, for the polynomial type potentials, with those of \cite{ULRI09,RIUL10,KT08,KT10}. 
We will also make use of the logarithmic time variable $\tau$
\begin{equation}
 \frac{d\tau}{dt}=H\quad,\quad H^{\prime}=-(1+q)H,
\label{TAU}
\end{equation}
so that $\tau\rightarrow -\infty$, as $t\rightarrow0^{+}$, and denote differentiation with respect to $\tau$ by a prime.
Using these variables, the system of ODEs governing the background dynamics becomes
\begin{equation}\label{syst-ode}
 \begin{aligned}
  \Psi^{\prime}_{A}&=(q-2)\Psi_{A}-n\sqrt{6}\left[\Phi^{2n-1}_{A}\frac{\partial\Phi_{A}}{\partial\phi_{A}}+\Theta^{2n-1}_{A}\frac{\partial\Theta}{\partial\phi_{A}}\right] \\
  \Phi^{\prime}_{A}&=\frac{1}{n}(q+1)\Phi_{A}+\sqrt{6}\frac{\partial\Phi_{A}}{\partial\phi_{A}}\Psi_{A} \\
    \Theta^{\prime}&=\frac{1}{n}(q+1)\Theta+\sqrt{6}\sum^{N}_{A=1}\frac{\partial\Theta}{\partial\phi_{A}}\Psi_{A} 
\end{aligned}
\end{equation}
subject to the Friedman constraint
\begin{equation}
\label{Fried-constraint}
K=1-\sum^{N}_{A=1}\Psi^{2}_{A}-\sum^{N}_{A=1}\Phi^{2n}_{A}-\Theta^{2n}
\end{equation}
and with
\begin{equation}
 q=2\sum^{N}_{A=1}\Psi^{2}_{A}-\sum^{N}_{A=1}\Phi^{2n}_{A}-\Theta^{2n}.
\end{equation}
These equations will allow us to treat, in a unified way, various families of scalar field potentials given that the system 
remains autonomous, which is the case for exponential potentials and the polynomial potentials' approximation. 
In general, if the scalar fields do not interact with each other in flat FLRW models then 
$\left(\Psi_{1},..,\Psi_{N},\Phi_{1},..,\Phi_{N}\right)\in[-1,1]^{N}\times[-1,1]^{N}$, but for the models 
under consideration we will only need to take the dynamical system state space
\begin{equation}\label{FLat_InvSet}
 \left\{\left(\Psi_{1},..,\Psi_{N},\Phi_{1},..,\Phi_{N}\right)\in[-1,0]^{N}\times[0,1]^{N}\,:\,\sum^{N}_{A=1}\Psi^{2}_{A}+\sum^{N}_{A=1}\Phi^{2n}_{A}=1\right\}
\end{equation}
with
$$
q=2-3\sum^{N}_{A=1}\Phi^{2n}_{A}=3\sum^{N}_{A=1}\Psi^{2}_{A}-1,
$$
and it is straightforward to get:
\begin{lem}\label{Lemma_1}
{\it For $N$ non-interacting scalar fields in flat backgrounds with $\Theta=K=0$, the 
 fixed points $\mathcal{P}$ and $\mathcal{Q}$ of the system (\ref{syst-ode}), when they exist, are given by}
\begin{equation*}
\mathcal{P}:~~ \begin{array}{cc}
            {\displaystyle \frac{\Phi^{2n}_{A}}{\sum^{N}_{B=1}\Phi^{2n}_{B}}\frac{\partial \Phi_A}{\partial \phi_A}= -\frac{\sqrt{6}}{2n}\Psi_A\Phi_A}\quad\text{and}\quad & {\displaystyle\frac{\Psi^{2}_{A}}{\sum^{N}_{B=1}\Psi^{2}_{B}}\frac{\partial \Phi_A}{\partial \phi_A}=-\frac{\sqrt{6}}{2n}\Psi_A\Phi_A} \\
            \end{array}
\end{equation*}
{\it and}
\begin{equation*}
{\cal Q}:~~\Phi_{A}=\Psi_{A}=0
\end{equation*}
{\it with} $A=1,..,N$.
\end{lem}
The fixed points $\mathcal{P}$ correspond to physical solutions depending on the potential, while the point $\mathcal{Q}$ is unphysical. 
For a single scalar field, the Friedman constraint \eqref{Fried-constraint} reads
\begin{equation}\label{FlatFC}
 \Psi^{2}+\Phi^{2n}=1
\end{equation}
and the linearised matrix of the system \eqref{syst-ode}, at $\mathcal{P}$, is
\begin{equation}\label{LinMat}
 \begin{pmatrix}
 9\Psi^{2}_{\mathcal{P}}-3 & \frac{\Phi^{2n}_{\mathcal{P}}}{\Psi_{\mathcal{P}}\Phi_{\mathcal{P}}}\left(3(2n-1)\Psi^{2}_{\mathcal{P}}+\frac{2n^2}{\Phi_{\mathcal{P}}}\left(\frac{\partial^{2}\Phi}{\partial\phi^{2}}\right)_{\mathcal{P}}\right) \\
  \frac{3}{n}\Psi_{\mathcal{P}}\Phi_{\mathcal{P}}   &  \frac{3}{n}\Psi^{2}_{\mathcal{P}}-\frac{2n}{\Phi_{\mathcal{P}}}\left(\frac{\partial^{2}\Phi}{\partial\phi^{2}}\right)_{\mathcal{P}} \\
 \end{pmatrix} 
\end{equation}
with characteristic polynomial
\begin{equation*}
 \omega^{2}-\left\{\left(9+\frac{3}{n}\right)\Psi^{2}_{\mathcal{P}}-3-2n\frac{1}{\Phi_{\mathcal{P}}}\left(\frac{\partial^{2}\Phi}{\partial\phi^{2}}\right)_{\mathcal{P}}\right\}\omega 
           +\frac{18}{n}\Psi^{2}_{\mathcal{P}}\left\{1-(n+1)\Phi^{2n}_{\mathcal{P}}-\frac{2n^{2}}{3\Phi_{\mathcal{P}}}\left(\frac{\partial^{2}\Phi}{\partial\phi^{2}}\right)_{\mathcal{P}}\right\}.
\end{equation*}
The eigenvalues of the matrix \eqref{LinMat} are
\begin{small}
\begin{equation*}
 \begin{aligned}
 \omega^{\pm}_{\mathcal{P}}=&\frac{3}{2}\left[\left(3+\frac{1}{n}\right)\Psi^{2}_{\mathcal{P}}-1-\frac{2n}{3\Phi}\left(\frac{\partial^{2}\Phi}{\partial\phi^{2}}\right)_{\mathcal{P}}\right] \\
              &\pm\frac{3}{2}\sqrt{\left[1+\frac{2n}{3\Phi}\left(\frac{\partial^{2}\Phi}{\partial\phi^{2}}\right)_{\mathcal{P}}\right]^2
                       +2\Psi^{2}\left[\left(1-\frac{1}{n}\right)-\frac{1}{3}\left(\frac{2}{n}-\frac{1}{n^{2}}-1\right)\Psi^{2}_{\mathcal{P}}+\frac{2}{3}(n-1)\frac{2n}{3\Phi}\left(\frac{\partial^{2}\Phi}{\partial\phi^{2}}\right)_{\mathcal{P}}\right]}\;.
\end{aligned}
\end{equation*}
\end{small}
If we denote the respective eigenvectors by $\left(\delta\Psi\;\delta\Phi\right)^{T}_{\pm}$, the general solution to the perturbations around $\mathcal{P}$ reads
\begin{equation*}
 \begin{pmatrix}
  \delta\Psi \\
  \delta\Phi
 \end{pmatrix}
=C_{-}\begin{pmatrix}
  \delta\Psi \\
  \delta\Phi
 \end{pmatrix}_{-}e^{\omega_{-}\tau}+C_{+}\begin{pmatrix}
  \delta\Psi \\
  \delta\Phi
 \end{pmatrix}_{+}e^{\omega_{+}\tau}.
\end{equation*}
However, \eqref{FlatFC} implies, to linear order, that
\begin{equation*}
 \Psi_{\mathcal{P}}\delta\Psi+n\Phi^{2n-1}_{\mathcal{P}}\delta\Phi=0\Leftrightarrow \begin{pmatrix}
										    \Psi_{\mathcal{P}} & n\Phi^{2n-1}_{\mathcal{P}} 
										  \end{pmatrix}
										    \begin{pmatrix}
										    \delta\Psi  \\
										    \delta\Phi   
										   \end{pmatrix}=0
\end{equation*}
and the evolution of linear perturbations around the fixed points $\mathcal{P}$ reduces to a single equation 
\begin{equation}\label{Single_LinMat}
 \delta\Phi^{\prime}=\left(-3\Phi^{2n}+\frac{3}{n}\Psi^{2}-\frac{2n}{\Phi_{\mathcal{P}}}\left(\frac{\partial^{2}\Phi}{\partial\phi^{2}}\right)_{\mathcal{P}}\right)\delta\Phi.
\end{equation}
Then, there is a single eigenvalue solution
\begin{equation}\label{EigenValue}
 \omega^{-}_{\mathcal{P}}=-3\left(1+\frac{1}{n}\right)\Phi^{2n}_{\mathcal{P}}+\frac{3}{n}-\frac{2n}{\Phi_{\mathcal{P}}}\left(\frac{\partial^{2}\Phi}{\partial\phi^{2}}\right)_{\mathcal{P}}
\end{equation}
which is proportional to the eigenvector $\left(\delta\Psi\;\delta\Phi\right)^{T}_{-}$. 

With the above, it is easy to show that,
at $\mathcal{P}$, the single scalar field solutions are inflationary if and only if
\begin{equation}\label{Inf_Cond}
\Phi^{2n}_{\mathcal{P}}>\frac{2}{3}\Leftrightarrow \Psi^{2}_{\mathcal{P}}<\frac{1}{3}\;. 
\end{equation}
We note that this result does not involve a specific class of potentials.  

We shall now revise the particular examples
of exponential and polynomial potentials which will be useful in the stability analysis of Section 5.
\subsection{Exponential Potentials}\label{subsection_4}
Using the above framework, we now review the flat (assisted) power-law solutions due to exponential potentials 
\begin{equation}\label{Ind_ExpPot}
\mathcal{V}_{A}=\Lambda e^{\lambda_{A}\phi_{A}} 
\end{equation}
where $\lambda_{A}$ and $\Lambda$ are positive constants. For such potentials, the Hubble normalized variables (\ref{ENV}) are defined with $n=1$, the zero curvature invariant set \eqref{FLat_InvSet} is a higher dimensional sphere $\mathbb{S}^{N}$ and
\begin{equation}
 \frac{\partial^{p}\Phi_{A}}{\partial\phi^{p}_{A}}=\left(\frac{\lambda_{A}}{2}\right)^{p}\Phi_{A},\;
\end{equation}
for any $p\in \mathbb{N}$.
We will now treat separately the single scalar field and the two scalar fields cases.
 \subsubsection{Power-law Inflation}
If only one scalar field is present, Lemma \ref{Lemma_1} implies
\begin{equation}
 \mathcal{P}:~~ \Phi\left(\Psi+\frac{\lambda}{\sqrt{6}}\right)=0
\end{equation}
satisfying \eqref{FlatFC}. Therefore, there are two fixed points $\left(\Psi_{\mathcal{P}},\Phi_{\mathcal{P}}\right)\in[-1,0]\times[0,1]$ in $\mathbb{S}^{1}$ such that

\begin{equation}\label{Single_P0}
 \mathcal{P}_{0}:~~ \left(\Psi,\Phi\right)=\left(-1,0\right) 
\end{equation}
\begin{equation}\label{Single_P1}
 \mathcal{P}_{1}:~~ \left(\Psi,\Phi\right)=\left(-\frac{\lambda}{\sqrt{6}},\frac{\sqrt{6-\lambda^{2}}}{\sqrt{6}}\right)\quad\text{with}\quad 0<\lambda<\sqrt{6}.
\end{equation}
Also, at $\mathcal{P}$, we have
\begin{equation*}
\left(\frac{\partial^{2}\Phi}{\partial\phi^{2}}\right)_{\mathcal{P}}=-\frac{\sqrt{6}\lambda}{4}\Psi\Phi,
\end{equation*}
so that, from \eqref{EigenValue}, the eigenvalues are
\begin{equation}
\omega^{-}_{\mathcal{P}_0}=3+\frac{\sqrt{6}}{2}\lambda\quad\text{and}\quad\omega^{-}_{\mathcal{P}_1}=-\frac{6-\lambda^{2}}{2}.
\end{equation}
The point $\mathcal{P}_{0}$ corresponds to the well-known massless scalar field solution, which is the early time attractor for the system and is a {\it source} for all $\lambda\in(0,\sqrt{6})$. The point $\mathcal{P}_{1}$ is a {\it sink}, with the deceleration parameter given by
\begin{equation*}
q_{\mathcal{P}_{1}}=\frac{\lambda^{2}-2}{2}\;
\end{equation*}
and, by \eqref{Inf_Cond}, the solution is inflationary if and only if $q_{\mathcal{P}_{1}}<0$, i.e. for
\begin{equation*}
 0<\lambda<\sqrt{2}\;,
\end{equation*}
which corresponds to the flat homogeneous and isotropic power-law inflationary solution found by Halliwell \cite{Hal87}. 
This solution is also known to be a future attractor for ever expanding scalar field Bianchi models, 
see e.g. \cite{Col03} for details and related references.
 \subsubsection{Assisted power-law Inflation}
For two scalar fields $\phi_1$ and $\phi_2$ with independent exponential potentials, Lemma \ref{Lemma_1} implies
\begin{equation*}
 \mathcal{P}:~~ \Phi_{A}\Psi_{A}\left(\frac{\Psi_{A}}{\Psi^{2}}\frac{\lambda_{A}}{2}+\frac{\sqrt{6}}{2}\right)=0\quad\text{and}\quad\Phi_{A}\left(\frac{\Phi^{2}_{A}}{\Phi^{2}}\frac{\lambda_{A}}{2}+\frac{\sqrt{6}}{2}\Psi_{A}\right)=0,\quad A=1,2
\end{equation*}
which leads to the four fixed points
\begin{equation}\label{P0_APLI}
 \mathcal{P}_{0}:~~ \left(\Psi_1,\Psi_2,\Phi_1,\Phi_2\right)=\left(-\Psi_{0},-\sqrt{1-\Psi^{2}_{0}},0,0\right)\;,\quad 0\leq\Psi_0\leq1
\end{equation}
\begin{equation}\label{P1_APLI}
 \mathcal{P}_{1}:~~ \left(\Psi_1, \Psi_2, \Phi_1, \Phi_2\right)=\left(-\frac{\lambda_{1}}{\sqrt{6}},0, \frac{\sqrt{6-\lambda^{2}_{1}}}{\sqrt{6}}, 0 \right)\;,\quad 0<\lambda_1<\sqrt{6}
\end{equation}
\begin{equation}\label{P2_APLI}
 \mathcal{P}_{2}:~~ \left(\Psi_1, \Psi_2, \Phi_1, \Phi_2\right)=\left(0,-\frac{\lambda_{2}}{\sqrt{6}},0, \frac{\sqrt{6-\lambda^{2}_{2}}}{\sqrt{6}}\right)\;,\quad 0<\lambda_2<\sqrt{6}
\end{equation}
\begin{small}
 \begin{equation}\label{P3_APLI}
\begin{aligned}
\mathcal{P}_{3}:~~ \left(\Psi_1, \Psi_2, \Phi_1, \Phi_2\right)=\left(-\frac{\lambda^2}{\sqrt{6}\lambda_1}, -\frac{\lambda^2}{\sqrt{6}\lambda_2}, \frac{\sqrt{\lambda^{2}(6-\lambda^{2})}}{\sqrt{6}\lambda_1}, \frac{\sqrt{\lambda^{2}(6-\lambda^{2})}}{\sqrt{6}\lambda_2} \right)\;,\quad  \frac{1}{\lambda^{2}}=\frac{1}{\lambda^{2}_{1}}+\frac{1}{\lambda^{2}_{2}}
\end{aligned}
\end{equation}\
\end{small}
and the linearised matrix of the system \eqref{syst-ode} at $\mathcal{P}$ is
\begin{equation*}
\begin{pmatrix}
 9\Psi^{2}_{1}+3\left(\Psi^{2}_{2}-1\right) & 6\Psi_{1}\Psi_{2} & -\sqrt{6}\lambda_{1}\Phi_{1} & 0 \\
  6\Psi_{1}\Psi_{2} & 9\Psi^{2}_{2}+3\left(\Psi^{2}_{1}-1\right) & 0 & -\sqrt{6}\lambda_{2}\Phi_{2} \\
 \left(\frac{\sqrt{6}}{2}\lambda_{1}+6\Psi_{1}\right)\Phi_{1} & 6\Psi_{2}\Phi_{1} & \frac{\sqrt{6}}{2}\lambda_{1}\Psi_{1}+3\left(\Psi^{2}_{1}+\Psi^{2}_{2}\right) & 0 \\
 6\Psi_{1}\Phi_{2}& \left(\frac{\sqrt{6}}{2}\lambda_{2}+6\Psi_{2}\right)\Phi_{2} & 0 & \frac{\sqrt{6}}{2}\lambda_{2}\Psi_{2}+3\left(\Psi^{2}_{1}+\Psi^{2}_{2}\right)\\
\end{pmatrix}_{\mathcal{P}}
 \end{equation*}
with eigenvalues 
\begin{equation}\label{P0_Eigen}
  \omega(\mathcal{P}_{0})=6\;,\;0\;,\;3-\frac{\sqrt{6}}{2}\lambda_{1}\Psi_{0}\;,\;3-\frac{\sqrt{6}}{2}\lambda_{2}\sqrt{1-\Psi^{2}_{0}}\quad\text{where}\quad 0\leq\Psi_0\leq1
\end{equation}
\begin{equation}\label{P12_Eigen}
  \omega(\mathcal{P}_{1,2})=\frac{\lambda^{2}_{1,2}}{2}\;,\;\lambda^{2}_{1,2}\;,\;-\frac{\left(6-\lambda^{2}_{1,2}\right)}{2}\;,\;-\frac{\left(6-\lambda^{2}_{1,2}\right)}{2}
\end{equation}
\begin{equation}\label{P3_Eigen}
  \omega(\mathcal{P}_{3})=\lambda^{2}\;,\;\frac{\lambda^{2}-6}{2}\;,\;\frac{1}{4}\left\{(\lambda^{2}-6)\pm\sqrt{(\lambda^{2}-6)+8\lambda^{2}(\lambda^{2}-6)}\right\}\;.
\end{equation}
Thus, in $\mathbb{S}^{2}$, the point $\mathcal{P}_{0}$ is a local source and corresponds to the massless scalar field solution which is the early time attractor. The points $\mathcal{P}_{1}$ and $\mathcal{P}_{2}$ are saddles which correspond to single power law solutions where either $\phi_{1}$ dominates over $\phi_{2}$ or vice-versa. As before, the Friedman constraint can be used to eliminate the unphysical radial direction, which 
corresponds to the positive eigenvalue of $\mathcal{P}_{3}$. 
Therefore, $\mathcal{P}_{3}$, having negative eigenvalues, becomes the stable late-time attractor and corresponds to the assisted power-law solution. 

The results of this section will be used in Section 5.2. 
\subsection{Polynomial Potentials}\label{subsection_5}
\begin{figure}[h!]
\centering
\begin{minipage}{\textwidth/2-1pc}
\includegraphics[width=\textwidth]{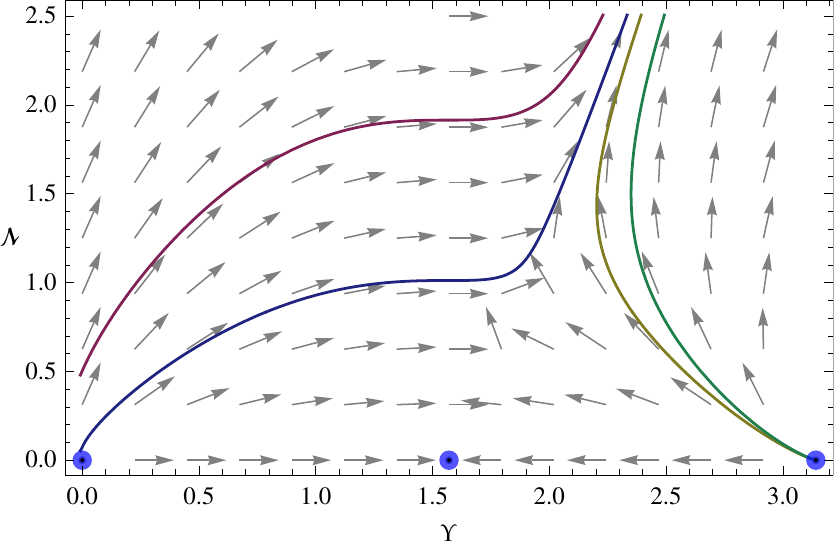}
\caption{\label{Fig1}Phase space for a quadratic potential $n=1$ and $\mathcal{N}=0$.}
\end{minipage}\hspace{2pc}%
\begin{minipage}{\textwidth/2-1pc}
\includegraphics[width=\textwidth]{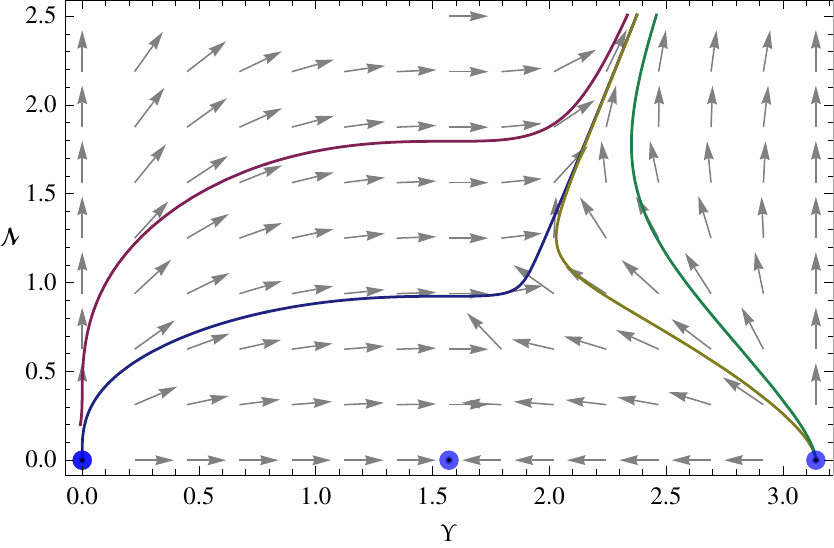}
\caption{\label{Fig2}Phase space for a quartic potential $n=2$ and $\mathcal{N}=0$.}
\end{minipage} 
\end{figure}

\begin{figure}[h!]
\centering
\begin{minipage}{\textwidth/2-1pc}
\includegraphics[width=\textwidth]{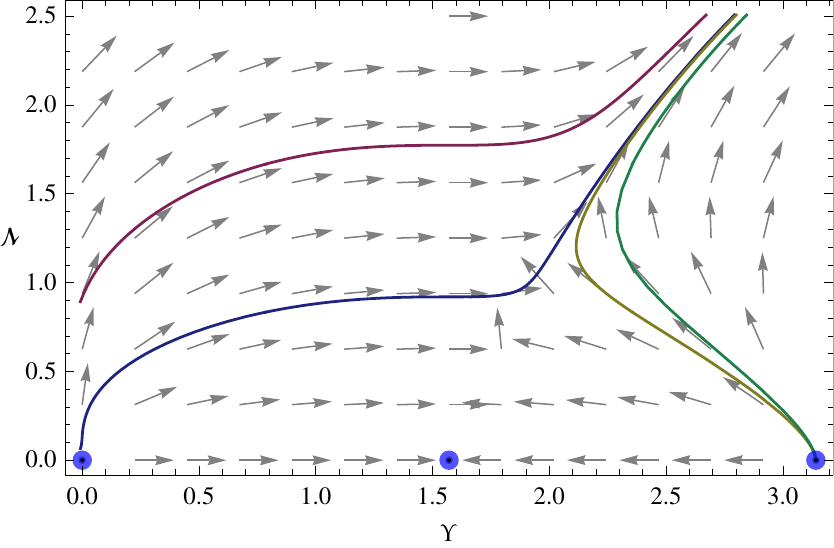}
\caption{\label{Fig3}Phase space for new inflation $n=2$ and $\mathcal{N}=1$.}
\end{minipage}\hspace{2pc}%
\begin{minipage}{\textwidth/2-1pc}
\includegraphics[width=\textwidth]{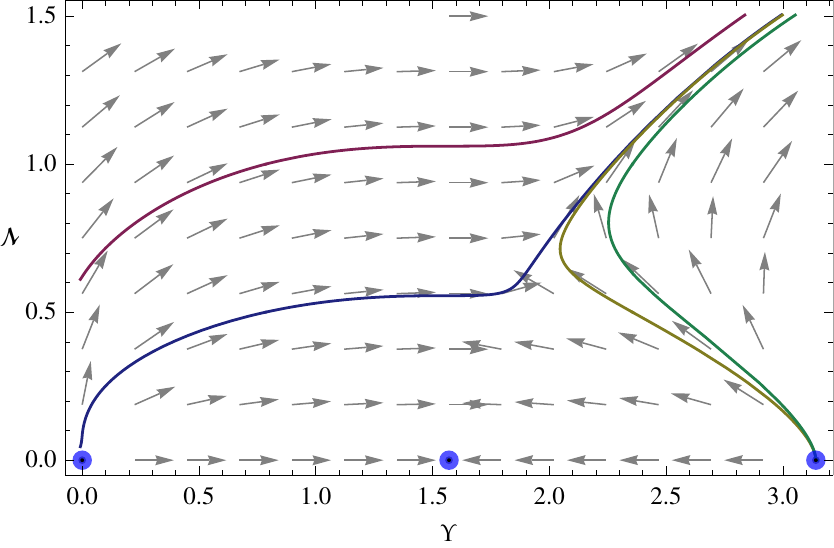}
\caption{\label{Fig4}Phase space for new inflation $n=2$ and $\mathcal{N}=4$.}
\end{minipage} 
\end{figure}

Recently, the expansion-normalized variables defined in \cite{ULRI09,RIUL10,KT08,KT10} were used to study the dynamical properties of scalar field cosmologies with potentials given by 
\begin{equation}
\label{GenHarmPot}
{\cal V}=C\frac{(\phi^2-v^2)^n}{2n}, 
\end{equation}
where $C>0$ and $v\ge 0$ are constants. Models with $v=0$ are called {\em chaotic inflation} and with $v>0$ {\em new inflation}, 
see e.g. \cite{KT10,Lin08}. Contrary to the case of exponential potentials, where the reduced dynamical system \eqref{syst-ode} 
is autonomous, for the potentials \eqref{GenHarmPot} one gets a non-autonomous system, in general. In order to turn the system 
autonomous, a new Hubble normalised variable must be introduced. An appropriate choice is
\begin{equation}\label{M}
 \mathcal{M}:=\left(6n\right)^{\frac{n-1}{2n}}n^{\frac{1}{2}}\left(\frac{C}{H^{2}}\right)^{\frac{1}{2n}}\;,
\end{equation}
which, in ever expanding models, is a monotone and growing function obeying the evolution equation
\begin{equation}\label{EvolM}
 \mathcal{M}^{\prime}=\frac{3}{n}\Psi^{2}\mathcal{M}\;.
\end{equation}
From \eqref{GenHarmPot} and \eqref{M} it follows that
\begin{equation}\label{2ndDerGenHarPot}
 \begin{aligned}
  \frac{\partial\Phi}{\partial\phi}
                     =\frac{\mathcal{M}}{\sqrt{6}n^2}\frac{\sqrt{n^{2}\Phi^2+\mathcal{M}^{2}\mathcal{N}^{2}}}{\Phi}\quad\text{and}\quad  \frac{\partial^{2}\Phi}{\partial\phi^{2}}&=-\left(\frac{\mathcal{M}}{\sqrt{6}n}\right)^{2}\frac{\mathcal{M}^{2}\mathcal{N}^{2}}{n\Phi^{3}}\;,
\end{aligned}
\end{equation}
where
\begin{equation}
 \mathcal{N}:=\frac{v}{\sqrt{6}}\;.
\end{equation}
To get a better picture of the state space of the new dynamical system, it is useful to make a change of variables and turn the above system into a 2-dimensional system. For $n=1,2$ and $v=0$ this was done in \cite{RIUL10}, where the new variable $\Upsilon$ was defined as
\begin{equation}
\Psi=\cos{(\Upsilon)}\quad,\quad\Phi=|\sin{(\Upsilon)}|^{\frac{1}{n}}\;,
\end{equation}
with $(\Upsilon,\mathcal{M})\in[\frac{\pi}{2},\pi]\times[0,+\infty)$. The dynamical system \eqref{syst-ode} coupled to \eqref{EvolM} then reads
\begin{equation}
 \begin{aligned}
\Upsilon^{\prime}&=\frac{|\sin{(\Upsilon)}|^{\frac{2(n-1)}{n}}}{\sin{(\Upsilon)}}\left[3\cos{(\Upsilon)}|\sin{(\Upsilon)}|^{\frac{2}{n}}+\frac{\mathcal{M}}{n}\sqrt{n^{2}|\sin{(\Upsilon)}|^{\frac{2}{n}}+\mathcal{M}^{2}\mathcal{N}^{2}}\right] \\
             \mathcal{M}^{\prime}&=\frac{3}{n}\mathcal{M}\cos^{2}{(\Upsilon)}\;.
 \end{aligned}
\end{equation}
The fixed points of this system are located at $\Upsilon=\frac{\pi}{2},\pi\quad\text{and}\quad\mathcal{M}=0$, and are 
independent of $n$. The point $(0,\pi)$ corresponds to the massless scalar field solutions $\Psi=-1$ and is unstable, 
while the point $(0,\frac{\pi}{2})$ is a saddle and corresponds to the potential dominated solutions $\Phi=1$. 
There are also heteroclinic curves connecting the unstable point with the saddle point along the stable direction i.e., 
along the $\mathcal{M}=0$ axis. Along the unstable direction of the saddle point departs a curve, which for small values 
of $\mathcal{M}$, acts as an attractor trajectory in the phase space, see \cite{RIUL10} for details when $\mathcal{N}=0$ 
(see also Figs. \ref{Fig1}, \ref{Fig2}, \ref{Fig3}, \ref{Fig4}).
The approach of \cite{ULRI09,KT10} consists in reducing the 3-dimensional system, obtained by coupling \eqref{syst-ode} with 
\eqref{EvolM}, to a 2-dimensional system with state vector $(\Psi,\Phi)$, by considering $\mathcal{M}$ as a control parameter. 
 \subsubsection{Chaotic Inflation $\mathcal{N}=0$}
For potentials having $\mathcal{N}=0$, Lemma \ref{Lemma_1} and \eqref{2ndDerGenHarPot} give
\begin{equation}\label{FP_ChaoticInf}
\mathcal{P}:\,\,  \ \Psi\Phi=-\frac{\mathcal{M}}{3}\quad\text{with}\quad \frac{\partial^{2}\Phi}{\partial\phi^{2}}=0
\end{equation}
subject to the Friedman constraint \eqref{FlatFC}. Then \eqref{EigenValue} gives
\begin{equation}\label{Eigenvalues_ChaoticInf}
 \omega_{-}=-3\left(1+\frac{1}{n}\right)\Phi^{2n}_{\mathcal{P}}+\frac{3}{n}
\end{equation}
and the fixed points $\mathcal{P}$ are stable if $\omega_{-}<0$, i.e. if
\begin{equation}\label{StabCond_ChaoticInf}
 \Phi^{2n}_{\mathcal{P}}>\frac{1}{1+n}\;,
\end{equation}
which contains the inflationary solutions \eqref{Inf_Cond} for all $n$. 
\\\\
{\bf (i) Quadratic Potential}
\\\\
For a quadratic potential $n=1$ and \eqref{M} reads
\begin{equation}
\mathcal{M}:=\frac{m}{H}.
\end{equation}
where $m=C> 0$. In this case, the fixed points $\left(\Psi_{\mathcal{P}},\Phi_{\mathcal{P}}\right)\in[-1,0]\times[0,1]$ in $\mathbb{S}^{1}$ are given by condition \eqref{FP_ChaoticInf} subject to \eqref{FlatFC} as
\begin{equation}\label{P0_QuadraticPot}
 \mathcal{P}_{0}:~~ \left(\Psi\;,\;\Phi\right)=\left(-\sqrt{\frac{1}{2}\left(1+\sqrt{1-\frac{4}{9}\mathcal{M}^{2}}\right)}\;,\;\sqrt{\frac{1}{2}\left(1-\sqrt{1-\frac{4}{9}\mathcal{M}^{2}}\right)}\right)
\end{equation}
\begin{equation}\label{P1_QuadraticPot}
 \mathcal{P}_{1}:~~ \left(\Psi\;,\;\Phi\right)=\left(-\sqrt{\frac{1}{2}\left(1-\sqrt{1-\frac{4}{9}\mathcal{M}^{2}}\right)}\;,\;\sqrt{\frac{1}{2}\left(1+\sqrt{1-\frac{4}{9}\mathcal{M}^{2}}\right)}\right)
\end{equation}
with eigenvalues \eqref{Eigenvalues_ChaoticInf}
\begin{equation}
 \omega^{-}_{\mathcal{P}_{0}}=3\sqrt{1-\frac{4}{9}\mathcal{M}^{2}}\quad,\quad \omega^{-}_{\mathcal{P}_{1}}=-3\sqrt{1-\frac{4}{9}\mathcal{M}^{2}}\;.
\end{equation}
Thus, in this case, the fixed points exist in the unitary circunference for $0\leq\mathcal{M}\leq\frac{3}{2}$. For $\mathcal{M}<\frac{3}{2}$, $\mathcal{P}_{0}$ is the local source which, at $\mathcal{M}=0$, represents the massless scalar field early attractor and $\mathcal{P}_{1}$ is the future attractor. At $(\Psi,\Phi)=(-\frac{1}{\sqrt{2}},\frac{1}{\sqrt{2}})$, the fixed points have zero eigenvalues. Moreover, from \eqref{Inf_Cond}, and as was shown in \cite{ULRI09}, the future attractor $\mathcal{P}_{1}$ is inflationary if and only if 
\begin{equation}
 \mathcal{M}<\sqrt{2}\;.
\end{equation}
For $\mathcal{M}=\sqrt{2}$, the attractor point ceases to be inflationary and the value of $\phi$ at this point 
(which represents the end of inflation) corresponds to that of the slow-roll approximation.
\\\\
{\bf (ii) Quartic Potential}
\\\\
For a quartic potential one has $n=2$ and \eqref{M} reads
\begin{equation*}
\mathcal{M}=\frac{\sqrt{2\sqrt{12}}\lambda}{\sqrt{H}},
\end{equation*}
with $\lambda^{4}=C\ge 0$, and the fixed points, given by condition \eqref{FP_ChaoticInf} subject to the flat Friedman 
constraint \eqref{FlatFC}
\begin{equation*}
 \Psi^{2}+\Phi^{4}=1,
\end{equation*}
are the zeros of the cubic polynomial
\begin{equation*}
 f(\Psi^{2})=\Psi^{6}-\Psi^{4}+\left(\frac{\mathcal{M}}{3}\right)^{4}\;.
\end{equation*}
The discriminant of this polynomial is
\begin{equation*}
 \varDelta=\frac{1}{9^{3}}\left(\frac{\mathcal{M}^{4}}{3}\right)\left(\frac{\mathcal{M}^{4}}{12}-1\right),
\end{equation*}
so that, for $0<\mathcal{M}^{4}<12$, it follows that $\varDelta<0$ and there are three distinct real roots. If $\mathcal{M}^{4}=12$, then $\varDelta=0$ and there is a repeated real root, otherwise there are two complex roots. Now, setting
 \begin{equation}\label{FPCond_Quartic}
  \mathcal{M}^{4}=12\sin^{2}{\left(\chi\right)}\quad,\quad 0<\chi<\frac{\pi}{2},
 \end{equation}
the three distinct roots are explicitly given by
\begin{equation}\label{P_QuarticPot}
 \mathcal{P}: \Psi^{2}=\frac{1}{3}\left(1+2\cos{\left(\frac{2}{3}\chi+\frac{2}{3}\pi\,l\right)}\right),\quad l=0,\pm1\;.
\end{equation}
The $l=1$ solution is unphysical since $\Psi^{2}<0$. Denoting the $l=0$ and $l=-1$ solutions by $\mathcal{P}_{0}$ and $\mathcal{P}_{1}$, respectively, then we get from \eqref{Eigenvalues_ChaoticInf}
\begin{equation}\label{Eigenvalues_Quartic}
 \omega^{-}(\mathcal{P}_{0})=
-\frac{3}{2}+3\cos{\left(\frac{2}{3}\chi\right)}\quad,\quad \omega^{-}(\mathcal{P}_{1})=-\frac{3}{2}+3\cos{\left(\frac{2}{3}(\chi-\pi)\right)}.
\end{equation}
For all values of $\mathcal{M}$ for which there are fixed points, $\mathcal{P}_{0}$  is a source and at $\mathcal{M}=0\Leftrightarrow\chi=0$ the solution represents that of a massless scalar field with $\Psi^{2}_{\mathcal{P}_{0}}=1$. In turn, $\mathcal{P}_{1}$ is a sink and, at $\mathcal{M}=0$, represents a potential dominated solution $\Phi^{4}_{\mathcal{P}_{1}}=1$.

When $\mathcal{M}^{4}=12$, the discriminant $\Delta$ of the cubic equation is zero and its roots coincide, 
having $\Psi^{2}=\frac{2}{3}$, which by \eqref{Eigenvalues_Quartic} gives a saddle. Moreover, from equation \eqref{Inf_Cond}, $\mathcal{P}_{1}$ is inflationary whenever
\begin{equation}\label{QuarticPot_CondInf}
 \chi<\frac{\pi}{4}\quad\Leftrightarrow\mathcal{M}^{4}<6\;.
\end{equation}
%
 \subsubsection{New Inflation $\mathcal{N}>0$}
%
For potentials having $\mathcal{N}>0$, Lemma \ref{Lemma_1} and \eqref{2ndDerGenHarPot} give the fixed points
\begin{equation}\label{FP_NewInf}
\mathcal{P}:\,\,  \Psi\Phi=-\frac{\mathcal{M}}{3n}\frac{\sqrt{n^{2}\Phi^2+\mathcal{M}^{2}\mathcal{N}^{2}}}{\Phi}\quad\text{and}\quad \frac{\partial^{2}\Phi}{\partial\phi^{2}}=\frac{\mathcal{M}}{2n^{3}}\frac{\Psi\mathcal{M}^{2}\mathcal{N}^{2}}{\Phi\sqrt{n^{2}\Phi^2+\mathcal{M}^{2}\mathcal{N}^{2}}}
\end{equation}
subject to the Friedman constraint \eqref{FlatFC}. Since $\mathcal{N}>0$, we can write
\begin{equation}\label{2nDerHarPotV}
 \left(\frac{\partial^{2}\Phi}{\partial\phi^{2}}\right)_{\mathcal{P}}=-\frac{3}{2n^{2}}\Psi^{2}\Phi+\frac{\mathcal{M}^{2}}{6n^{2}\Phi}
\end{equation}
and then \eqref{EigenValue} gives
\begin{equation}\label{Eigenvalues_NewInf}
\omega_{-}=-3\left(1+\frac{2}{n}\right)\Phi^{2n}_{\mathcal{P}}+\frac{6}{n}-\frac{\mathcal{M}^{2}}{3n\Phi^{2}}
\end{equation}
so that the fixed points $\mathcal{P}$ are stable if $\omega_{-}<0$ which, as in the case of chaotic inflation, contains the inflationary solutions
\begin{equation}
 \omega_{-}<-2\left(1+\frac{1}{n}\right)-\frac{\mathcal{M}^{2}}{3n\Phi^{2}}\;.
\end{equation}
The case $n=2$, $C=\lambda^{\frac{1}{4}}$, was studied in \cite{KT10} and the corresponding dynamical system is given by
\begin{equation}
 \begin{aligned}
\Psi^{\prime}&=3\Psi^{3}-3\Psi-\Phi^{2}\mathcal{M}\sqrt{\Phi^{2}+\mathcal{N}^{2}\mathcal{M}^{2}} \\
\Phi\Phi^{\prime}&=\frac{1}{2}\left(3\Psi^{2}\Phi^{2}+\Psi\mathcal{M}\sqrt{\Phi^{2}+\mathcal{N}^{2}\mathcal{M}^{2}} \right)
 \end{aligned}
\end{equation}
subject to the flat Friedman constraint
\begin{equation}
 \Psi^{2}+\Phi^{4}=1.
\end{equation}
As in the quadratic case, considering the state vector $(\Psi,\Phi)$ with control parameter $\mathcal{M}$, the fixed points 
are solutions of
\begin{equation}
 \begin{aligned}
   3\Psi^{2}-3\Psi-\Phi^{2}\mathcal{M}\sqrt{\Phi^{2}+\mathcal{N}^{2}\mathcal{M}^{2}}&=0 \\
   3\Psi\Phi^{2}+\mathcal{M}\sqrt{\sqrt{1-\Psi^{2}}+\mathcal{N}^{2}\mathcal{M}^{2}}&=0   
 \end{aligned} 
\end{equation}
\begin{equation}
 3\Psi\sqrt{1-\Psi^{2}}+\mathcal{M}\sqrt{\sqrt{1-\Psi^{2}}+\mathcal{N}^{2}\mathcal{M}^{2}}=0.
\end{equation}
It is not possible, in general, to find explicitly the fixed points for this system. We shall then make a numerical stability analysis in Section~\ref{HArm}.
\section{Kinematic variables and scalar field sources}
The kinematical quantities associated with a timelike congruence in General Relativity were first introduced 
by Ehlers \cite{Ehl93} and Ellis \cite{Ell71}. Given a timelike vector field $\bmath{u}$, the unique tensors 
\begin{equation*}
h_{\alpha\beta}:=g_{\alpha\beta}+u_{\alpha}u_{\beta}
\end{equation*}
\begin{equation*}
 U_{\alpha\beta}=-u_{\alpha}u_{\beta}
\end{equation*}
project, at each point, tensors orthogonal and parallel to $\bmath{u}$, respectively. We will use the following notation:
\begin{equation*}
\dot{f}=u^{\sigma}\nabla_{\sigma}f\quad,\quad D_{\alpha}f=h^{\,\,\beta}_{\alpha}\nabla_{\beta}f
\end{equation*}
\begin{equation*}
 T_{<\alpha_{1}...\alpha_{p}>}=h^{\,\,\beta_{1}}_{\alpha_{1}}...h^{\,\,\beta_{p}}_{\alpha_{p}}T_{\beta_{1}...\beta_{p}}
\end{equation*}
so that the covariant derivative of a scalar field is decomposed into
\begin{equation*}
 \nabla_{\alpha}f=-u_{\alpha}\dot{f}+D_{\alpha}f\,.
\end{equation*}
The covariant derivative of $\bmath{u}$ can also be decomposed into its irreducible parts
\begin{equation*}
 \nabla_{\alpha}u_{\beta}=D_{\alpha}u_{\beta}-u_{\alpha}\dot{u}_{\beta}=\frac{1}{3}\theta h_{\alpha\beta}+\sigma_{\alpha\beta}+w_{\alpha\beta}-u_{\alpha}\dot{u}_{\beta}
\end{equation*}
with
\begin{equation}\label{trace}
\sigma_{\alpha\beta}=\sigma_{(\alpha\beta)};~~\sigma^{\alpha}_{\,\alpha}=0;\,\,\,\sigma^2=\frac{1}{2}\sigma_{\alpha\beta}\sigma^{\alpha\beta};
\,\,\,\sigma_{\alpha\beta}u^{\beta}=0;\,\,\,
\omega_{\alpha\beta}=\omega_{[\alpha\beta]};\,\,\,\omega_{\alpha\beta}u^{\beta}=0,
\end{equation}
where the curly (resp. squared) brackets denote symmetrization (resp. anti-symmetrization) of a tensor and
\begin{equation}
\begin{aligned}
&\theta=\nabla_{\alpha}u^{\alpha}\\
&\dot{u}_{\alpha}=\nabla_{\beta}u_{\alpha}u^{\beta}\\
&\sigma_{\alpha\beta}=\nabla_{(\beta}u_{\alpha)}-\frac{1}{3}\theta h_{\alpha\beta}+\dot{u}_{(\alpha}u_{\beta)}\\
& w_{\alpha\beta}=\nabla_{[\beta}u_{\alpha]}+\dot{u}_{[\alpha}u_{\beta]}.
\end{aligned}
\end{equation}
The tensor $\omega_{\alpha\beta}$ is called the vorticity, $\sigma_{\alpha\beta}$ the shear
and the expansion tensor is defined as
\begin{equation}\label{Exp_Tensor}
\theta_{\alpha\beta}=\sigma_{\alpha\beta}+\frac{1}{3}{\theta h_{\alpha\beta}}.
\end{equation}
In the following, we shall also use the {\it Hubble function} defined by
\begin{equation}
 H=\frac{1}{3}\theta.
\end{equation}
To deduce the propagation equations for the gauge-invariant perturbation variables in Section 4 it is useful to recall the following relations between commutators of spatial and time 
derivatives acting on scalars \cite{EBH90}
\begin{equation}\label{Ddot}
D_{\alpha}\dot{f}=\frac{1}{3}\theta D_{\alpha}f+\left(\sigma^{\,\beta}_{\alpha}+\omega^{\,\beta}_{\alpha}\right)D_{\beta}f+h^{\,\mu}_{\alpha}(\dot{D_{\mu}f})-\dot{f}\dot{u}_{\alpha}
\end{equation}
\begin{equation}
 D_{[\alpha}D_{\beta]}f=-\omega_{\alpha\beta}\dot{f}.
\end{equation}
The more general decomposition of the energy-momentum tensor field with respect to $\bmath{u}$,  is given by
\begin{equation}\label{imperfectfluid}
 T_{\alpha\beta}=\rho u_{\alpha}u_{\beta}+ph_{\alpha\beta}+2u_{(\alpha}q_{\beta)}+\pi_{\alpha\beta}\;,
\end{equation}
where $q_{\alpha}$ is the energy-transfer function and $\pi_{\alpha\beta}$ is the anisotropic stress with $u^{\alpha}q_{\alpha}=0$, 
$u^{\alpha}\pi_{\alpha\beta}=0$, $\pi_{\alpha\beta}=\pi_{(\alpha\beta)}$, $\pi^{\alpha}_{\,\,\,\alpha}=0$. 
In the multicomponent case, we assume that the total matter energy-momentum tensor is the sum of the individual energy-momentum tensors for the components plus an interaction term $\Pi$ between these components: 
\begin{equation}\label{totalEM}
 T_{\alpha\beta}:=\sum^{N}_{A=1}T^{A}_{\alpha\beta}+g_{\alpha\beta}\Pi.
\end{equation}
Moreover, given the preferred future directed time-like vector field $\bmath{u}$, 
the EFEs are expressed through the  Ricci identities applied to $\bmath{u}$ and 
the Bianchi identities in terms of the kinematic quantities, see e.g. \cite{WE97}. 
\subsection{Scalar fields}
It was shown by Madsen \cite{Mad88} that, if we require $\phi_{A}$ to be locally constant on a spacelike hypersurface, 
$D^{\mu}_{A}\phi_{A}=0$ and $\nabla^{\mu}\phi_{A}\neq0$, such that $\nabla^{\mu}\phi_{A}$ defines uniquely a time-like vector field 
orthogonal to the surfaces $\phi_{A}=const$ with
\begin{equation}
(\nabla_{\lambda}\phi_{A})(\nabla^{\lambda}\phi_{A})<0.
\end{equation}
From the local decomposition of the covariant derivative ($\nabla^{\mu}\phi_{A}=-u^{\mu}_{A}\psi_{A}$), we find that
\begin{equation}\label{velocityscalarfield}
 u^{\mu}_{A}:=-\frac{1}{\psi_{A}}\left(\nabla^{\mu}\phi_{A}\right)
\end{equation}
is a unitary time-like vector field, with $\psi_{A}$ the momentum-density defined by
\begin{equation}
\label{psiA}
\psi_{A}:=\dot{\phi}_{A}=u^{\lambda}_{A}\nabla_{\lambda}\phi_{A}\;.
\end{equation}
Due to the uniqueness of $u^{\mu}_{A}$, we can use the $1+3$ covariant decomposition, in this case, taking the local projector on 
the spacelike hypersurfaces of constant $\phi_{A}$, in the form
\begin{equation}
h^{A}_{\alpha\beta}\equiv g_{\alpha\beta}+\frac{1}{\psi^{2}_{A}}\left(\nabla_{\alpha}\phi_{A}\right)\left(\nabla_{\beta}\phi_{A}\right).
\end{equation} 
Then, the energy-momentum tensor of each scalar field has the perfect fluid form
\begin{equation}\label{E-Mscalarfield}
T^{A}_{\mu\nu}=\left[\frac{1}{2}\psi^{2}_{A}+\mathcal{V}(\phi_{A})\right]u^{A}_{\mu}u^{A}_{\nu}+\left[\frac{1}{2}\psi^{2}_{A}-\mathcal{V}(\phi_{A})\right]h^{A}_{\mu\nu}
\end{equation}
with the identifications
\begin{equation}
\begin{aligned}
 \rho_{A}&=\frac{1}{2}\psi^{2}_{A}+\mathcal{V}_{A} \\
    p_{A}&=\frac{1}{2}\psi^{2}_{A}-\mathcal{V}_{A}
\end{aligned}
\end{equation} 
and, from the total energy-momentum tensor \eqref{totalEM}, we have $\Pi=-\mathcal{W}$. 
Finally, decomposing each $\bmath{u}_{A}$ into components orthogonal and parallel to $\mathbf{u}$, 
it follows that \cite{EU99}
\begin{equation}\label{linmulti}
 u^{\mu}_{A}=\Gamma_{A}\left(u^{\mu}+v^{\mu}_{A}\right)\quad,\quad\Gamma_{A}=\frac{1}{\sqrt{1-v^{2}_{{A}}}}\quad,\quad u^{\mu}v_{\mu}=0\;.
\end{equation}
\subsection{Characterization of FLRW models}
The particular case of FLRW models is characterized by
\begin{equation}\label{Kin_FL}
\dot{u}_{\mu}=0,\quad \sigma_{\mu\nu}=0,\quad\omega_{\mu\nu}=0,
\end{equation}
and by the fact that spatial gradients of scalars are zero, in particular
\begin{equation}
D_{\mu}\phi=0\quad,\quad D_{\mu}\psi=0,\quad, \quad D_{\mu}H=0.
\end{equation}
Furthermore, the symmetry of the spacetime forces the energy-momentum tensor \eqref{imperfectfluid} 
to have the algebraic form of a perfect fluid with
\begin{equation}
 q_{\alpha}=0,\quad\pi_{\alpha\beta}=0.
\end{equation}
\section{Evolution of inhomogeneity variables}
In this section, we shall construct a system of differential equations governing the dynamical 
behaviour of linear scalar perturbations of FLRW models with $N$ interacting scalar fields which 
generalises the system derived in \cite{BED92}. We shall take the perturbative covariant and 
gauge-invariant approach \cite{EB89,EBH90,BDE92} which is constructed through the 1+3 covariant formalism, 
see e.g. \cite{EvH98} and references therein. 
\subsection{Covariant and gauge-invariant variables}
We will use the following definitions for the covariant and gauge-invariant variables \cite{BED92}
\begin{equation}
\label{deltaalpha}
\Delta_{\alpha}:=a(t)\frac{D_{\alpha}\psi}{\psi},\quad \Delta^{A}_{\alpha}:=a(t)\frac{D_{\alpha}\psi_{A}}{\psi_{A}},\quad \mathcal{Z}_{\alpha}:=a(t)D_{\alpha}\theta,\quad  v^{A}_{\alpha}=-\frac{D_{\alpha}\phi_{A}}{\psi_{A}}
\end{equation}
which represent, respectively, the total and each scalar field {\it comoving fractional momentum-density spatial gradients}, 
the {\it comoving spatial gradient of the expansion}, and the {\it velocity perturbations}. 
From (\ref{psiA}), we can define the {\it effective scalar field momentum-density}
\begin{equation}
 \psi^{2}=\sum^{N}_{A=1}\psi^{2}_{A}
\end{equation}
which leads to the following relations between the perturbations variables 
\begin{equation}\label{TotDens_SingDens}
 \Delta_{\alpha}=\frac{1}{\psi^2}\sum^{N}_{A=1}\psi^{2}_{A}\Delta^{A}_{\alpha}.
\end{equation}
The variables (\ref{deltaalpha})-(\ref{TotDens_SingDens}) contain information about three types of inhomogeneities and, similarly to the standard non-local decomposition, we follow \cite{EBH90} defining a local decomposition for comoving vector gradients as
\begin{equation}
 a(t)D_{\alpha}X_{\beta}:=X_{\alpha\beta}=\frac{1}{3}Xh_{\alpha\beta}+\Sigma_{\alpha\beta}+\Omega_{\alpha\beta}\;,
\end{equation}
where $X=a(t)D^{\alpha}X_{\alpha}$, $\Sigma_{\alpha\beta}=a(t)D_{(\alpha}X_{\beta)}$ and $\Omega_{\alpha\beta}=a(t)D_{[\alpha}X_{\beta]}$.
In this way, local scalar variables can be obtained by taking the divergence of quantities (\ref{deltaalpha}) as
\begin{equation}\label{GIscalarfields}
 \Delta:=a(t)D^{\alpha}\Delta_{\alpha},\quad \mathcal{Z}:=a(t)D^{\alpha}\mathcal{Z}_{\alpha},\quad v_{A}:=a(t)D^{\alpha}v^{A}_{\alpha}.
\end{equation}
Also, we shall refer to the cosmological model with a self-interacting scalar field of potential $\mathcal{V}(\phi)$ as 
{\it close to a FLRW-nonlinear scalar field universe} in some open set if, for some suitably small constants $\varepsilon_1\ll 1$, $\varepsilon_2\ll 1$ and 
$\varepsilon_3\ll 1$, the following inequalities hold
\begin{equation}
\frac{|D_{\alpha}\psi|}{H\psi}<\varepsilon_1,\qquad\frac{|D_{\alpha}\phi|}{H^{2}\psi}<\varepsilon_2,\qquad \frac{|D_{\alpha}H|}{H^2}<\varepsilon_3,
\end{equation}
where $|D_{\alpha}\psi|=(D_{\alpha}\psi D^{\alpha}\psi)^{\frac{1}{2}}$. We note that the constants $\varepsilon_1$, $\varepsilon_2$ and $\varepsilon_3$ are taken to be different since the perturbation variables don't have the same dimensions.
\subsection{Linearised equations}
Let $\bmath{u}$ be a time-like future-directed vector-field associated with the 4-velocity field of the total matter and 
$\bmath{u}_{A}$ the orthogonal vectors to the surfaces $\phi_{A}=const.$, which are tilted with respect to $\bmath{u}$ 
by a small angle so that in \eqref{linmulti}
\begin{equation*}
 \Gamma_{A}\approx1.
\end{equation*}
Then, to first order, the relation between each $\bmath{u}_{A}$ and $\bmath{u}$ in the local rest frame defined by the 
latter vector field is given by \cite{Dun91}
\begin{equation}
 \bmath{u}_{A}\approx\bmath{u}+\bmath{v}_{A}.
\end{equation}
By a small angle it is meant that $\bmath{u}_{A}$ is time-like, which validates the space-like vector field $\bmath{v}_{A}$ as being a small deviation from the background solution. Thus, in a FLRW background, $\bmath{v}_{A}=0$, so that $\bmath{v}_{A}$ will be a gauge-invariant perturbation variable. The total energy-momentum tensor for $N$ minimally coupled scalar-fields, in this frame, is given by \cite{Dun91,DBE92}
\begin{equation}
 T_{\alpha\beta}=\rho u_{\alpha}u_{\beta}+ph_{\alpha\beta}+2u_{(\alpha}q_{\beta)}+\pi_{\alpha\beta}-g_{\alpha\beta}\mathcal{W},
\end{equation}
where ${\cal W}=-\Pi$, $u_\alpha$ denotes the components of $\bmath{u}$ and, to first order,
\begin{equation}
 \rho=\sum^{N}_{A=1}\left[\frac{1}{2}\psi^{2}_{A}+\mathcal{V}_{A}\right],\quad p=\sum^{N}_{A=1}\left[\frac{1}{2}\psi^{2}_{A}-\mathcal{V}_{A}\right],
\quad q_{\alpha}=\sum^{N}_{A=1}\psi^{2}_{A}v^{A}_{\alpha},\quad\pi_{\alpha\beta}=0.
\end{equation}
Also, to first order, we have that
\begin{equation}\label{evol_q}
\begin{aligned}
 \dot{q}^{<\mu>}&=\sum^{N}_{A=1}\left[2\frac{\dot{\psi}_{A}}{\psi_{A}}v^{\mu}_{A}+\dot{v}^{<\mu>}_{A}\right]\psi^{2}_{A} \\
                &=-6H\sum^{N}_{A=1}\psi^{2}_{A}v^{\mu}_{A}-2\sum^{N}_{A=1}\psi_{A}\left[\frac{d\mathcal{V}_{A}}{d\phi_{A}}+\frac{\partial\mathcal{W}}{\partial\phi_{A}}\right]v^{\mu}_{A}+\sum^{N}_{A=1}\psi^{2}_{A}\dot{v}^{<\mu>}_{A} 
\end{aligned}
\end{equation} 
and
\begin{equation}
 \nabla^{\mu}\mathcal{W}=\sum^{N}_{A=1}\frac{\partial\mathcal{W}}{\partial\phi_{A}}\left[-u^{\mu}\psi_{A}+D_{\mu}\phi_{A}\right]=-\sum^{N}_{A=1}\psi_{A}\frac{\partial\mathcal{W}}{\partial\phi_{A}}\left[u^{\mu}+v^{\mu}_{A}\right].
\end{equation}
Then, the exact linearised evolution equations around a FLRW-scalar field model in the frame defined by $\bmath{u}$, are given by a wave equation in the $1+3$ covariant form for the effective momentum-density
\begin{equation}\label{linear_wave_eq_eff_field}
\dot{\psi}=-3H\psi-\frac{1}{\psi}\sum^{N}_{A=1}\psi_{A}\left[\frac{d\mathcal{V}_{A}}{d\phi_{A}}+\frac{\partial\mathcal{W}}{\partial\phi_{A}}\right]-\frac{1}{\psi}\sum^{N}_{A=1}\psi^{2}_{A}v_{A}\quad,\quad\psi\neq0\;. 
\end{equation}
The momentum conservation equation is
\begin{equation*}
 \psi^2\dot{u}^{\mu}=-\psi(D^{\mu}\psi)-\sum^{N}_{A=1}\psi_{A}\left[\frac{d\mathcal{V}_{A}}{d\phi_{A}}+\frac{\partial\mathcal{W}}{\partial\phi_{A}}\right]v^{\mu}_{A}-4H\sum^{N}_{A=1}\psi^{2}_{A}v^{\mu}_{A}-\dot{q}^{<\mu>}
\end{equation*}
which, after multiplying by the scale factor $a(t)$ and using \eqref{evol_q}, simplifies to
\begin{equation}\label{TOTMcons}
 a(t)\dot{u}^{\mu}=-\Delta^{\mu}+\frac{a(t)}{\psi^2}\sum^{N}_{A=1}\psi_{A}\left[\frac{d\mathcal{V}_{A}}{d\phi_{A}}+\frac{\partial\mathcal{W}}{\partial\phi_{A}}\right]v^{\mu}_{A}+2H\frac{a(t)}{\psi^2}\sum^{N}_{A=1}\psi^{2}_{A}v^{\mu}_{A}-\frac{a(t)}{\psi^2}\sum^{N}_{A=1}\psi^{2}_{A}\dot{v}^{<\mu>}_{A}\;.
\end{equation}
In turn, the linearised Raychaudhuri equation for scalar fields is (see also e.g. \cite{BED92})
\begin{equation}\label{Raychauduriscalarfields}
 3\dot{H}=-3H^{2}-\psi^{2}+\sum^{N}_{A=1}\mathcal{V}_{A}+\mathcal{W}+D_{\sigma}\dot{u}^{\sigma}.
\end{equation}
The $N$ linearised wave equations in the 1+3 covariant form, for each scalar field, are
\begin{equation}\label{K-G3-1}
 \dot{\psi}_{A}=-3H\psi_{A}-\frac{\partial \mathcal{V}_{A}}{\partial\phi_{A}}-\frac{\partial\mathcal{W}}{\partial\phi_{A}}-\psi_{A}v_{A}\quad,\quad\psi_{A}\neq0\quad,\quad A=1,...,N
\end{equation}
and the first order equation associated with the momentum conservation equation, for each scalar field, is
\begin{equation}\label{lmcescalarfields}
a(t)\dot{u}_{\alpha}+\Delta^{A}_{\alpha}+\left\{2\frac{\dot{\psi}_{A}}{\psi_{A}}+4H+\frac{1}{\psi_{A}}\left[\frac{\partial \mathcal{W}}{\partial\phi_{A}}+\frac{d\mathcal{V}_{A}}{d\phi_{A}}\right]\right\}a(t)v^{A}_{\alpha}+a(t)\dot{v}^{A}_{<\alpha>}=0\;.
\end{equation}
\subsubsection{Evolution equation for $\Delta_\mu$}
To obtain the evolution equation for $\Delta_{\mu}$, we take spatial gradients of equation \eqref{linear_wave_eq_eff_field} 
and keep the first order terms to get
\begin{small}\begin{equation}\label{spat_grad_eff_field}
\begin{aligned}
 a(t)\frac{D_{\mu}\dot{\psi}}{\psi}=&-\mathcal{Z}_{\mu}-\left[3H-\frac{1}{\psi^2}\sum^{N}_{A=1}\psi_{A}\left(\frac{d\mathcal{V}_{A}}{d\phi_{A}}+\frac{\partial\mathcal{W}}{\partial\phi_{A}}\right)\right]\Delta_{\mu}-\frac{1}{\psi^2}\sum^{N}_{A=1}\psi_{A}\left(\frac{d\mathcal{V}_{A}}{d\phi_{A}}+\frac{\partial\mathcal{W}}{\partial\phi_{A}}\right)\Delta^{A}_{\mu} \\
                                    &-\frac{a(t)}{\psi^2}D_{\mu}\sum_{A=1}\psi^{2}_{A}v_{A}+\frac{a(t)}{\psi^{2}}\sum^{N}_{A=1}\psi^{2}_{A}\frac{d^{2}\mathcal{V}_{A}}{d\phi^{2}_{A}}v^{A}_{\mu}+\frac{a(t)}{\psi^{2}}\sum^{N}_{A,B=1}\psi_{A}\psi_{B}\frac{\partial^{2}\mathcal{W}}{\partial\phi_{B}\partial\phi_{A}}v^{B}_{\mu}.
\end{aligned}
\end{equation}              \end{small}
Then, using the linearised relation $(\ref{Ddot})$ for the effective momentum-density variable $\psi$, we find 
\begin{equation*}
 D_{\mu}\dot{\psi}=H D_{\mu}\psi+h_{\mu}^{\,\,\nu}\dot{\left(D_{\nu}\psi\right)}-\dot{\psi}\dot{u}_{\mu}
\end{equation*}
together with the relation
\begin{equation*}
 a(t)\frac{h_{\mu}^{\,\,\nu}\dot{\left(D_{\nu}\psi\right)}}{\psi}=\dot{\Delta}_{<\mu>}-\left[H-\frac{\dot{\psi}}{\psi}\right]\Delta_{\mu}
\end{equation*}
giving 
\begin{equation*}
 a(t)\frac{D_{\mu}\dot{\psi}}{\psi}=\dot{\Delta}_{<\mu>}+\frac{\dot{\psi}}{\psi}\Delta_{\mu}-a(t)\frac{\dot{\psi}}{\psi}\dot{u}_{\mu}\;,
\end{equation*}
which, after inserting into \eqref{spat_grad_eff_field}, finally gives
\begin{small}\begin{equation}\label{Deltavector}
\begin{aligned}
 \dot{\Delta}_{<\mu>}=&-\mathcal{Z}_{\mu}-\left[\frac{\dot{\psi}}{\psi}+3H-\frac{1}{\psi^2}\sum^{N}_{A=1}\psi_{A}\left(\frac{d\mathcal{V}_{A}}{d\phi_{A}}+\frac{\partial\mathcal{W}}{\partial\phi_{A}}\right)\right]\Delta_{\mu}-\frac{1}{\psi^2}\sum^{N}_{A=1}\psi_{A}\left(\frac{d\mathcal{V}_{A}}{d\phi_{A}}+\frac{\partial\mathcal{W}}{\partial\phi_{A}}\right)\Delta^{A}_{\mu} \\
                                    &-\frac{a(t)}{\psi^2}D_{\mu}\sum^{N}_{A=1}\psi^{2}_{A}v_{A}+\frac{a(t)}{\psi^{2}}\sum^{N}_{A=1}\psi^{2}_{A}\frac{d^{2}\mathcal{V}_{A}}{d\phi^{2}_{A}}v^{A}_{\mu}+\frac{a(t)}{\psi^{2}}\sum^{N}_{A,B=1}\psi_{A}\psi_{B}\frac{\partial^{2}\mathcal{W}}{\partial\phi_{B}\partial\phi_{A}}v^{B}_{\mu}+a(t)\frac{\dot{\psi}}{\psi}\dot{u}_{\mu}.
\end{aligned}
\end{equation}              \end{small}

\subsubsection{Evolution equation for $\mathcal{Z}_\mu$}
The evolution equation for the perturbation variable $\mathcal{Z}_{\mu}$ is found by taking spatial gradients of the 
linearized Raychaudhuri equation $(\ref{Raychauduriscalarfields})$ which, after multiplication by the scale factor $a(t)$, 
reads
\begin{equation}\label{SG_EVOL_Hubble}
 3a(t)D_{\mu}\dot{H}=-2H\mathcal{Z}_{\mu}-2\psi^2\Delta_{\mu}-a(t)\sum^{N}_{A=1}\psi_{A}\left(\frac{d\mathcal{V}_{A}}{d\phi_{A}}+\frac{\partial\mathcal{W}}{\partial\phi_{A}}\right)v^{A}_{\mu}+a(t)D_{\mu}D_{\sigma}\dot{u}^{\sigma}.
\end{equation}
Now, using $(\ref{Ddot})$ we get
\begin{equation*}
 D_{\mu}\dot{H}=H D_{\mu}H+h_{\mu}^{\,\,\nu}\dot{\left(D_{\nu}H\right)}-\dot{H}\dot{u}_{\mu}\;,
\end{equation*}
which together with
\begin{equation*}
3 a(t) h_{\mu}^{\,\,\nu}\dot{\left(D_{\nu}H\right)}=\dot{\mathcal{Z}}_{<\mu>}-H\mathcal{Z}_{\mu}\;,
\end{equation*}
gives
\begin{equation*}
3a(t) D_{\mu}\dot{H}=\dot{\mathcal{Z}}_{<\mu>}-3\dot{H}a(t)\dot{u}_{\mu}\;.
\end{equation*}
Finally, inserting the last equation into \eqref{SG_EVOL_Hubble} gives
\begin{equation}\label{Zscalar}
\dot{\mathcal{Z}}_{<\mu>}=-2H\mathcal{Z}_{\mu}-2\psi^2\Delta_{\mu}-a(t)\sum^{N}_{A=1}\psi_{A}\left(\frac{d\mathcal{V}_{A}}{d\phi_{A}}+\frac{\partial\mathcal{W}}{\partial\phi_{A}}\right)v^{A}_{\mu}+a(t)D_{\mu}D_{\sigma}\dot{u}^{\sigma}+3\dot{H}a(t)\dot{u}_{\mu}\;.
\end{equation}
\subsubsection{Evolution equation for $\Delta^{A}_\mu$ and $v^{A}_\mu$}
The evolution equation for each variable $\Delta^{A}_{\mu}$ is obtained by taking spatial gradients of equation $(\ref{K-G3-1})$ and keeping first order terms as
\begin{equation*}
 a(t)\frac{D_{\mu}\dot{\psi}_{A}}{\psi_{A}}=-3H\Delta^{A}_{\mu}-\mathcal{Z}_{\mu}-a(t)D_{\mu}v_{A}+a(t)\frac{d^{2}\mathcal{V}_{A}}{d\phi^{2}_{A}}v^{A}_{\mu}+a(t)\sum^{N}_{B=1}\frac{\psi_{B}}{\psi_{A}}\frac{\partial^{2}\mathcal{W}}{\partial\phi_{B}\partial\phi_{A}}v^{B}_{\mu}
\end{equation*}
and
\begin{equation}\label{densgradscalarfields}
\begin{aligned}
 \dot{\Delta}^{A}_{<\mu>}=&-\left[\frac{\dot{\psi}_{A}}{\psi_{A}}+3H\right]\Delta^{A}_{\mu}-\mathcal{Z}_{\mu}-a(t)D_{\mu}D_{\sigma}v^{\sigma}_{A}+a(t)\frac{d^{2}\mathcal{V}_{A}}{d\phi^{2}_{A}}v^{A}_{\mu} \\
                          &+a(t)\sum^{N}_{B=1}\frac{\psi_{B}}{\psi_{A}}\frac{\partial^{2}\mathcal{W}}{\partial\phi_{B}\partial\phi_{A}}v^{B}_{\mu}+a(t)\frac{\dot{\psi}_{A}}{\psi_{A}}\dot{u}_{\mu}.
\end{aligned}
\end{equation}
To get the evolution equation for the velocity perturbation variables, we can use the relation \eqref{Ddot} 
for each scalar field $\phi_{A}$ and get
\begin{equation}\label{densgradsMD}
 a(t)\dot{v}^{A}_{<\mu>}=-a(t)\left[H+\frac{\dot{\psi}_{A}}{\psi_{A}}\right]v^{A}_{\mu}-\Delta^{A}_{\mu}-a(t)\dot{u}_{\mu}\quad,\quad\psi_{A}\neq 0.
\end{equation}
This equation is identical to the momentum conservation equation for the $A$th component \eqref{lmcescalarfields}, 
after using the background nonlinear wave equation in the $1+3$ form.

\subsection{Equations in the Energy frame}
In order to close the system of evolution and constraint equations, we need to fix the frame for which we are constructing perturbation variables. Furthermore, this choice of frame must ensure that the perturbation variables are gauge invariant. 
A suitable choice is the energy frame defined through
\begin{equation}\label{EFcond}
 q^{\mu}=0\quad,\quad\dot{q}^{<\mu>}=0.
\end{equation}
Thus, if we choose $\bmath{u}=\bmath{u}_{E}$ to be the energy frame, then
\begin{equation*}
\sum^{N}_{A=1}\psi^{2}_{A}v^{\mu}_{A}=0
\end{equation*}
and
\begin{equation*}
 \sum_{A=1}^{N}\psi^{2}_{A}\dot{v}^{<\mu>}_{A}=2\sum_{A=1}^{N}\psi_{A}\left[\frac{d\mathcal{V}_{A}}{d\phi_{A}}+\frac{\partial\mathcal{W}}{\partial\phi_{A}}\right]v^{\mu}_{A}\;.
\end{equation*}
In this frame, \eqref{TOTMcons} reads
\begin{equation}\label{ACEF}
 a(t)\dot{u}^{\mu}=-\Delta^{\mu}-\frac{a(t)}{\psi^2}\sum^{N}_{A=1}\psi_{A}\left[\frac{d\mathcal{V}_{A}}{d\phi_{A}}+\frac{\partial\mathcal{W}}{\partial\phi_{A}}\right]v^{\mu}_{A}
\end{equation}
and the first order equation for the divergence of the acceleration is 
\begin{equation}\label{DACEF}
 a^{2}D_{\sigma}\dot{u}^{\sigma}=-\Delta-\frac{a(t)}{\psi^2}\sum^{N}_{A=1}\psi_{A}\left[\frac{d\mathcal{V}_{A}}{d\phi_{A}}+\frac{\partial\mathcal{W}}{\partial\phi_{A}}\right]v_{A}\;.
\end{equation}
By using \eqref{ACEF} and \eqref{DACEF} into \eqref{Deltavector}, \eqref{Zscalar}, \eqref{densgradscalarfields} and \eqref{densgradsMD}, and after multiplying the resulting equation by the scale factor and taking the spatial divergence 
of these equations, we finally obtain:
\begin{prop}
The evolution of first order scalar perturbations on FLRW-scalar fields background with arbitrary smooth potentials in the energy frame, is given by the following system of equations for the variables $\Delta$, $\mathcal{Z}$, $\Delta_{A}$ and $v_{A}$:
\begin{equation}\label{scalarDelta}
\begin{aligned}
 \dot{\Delta}=&-\mathcal{Z}+\left[3H+\frac{3}{\psi^2}\sum^{N}_{A=1}\psi_{A}\left(\frac{d\mathcal{V}_{A}}{d\phi_{A}}+\frac{\partial\mathcal{W}}{\partial\phi_{A}}\right)\right]\Delta-\frac{1}{\psi^2}\sum^{N}_{A=1}\psi_{A}\left(\frac{d\mathcal{V}_{A}}{d\phi_{A}}+\frac{\partial\mathcal{W}}{\partial\phi_{A}}\right)\Delta_{A} \\
                      &+a(t)\left[3H+\frac{1}{\psi^2}\sum^{N}_{C=1}\psi_{C}\left(\frac{d\mathcal{V}_{C}}{d\phi_{C}}+\frac{\partial\mathcal{W}}{\partial\phi_{C}}\right)\right]\frac{1}{\psi^2}\sum^{N}_{A=1}\psi_{A}\left(\frac{d\mathcal{V}_{A}}{d\phi_{A}}+\frac{\partial\mathcal{W}}{\partial\phi_{A}}\right)v_{A} \\
                      &+\frac{a(t)}{\psi^{2}}\sum^{N}_{A=1}\psi^{2}_{A}\frac{d^{2}\mathcal{V}_{A}}{d\phi^{2}_{A}}v_{A}+\frac{a(t)}{\psi^{2}}\sum^{N}_{A,B=1}\psi_{A}\psi_{B}\frac{\partial^{2}\mathcal{W}}{\partial\phi_{B}\partial\phi_{A}}v_{B}
\end{aligned}
\end{equation}

\begin{equation}\label{scalarZ}
\dot{\mathcal{Z}}=-2H\mathcal{Z}-\left[3\dot{H}+2\psi^2+D^{2}\right]\Delta-\left[\psi^2+3\dot{H}+D^{2}\right]\frac{a(t)}{\psi^2}\sum^{N}_{A=1}\psi_{A}\left(\frac{d\mathcal{V}_{A}}{d\phi_{A}}+\frac{\partial\mathcal{W}}{\partial\phi_{A}}\right)v_{A}
\end{equation}

\begin{equation}
\begin{aligned}
 \dot{\Delta}_{A}=&-\mathcal{Z}+\left[3H+\frac{1}{\psi_{A}}\left(\frac{d\mathcal{V}_{A}}{d\phi_{A}}+\frac{\partial\mathcal{W}}{\partial\phi_{A}}\right)\right]\Delta-\left[\frac{1}{\psi_{A}}\left(\frac{d\mathcal{V}_{A}}{d\phi_{A}}+\frac{\partial\mathcal{W}}{\partial\phi_{A}}\right)\right]\Delta_{A} \\
                  &-a(t)D^{2}v_{A}+a(t)\frac{d^{2}\mathcal{V}_{A}}{d\phi^{2}_{A}}v_{A}+a(t)\sum^{N}_{B=1}\frac{\psi_{B}}{\psi_{A}}\frac{\partial^{2}\mathcal{W}}{\partial\phi_{B}\partial\phi_{A}}v_{B} \\
                  &+\left[3H+\frac{1}{\psi_{A}}\left(\frac{d\mathcal{V}_{A}}{d\phi_{A}}+\frac{\partial\mathcal{W}}{\partial\phi_{A}}\right)\right]\frac{a(t)}{\psi^2}\sum^{N}_{B=1}\psi_{B}\left[\frac{d\mathcal{V}_{B}}{d\phi_{B}}+\frac{\partial\mathcal{W}}{\partial\phi_{B}}\right]v_{B}
\end{aligned}
\end{equation}

\begin{equation}\label{scalarv}
 a(t)\dot{v}_{A}=\Delta-\Delta_{A}+a(t)\left[2H+\frac{1}{\psi_{A}}\left(\frac{d\mathcal{V}_{A}}{d\phi_{A}}+\frac{\partial\mathcal{W}}{\partial\phi_{A}}\right)\right]v_{A}+\frac{a(t)}{\psi^2}\sum^{N}_{B=1}\psi_{B}\left[\frac{d\mathcal{V}_{B}}{d\phi_{B}}+\frac{\partial\mathcal{W}}{\partial\phi_{B}}\right]v_{B},
\end{equation}
together with the background equations \eqref{K-G3-1background}-\eqref{Friedmann}.
\end{prop}
\subsection{Equations in Relative variables}
In order to simplify the notation and the calculations in the multiple scalar field case, let
\begin{equation}
 \alpha_{A}:=\frac{1}{\psi_A}\left(\frac{d\mathcal{V}_{A}}{d\phi_{A}}+\frac{\partial\mathcal{W}}{\partial\phi_{A}}\right)\quad,\quad \beta^{2}_{A}=\frac{\psi^{2}_{A}}{\psi^{2}}
\end{equation}
so that
\begin{equation}\label{Der_alpha}
 \dot{\alpha}_{A}=\left(3H+\alpha_{A}\right)\alpha_{A}+\frac{d^{2}\mathcal{V}_{A}}{d\phi^{2}_{A}}+\frac{\partial^{2}\mathcal{W}}{\partial\phi^{2}_{A}}+\sum^{N}_{B\neq A}\frac{\psi_{B}}{\psi_{A}}\frac{\partial^{2}\mathcal{W}}{\partial\phi_B\partial\phi_A}
\end{equation}
\begin{equation}\label{Der_beta}
\begin{aligned}
 \dot{\left(\beta^{2}_{A}\right)}
                    &=-2\sum^{N}_{B\neq A}\left(\alpha_{A}-\alpha_{B}\right)\beta^{2}_{A}\beta^{2}_{B} 
\end{aligned}
\end{equation}
and, from the fact that $\beta_{A}$ verifies 
${\displaystyle \sum^{N}_{A=1}\beta^{2}_{A}=1}$, we get
\begin{equation}
 \sum^{N}_{A=1}\dot{\left(\beta^{2}_{A}\right)}=0\;.
\end{equation}
Defining {\em relative perturbation variables} as
\begin{equation}
 \Delta_{[AB]}=\Delta_{A}-\Delta_{B}\quad,\quad v_{[AB]}=v_{A}-v_{B},
\end{equation}
then \eqref{TotDens_SingDens} reads 
\begin{equation}\label{RelDeltas}
 \Delta_{A}=\Delta+\sum^{N}_{B\neq A}\beta^{2}_{B}\Delta_{[AB]}.
\end{equation}
For the relative velocity perturbation variables, using the energy-frame condition \eqref{EFcond}, it follows that
\begin{equation}
 v_{A}=\sum^{N}_{B\neq A}\beta^{2}_{B}v_{[AB]},
\end{equation}
and, using the above variables, we obtain from equations \eqref{scalarDelta} and \eqref{scalarZ}, 
\begin{eqnarray}\label{FO_Delta}
 \dot{\Delta}&=&A(t)\Delta-\mathcal{Z}-\sum^{N}_{A=1}\sum^{N}_{B>A}B_{AB}(t)\Delta_{[AB]}+a(t)\sum^{N}_{A=1}\sum^{N}_{B>A}C_{AB}(t)v_{[AB]}\\
\label{FO_Z}
 \dot{\mathcal{Z}}&=&-2H\mathcal{Z}-\left[3\dot{H}+2\psi^2+D^{2}\right]\Delta \\
                   &-& a(t)\sum^{N}_{A=1}\sum^{N}_{B>A}\left(\alpha_{A}-\alpha_{B}\right)\beta^{2}_{A}\beta^{2}_{B}\left[\psi^2+3\dot{H}+D^{2}\right]v_{[AB]}\;.
\end{eqnarray}
To get the evolution equation for $\Delta_{[AB]}$ we take the difference $\dot{\Delta}_{[AB]}=\dot{\Delta}_A-\dot{\Delta}_B$, which upon using \eqref{RelDeltas} and the following relation
\begin{equation*}
\begin{aligned}
 (\alpha_A-\alpha_B)\Delta-\alpha_A\Delta_A+\alpha_B\Delta_B=&-\left(\alpha_{A}\beta^{2}_{B}+\alpha_{B}\beta^{2}_{A}\right)\Delta_{[AB]} \\
                                                            &-\alpha_{A}\sum^{N}_{C\neq A,B}\beta^{2}_{C}\Delta_{[AC]}-\alpha_{B}\sum^{N}_{C\neq A,B}\beta^{2}_{C}\Delta_{[BC]}  
\end{aligned}
\end{equation*}
gives
\begin{equation}\label{scalarDeltasys3}
\begin{aligned}
 \dot{\Delta}_{[AB]}=&-D_{AB}(t)\Delta_{[AB]}+\sum^{N}_{C\neq A,B}\left(\alpha_{A}\beta^{2}_{C}\Delta_{[AC]}-\alpha_{B}\beta^{2}_{C}\Delta_{[BC]}\right) \\
                     &+a(t)\sum^{N}_{C\neq A,B}\left[F_{AC}(t)v_{[AC]}-F_{BC}(t)v_{[BC]}\right] \\
                     &+a(t)\sum^{N}_{C\neq A,B}\sum^{N}_{D\neq A,B,C}\left(\frac{\psi_{C}}{\psi_{A}}\frac{\partial^{2}\mathcal{W}}{\partial\phi_{C}\partial\phi_{A}}-\frac{\psi_{C}}{\psi_{B}}\frac{\partial^{2}\mathcal{W}}{\partial\phi_{C}\partial\phi_{B}}\right)\beta^{2}_{D}v_{[CD]} \\ 
\end{aligned}
\end{equation}            
as well as
\begin{equation}\label{scalarvsys3}
 a\dot{v}_{[AB]}=-\Delta_{[AB]}+a(t)\left(2H+D_{AB}\right)v_{[AB]}
                 +a(t)\sum^{N}_{C\neq A,B}\left(\alpha_{A}\beta^{2}_{C}v_{[AC]}-\alpha_{B}\beta^{2}_{C}v_{[BC]}\right),
\end{equation}
where
\begin{equation}\label{CoefPERT}
\begin{aligned}
      A(t)=&3H+2\left(\alpha_{A}\beta^{2}_{A}+\alpha_{B}\beta^{2}_{B}\right)+2\sum^{N}_{C\neq A,B}\alpha_{C}\beta^{2}_{C} \\
 B_{AB}(t)=&\left(\alpha_{A}-\alpha_{B}\right)\beta^{2}_{A}\beta^{2}_{B} \\
C_{AB}(t)
       =&\left(3H+\left(\alpha_{A}\beta^{2}_{A}+\alpha_{B}\beta^{2}_{B}\right)\right)B_{AB} \\
        &+\left[\left(\frac{d^{2}\mathcal{V}}{d\phi^{2}_{A}}-\frac{d^{2}\mathcal{V}}{d\phi^{2}_{B}}\right)+\left(\frac{\partial^{2}\mathcal{W}}{\partial\phi^{2}_{A}}-\frac{\partial^{2}\mathcal{W}}{\partial\phi^{2}_{B}}\right)-\left(\frac{\psi_{A}}{\psi_{B}}-\frac{\psi_{B}}{\psi_{A}}\right)\frac{\partial^{2}\mathcal{W}}{\partial\phi_{B}\partial\phi_{A}}\right]\beta^{2}_{A}\beta^{2}_{B} \\      
          &+\sum^{N}_{C\neq A,B}\left[\alpha_{C}\beta^{2}_{C}B_{AB}+\left(\frac{\psi_{C}}{\psi_{A}}\frac{\partial^{2}\mathcal{W}}{\partial\phi_{A}\partial\phi_{C}}-\frac{\psi_{C}}{\psi_{B}}\frac{\partial^{2}\mathcal{W}}{\partial\phi_{B}\partial\phi_{C}}\right)\right] \\
 D_{AB}(t)=&\alpha_{A}\beta^{2}_{B}+\alpha_{B}\beta^{2}_{A} \\
 E_{AB}(t)=&\left(\frac{d^{2}\mathcal{V}}{d\phi^{2}_{A}}+\frac{\partial^{2}\mathcal{W}}{\partial\phi^{2}_{A}}-\frac{\psi_{A}}    		{\psi_{B}}\frac{\partial^{2}\mathcal{W}}{\partial\phi_{A}\partial\phi_{B}}\right)\beta^{2}_{B}+\left(\frac{d^{2}\mathcal{V}}{d\phi^{2}_{B}}+\frac{\partial^{2}\mathcal{W}}{\partial\phi^{2}_{B}}-\frac{\psi_{B}} {\psi_{A}}\frac{\partial^{2}\mathcal{W}}{\partial\phi_{B}\partial\phi_{A}}\right)\beta^{2}_{A} 	\\
           &+\left(\alpha_{A}-\alpha_{B}\right)B_{AB} \\
 F_{AB}(t)=&\alpha_{A}\left(\alpha_{A}+\alpha_{B}\right)\beta^{2}_{A}\beta^{2}_{B}.
\end{aligned}
\end{equation}
We note that the system (\ref{scalarDeltasys3})-(\ref{scalarvsys3}) is closely related to the system obtained by 
Bruni et al. in \cite{BED92,DBE92}.

Now, as usual, we can decouple the evolution equation for $\mathcal{Z}$ by differentiating \eqref{FO_Delta} with respect to time $t$ and using \eqref{FO_Z}, 
to obtain
\begin{equation}\label{ddotDelta}
\begin{aligned}
 \ddot{\Delta}=&\left(A-2H\right)\dot{\Delta}+\left(\dot{A}+2HA+3\dot{H}+2\psi^{2}+D^{2}\right)\Delta \\
               &-\sum^{N}_{A=1}\sum^{N}_{B>A}\left[\dot{B}_{AB}+\left(2H-D_{AB}\right)B_{AB}+C_{AB}\right]\Delta_{[AB]} \\
               &-\sum^{N}_{A=1}\sum^{N}_{B>A}\sum^{N}_{C\neq A,B}B_{AB}\left(\alpha_{A}\beta^{2}_{C}\Delta_{[AC]}-\alpha_{B}\beta^{2}_{C}\Delta_{[BC]}\right) \\
               &+a(t)\sum^{N}_{A=1}\sum^{N}_{B>A}\left[\dot{C}_{AB}+\left(5H+D_{AB}\right)C_{AB}+B_{AB}\left(3\dot{H}+\psi^2-E_{AB}+2D^2\right)\right]v_{[AB]} \\
               &-a(t)\sum^{N}_{A=1}\sum^{N}_{B>A}\sum^{N}_{C\neq A,B}\left[\left(B_{AB}F_{AC}-C_{AB}\alpha_{A}\beta^{2}_{C}\right)v_{[AC]}+\left(B_{AB}F_{BC}+C_{AB}\alpha_{B}\beta^{2}_{C}\right)v_{[BC]}\right] \\
               &-a(t)\sum^{N}_{A=1}\sum^{N}_{B>A}\sum^{N}_{C\neq A,B}\sum^{N}_{D\neq A,B,C}B_{AB}\left(\frac{\psi_{C}}{\psi_{A}}\frac{\partial^{2}\mathcal{W}}{\partial\phi_{C}\partial\phi_{A}}-\frac{\psi_{C}}{\psi_{B}}\frac{\partial^{2}\mathcal{W}}{\partial\phi_{C}\partial\phi_{B}}\right)\beta^{2}_{D}v_{[CD]}.
\end{aligned}
\end{equation}
where the evolution equations for $A, B_{AB}$ and $C_{AB}$ are given in section $6.1$ of the appendix. Due to its generality, the system composed by equations (\ref{ddotDelta}) and (\ref{Aevol})-(\ref{Cevol}) is quite long. We shall see ahead a simplified version of this system in the case of two scalar fields.
\subsection{Decomposition into scalar harmonics and particular solutions}
A common procedure to analyse the PDE system of equations derived above is to transform it 
into a system of ODEs by doing a harmonic decomposition. 
This is done by expanding the first order gauge invariant scalars $\Delta$ in terms 
of scalar harmonics\footnote{The wave number $\text{n}$ shouldn't be confused with the exponent $n$ 
in the scalar field potentials.}  $Q_{(\text{n})}$
as \cite{Har67},
\begin{equation}
\Delta=\sum_{\text{n}}\Delta_{(\text{n})}Q_{(\text{n})}, 
\end{equation}
which are comoving eigenfunctions of the operator
\begin{equation}
 D^{2}Q_{(\text{n})}=-\frac{\text{n}^{2}}{a^{2}}Q_{(\text{n})},\quad \text{with}\quad \dot{Q}_{(\text{n})}=0.
\end{equation}  
This is a useful procedure that has been followed many times in the literature, 
such as in models of structure formation in the universe \cite{Haw66,BDE92} and 
in the study of perturbations of spherically symmetric models, see e.g. \cite{KES99} and references therein. 

Using the time variable $\tau$ defined in (\ref{TAU}) and the above harmonic decomposition, 
the system of equations \eqref{scalarDeltasys3}, \eqref{scalarvsys3} and \eqref{ddotDelta} reads 
\begin{small}
\begin{equation}\label{Syst}
\begin{aligned}
 &\Delta^{\prime\prime}=\left[q+2+2\left(\frac{\alpha_{A}}{H}\beta^{2}_{A}+\frac{\alpha_{B}}{H}\beta^{2}_{B}\right)+2\sum^{N}_{C\neq A,B}\beta^{2}_{C}\frac{\alpha_{C}}{H}\right]\Delta^{\prime}+\left[\frac{2HA+3HH^{\prime}+2\psi^2+HA^{\prime}}{H^{2}}-\frac{\text{n}^{2}}{a^{2}H^{2}}\right]\Delta \\
 	               &-\sum^{N}_{A=1}\sum^{N}_{B>A}\left[\frac{H B^{\prime}_{AB}+\left(2H-D_{AB}\right)B_{AB}+C_{AB}}{H^2}\right]\Delta_{[AB]} \\
                       &+\sum^{N}_{A=1}\sum^{N}_{B>A}\sum^{N}_{C\neq A,B}\frac{B_{AB}}{H}\beta^{2}_{C}\left(\frac{\alpha_{A}}{H}\Delta_{[AC]}-\frac{\alpha_{B}}{H}\Delta_{[BC]}\right) \\	
                       &+a(t)\sum^{N}_{A=1}\sum^{N}_{B>A}\left[\frac{H C^{\prime}_{AB}+\left(5H+D_{AB}\right)C_{AB}}{H^{2}}+B_{AB}\left(\frac{3H H^{\prime}+\psi^2-E_{AB}}{H^{2}}-2\frac{\text{q}^{2}}{a^{2}H^{2}}\right)\right]v_{[AB]} \\
                       &-a(t)\sum^{N}_{A=1}\sum^{N}_{B>A}\sum^{N}_{C\neq A,B}\left[\left(B_{AB}F_{AC}-C_{AB}\alpha_{A}\beta^{2}_{C}\right)v_{[AC]}+\left(B_{AB}F_{BC}+C_{AB}\alpha_{B}\beta^{2}_{C}\right)v_{[BC]}\right] \\
                       &-a(t)\sum^{N}_{A=1}\sum^{N}_{B>A}\sum^{N}_{C\neq A,B}\sum^{N}_{D\neq A,B,C}B_{AB}\left(\frac{\psi_{C}}{\psi_{A}}\frac{\partial^{2}\mathcal{W}}{\partial\phi_{C}\partial\phi_{A}}-\frac{\psi_{C}}{\psi_{B}}\frac{\partial^{2}\mathcal{W}}{\partial\phi_{C}\partial\phi_{B}}\right)\beta^{2}_{D}v_{[CD]} \\
 &\Delta^{\prime}_{[AB]}=-\frac{D_{AB}}{H}\Delta_{[AB]}+\left(\frac{E_{AB}}{H^{2}}+\frac{\text{q}^{2}} {a^{2}H^{2}}\right)v_{[AB]}+\sum^{N}_{C\neq A,B}\beta^{2}_{C}\left(\frac{\alpha_{A}}{H}\Delta_{[AC]}-\frac{\alpha_{B}}{H}\Delta_{[BC]}\right) \\
                        &+a(t)H\sum^{N}_{C\neq A,B}\left[\frac{F_{AC}}{H^{2}}v_{[AC]}-\frac{F_{BC}}{H^{2}}v_{[BC]}\right] \\
                        &+a(t)H\sum^{N}_{C\neq A,B}\left[\frac{1}{H^{2}}\sum^{N}_{D\neq A,B,C}\left(\frac{\psi_{C}}{\psi_{A}}\frac{\partial^{2}\mathcal{W}}{\partial\phi_{C}\partial\phi_{A}}-\frac{\psi_{C}}{\psi_{B}}\frac{\partial^{2}\mathcal{W}}{\partial\phi_{C}\partial\phi_{B}}\right)\beta^{2}_{D}v_{[CD]}\right] \\
 &aHv^{\prime}_{[AB]}=-\Delta_{[AB]}+a\left(2H+D_{AB}\right)v_{[AB}+a(t)H\sum^{N}_{C\neq A,B}\left(\frac{\alpha_{A}}{H}\beta^{2}_{C}v_{[AC]}-\frac{\alpha_{B}}{H}\beta^{2}_{C}v_{[BC]}\right),\;
\end{aligned}
\end{equation}
\end{small}
where $\text{q}$ is the wave number\footnote{The wave number $\text{q}$ shouldn't be confused with the deceleration parameter $q$.} associated to the velocity scalar perturbations. 
In the case where the only matter present is a single nonlinear scalar field $\phi$, then the relative density perturbations 
are identically zero $\Delta_{[AB]}=0$ and the velocity perturbations vanish since $D^{2}\phi=0$, by construction. 
In this case, the system \eqref{Syst} reduces to the linear second order ODE
\begin{equation}\label{2ndOrderODE}
\begin{aligned}
 \Delta^{\prime\prime}_{(\text{n})}&-\left[1-\frac{\dot{H}}{H^{2}}+\frac{2}{H\psi}\frac{d\mathcal{V}}{d\phi}\right]\Delta^{\prime}_{(\text{n})} \\
                                   &-\left[6+6\frac{\dot{H}}{H^{2}}+2\frac{\psi^{2}}{H^{2}}+\frac{2}{H\psi}\frac{d\mathcal{V}}{d\phi}\left(5+\frac{1}{H\psi}\frac{d\mathcal{V}}{d\phi}\right)+\frac{2}{H^{2}}\frac{d^{2}\mathcal{V}}{d\phi^{2}}-\frac{\text{n}^{2}}{a^{2}H^{2}}\right]\Delta_{(\text{n})}=0,
\end{aligned} 
\end{equation}
which depends on the background quantities $H$, $\psi$ and on the first and second derivatives of the potential $\mathcal{V}$ 
with respect to $\phi$ which, in turn, depends on time. 

In the case of a perfect fluid with a linear equation of state $p=(1-\gamma)\rho$, the analogous differential equation 
can be found in Chapter 14 of \cite{WE97}, as equation $(14.32)$. The case of dust in a flat background can be solved 
explicitly, with the particularity that the solution is independent of $\text{n}$, while for general $\gamma$, 
the solution can be written in terms of Bessel functions (see Goode \cite{Goo89} for the corresponding equations 
using Bardeen variables \cite{Bar80}).

It is also possible to get explicit solutions in the so-called {\it long-wavelength limit}, which amounts to consider solutions with wavelengths larger than the Hubble distance, or equivalently
\begin{equation}\label{LWL}
 \frac{\text{n}}{aH}\ll 1.
\end{equation}
For example, for the exact massless scalar field background solution, the general solution to the perturbation equations \eqref{2ndOrderODE} can be written in terms of special functions, depending on the wave number $\text{n}$, while in the long-wavelength limit the solution is
\begin{equation}
 \Delta(\tau)=\frac{C_{1}}{4}e^{4\tau}+C_{2}
\end{equation}
or, in terms of cosmic time $t$,
\begin{equation}\label{MSF_LWL}
 \Delta=C_{2}+\frac{3}{4}C_{1}t^{\frac{4}{3}}.
\end{equation}
Even when it is possible to solve explicitly the background equations for a given potential, 
the perturbed system is, in general, impossible to solve. In \cite{Zim97}, Zimdahl used the slow-roll 
approximation in order to simplify the coefficients and studied the behaviour of density inhomogeneities during slow-roll 
inflation. In what follows, we shall apply a dynamical systems' approach to perform a qualitative analysis of the evolution 
of density inhomogeneities.
\section{Dynamical systems' approach to density inhomogeneities}
We shall now use the system of equations derived in last section and employ a dynamical systems` approach.  
Our approach follows the methods of Woszczyna \cite{Wos92II,Wos92,Bru93,Wos95} to study the evolution of 
inhomogeneities, which was also used to study stability problems in a universe with dust and radiation \cite{BP94}, 
magnetized cosmologies \cite{HD00} and locally rotational symmetric Bianchi I models \cite{Dun93}. 
Following these works, we introduce the dimensionless variables
\begin{equation}\label{U}
 \mathcal{U}_{(\text{n})}:=\frac{\Delta^{\prime}_{(\text{n})}}{\Delta_{(\text{n})}}\quad,\quad  \mathcal{X}_{[AB]}:=\frac{\Delta_{[AB]}}{\Delta}\quad,\quad \mathcal{Y}_{[AB]}:=\frac{aHv_{[AB]}}{\Delta}\;,
\end{equation}
and we arrive at the following result:
\begin{prop}\label{PERTSYSTEM}
The evolution for the first order scalar perturbations on a FLRW-scalar fields background, with arbitrary smooth
potentials, is given by the following system of differential equations for the state vector $\left((\Psi_{A},\Phi_{A},\Theta_{A}),\mathcal{U},\mathcal{X}_{[AB]},\mathcal{Y}_{[AB]}\right)$:
\begin{eqnarray}\label{pert-system}
\begin{aligned}
\mathcal{U}^{\prime}=&-\mathcal{U}^{2}-\xi\mathcal{U}-\zeta+\sum^{N}_{A=1}\sum^{N}_{B>A}\gamma_{AB}\mathcal{X}_{[AB]}+\sum^{N}_{A=1}\sum^{N}_{B>A}\eta_{AB}\mathcal{Y}_{[AB]}  \\
                     &+\sum^{N}_{A=1}\sum^{N}_{B>A}\sum^{N}_{C\neq A,B}\frac{\Psi^{2}_{C}}{\Psi^{2}}\frac{B_{AB}}{H}\left(\frac{\alpha_{A}}{H}\mathcal{X}_{[AC]}-\frac{\alpha_{B}}{H}\mathcal{X}_{[BC]}\right) \\
                     &-\sum^{N}_{A=1}\sum^{N}_{B>A}\sum^{N}_{C\neq A,B}\left(\frac{\left(B_{AB}F_{AC}-C_{AB}\alpha_{A}\beta^{2}_{C}\right)}{H^{3}}\mathcal{Y}_{[AC]}+\frac{\left(B_{AB}F_{BC}-C_{AB}\alpha_{B}\beta^{2}_{C}\right)}{H^{3}}\mathcal{Y}_{[BC]}\right) \\
                     &-\sum^{N}_{A=1}\sum^{N}_{B>A}\sum^{N}_{C\neq A,B}\sum^{N}_{D\neq A,B,C}\frac{B_{AB}}{H^{3}}\left(\frac{\Psi_{C}}{\Psi_{A}}\frac{\partial^{2}\mathcal{W}}{\partial\phi_{C}\partial\phi_{A}}-\frac{\Psi_{C}}{\Psi_{B}}\frac{\partial^{2}\mathcal{W}}{\partial\phi_{C}\partial\phi_{B}}\right)\beta^{2}_{D}\mathcal{Y}_{[CD]} \\
\mathcal{X}^{\prime}_{[AB]}=&-\mathcal{U}\mathcal{X}_{[AB]}+\varsigma_{AB}\mathcal{X}_{[AB]}+\varpi_{AB}\mathcal{Y}_{[AB]}+\sum_{C\neq A,B}\beta^{2}_{C}\left(\frac{\alpha_{A}}{H}\mathcal{X}_{[AC]}-\frac{\alpha_{B}}{H}\mathcal{X}_{[BC]}\right) \\
                            &+\sum^{N}_{C\neq A,B}\left[\frac{F_{AC}}{H^{2}}\mathcal{Y}_{[AC]}-\frac{F_{BC}}{H^{2}}\mathcal{Y}_{[BC]}+\frac{1}{H^{2}}\sum^{N}_{D\neq A,B,C}\left(\frac{\Psi_{C}}{\Psi_{A}}\frac{\partial^{2}\mathcal{W}}{\partial\phi_{C}\partial\phi_{A}}-\frac{\Psi_{C}}{\Psi_{B}}\frac{\partial^{2}\mathcal{W}}{\partial\phi_{C}\partial\phi_{B}}\right)\beta^{2}_{D}\mathcal{Y}_{[CD]}\right] \\
\mathcal{Y}^{\prime}_{[AB]}=&-\mathcal{U}\mathcal{Y}_{[AB]}+\iota_{AB}\mathcal{Y}_{[AB]}-\mathcal{X}_{[AB]} \\
                            &+\sum^{N}_{C\neq A,B}\left(\frac{\alpha_{A}}{H}\beta^{2}_{C}\mathcal{Y}_{[AC]}-\frac{\alpha_{B}}{H}\beta^{2}_{C}\mathcal{Y}_{[BC]}\right) \\  
   \Psi^{\prime}_{A}=&(q-2)\Psi_{A}-n\sqrt{6}\left[\Phi^{2n-1}_{A}\frac{\partial\Phi_{A}}{\partial\phi_{A}}+\Theta^{2n-1}_{A}\frac{\partial\Theta}{\partial\phi_{A}}\right] \\
   \Phi^{\prime}_{A}=&\frac{1}{n}(q+1)\Phi_{A}+\sqrt{6}\frac{\partial\Phi_{A}}{\partial\phi_{A}}\Psi_{A} \\
     \Theta^{\prime}=&\frac{1}{n}(q+1)\Theta+\sqrt{6}\sum^{N}_{A=1}\frac{\partial\Theta}{\partial\phi_{A}}\Psi_{A}
\end{aligned}
  \end{eqnarray}
subject to the  background constraint equation
\begin{equation}
 \sum^{N}_{A=1}\Psi^{2}_{A}+\sum^{N}_{A=1}\Phi^{2n}_{A}+\Theta^{2n}=1-K,
\end{equation}
where
\begin{equation*}
 q=2\sum^{N}_{A=1}\Psi^{2}_{A}-\sum^{N}_{A=1}\Phi^{2n}_{A}-\Theta^{2n}
\end{equation*}
and the coefficients $\xi, \zeta, \gamma_{AB}, \eta_{AB}, \varsigma, \varpi$ and $\iota_{AB}$ are given in appendix.
\end{prop}
We note that, in terms of the quantities  \eqref{ENV}, we have
\begin{equation*}
\frac{d\mathcal{V}_{A}}{d\phi_{A}}=6nH^{2}\Phi^{2n-1}_{A}\frac{\partial\Phi_{A}}{\partial\phi_{A}}\quad, \quad \frac{d^{2}\mathcal{V}_{A}}{d\phi^{2}_{A}}=6nH^{2}\left[(2n-1)\Phi^{2(n-1)}_{A}\left(\frac{\partial\Phi_{A}}{\partial\phi_{A}}\right)^{2}+\Phi^{2n-1}_{A}\frac{\partial^{2}\Phi_{A}}{\partial\phi^{2}_{A}}\right]
\end{equation*}
\begin{equation}
\alpha_{A}=\frac{\sqrt{6}nH}{\Psi_{A}}\left[\Phi^{2n-1}_{A}\frac{\partial\Phi_{A}}{\partial\phi_{A}}+\Theta^{2n-1}\frac{\partial\Theta}{\partial\phi_{A}}\right]\quad,\quad \beta^{2}_{A}=\frac{\Psi^{2}_{A}}{\Psi^{2}}
\end{equation}
where 
$$\Psi^{2}=\sum^{N}_{A=1}\Psi^{2}_{A}$$ 
and
\begin{equation*}
\frac{\partial^{2}\mathcal{W}}{\partial\phi_{B}\partial\phi_{A}}=6nH^{2}\left[(2n-1)\Theta^{2(n-1)}\left(\frac{\partial\Theta}{\partial\phi_{B}}\right)\left(\frac{\partial\Theta}{\partial\phi_{A}}\right)+\Theta^{2n-1}\frac{\partial^{2}\Theta}{\partial\phi_{B}\partial\phi_{A}}\right] 
\end{equation*}
\begin{equation*}
 \frac{d^{3}\mathcal{V}_{A}}{d\phi^{3}_{A}}=6nH^{2}(2n-1)\left[3\Phi^{2(n-1)}\frac{\partial\Phi_{A}}{\partial\phi_{A}}\frac{\partial^{2}\Phi_{A}}{\partial\phi^{2}_{A}}+2(n-1)\Phi^{2\left(n-\frac{3}{2}\right)}_{A}\left(\frac{\partial\Phi_{A}}{\partial\phi_{A}}\right)^{3}\right]+6nH^{2}\Phi^{2n-1}\frac{\partial^{3}\Phi_{A}}{\partial\phi^{3}_{A}}
\end{equation*}
which may be substituted in the above equations.

As explained by Dunsby in Chapter 14 of \cite{WE97}, the variable $\mathcal{U}_{(\text{n})}$ 
should be viewed as  $\tan{(\theta_{(\text{n})})}$, where $0\leq\theta_{(\text{n})}<2\pi$ 
is the polar angle in the plane $(\Delta_{(\text{n})},\Delta^{\prime}_{(\text{n})})$. 
For scalar fields in a flat background, we consider the variables subset defined by
\begin{equation}
 \sum^{N}_{A=1}\Psi^{2}_{A}+\sum^{N}_{A=1}\Phi^{2}_{A}=1\;,\quad-\infty<\mathcal{U}_{\text{n}}<+\infty\;,\quad-\infty<\mathcal{X}_{[AB]}<+\infty\;,\quad-\infty<\mathcal{Y}_{[AB]}<+\infty
\end{equation}
and regard the state space as $\mathbb{S}^{N}\times\mathbb{S}^{1}\times\mathbb{R}^{N(N-1)}$. 
As we shall see ahead, in the case of assisted power-law inflation, the fixed points are restricted to the compact subset 
$\mathbb{S}^{N}\times\mathbb{S}^{1}$. 
When only one scalar field is present, we can use either the variable $\Psi$ or $\Phi$ since they are related through 
the flat Friedman constraint \eqref{FlatFC}, and the state space is regarded as the {\it cylinder} $[0,1]\times\mathbb{S}^{1}$. 

The use of the variable $\mathcal{U}_{(\text{n})}$ makes the analysis of the system's stability quite transparent: 
If an orbit is asymptotic to an equilibrium point, the perturbation approaches a stationary state either: 
decaying to zero if $\mathcal{U}<0$, growing if $\mathcal{U}>0$ or having a constant value $\mathcal{U}=0$. If the orbit is asymptotic to a periodic orbit in the cylinder, the perturbation propagates as waves (see pgs. 296-297 of Chapter 14 in \cite{WE97}). 

We shall now investigate, separately, the cases of one and two scalar fields.
\subsection{Single scalar field}
 In the case of a single scalar field, we obtain from Proposition \ref{PERTSYSTEM} the following result:\\

{\it The evolution of the phase of first order scalar perturbations of FLRW-nonlinear scalar field models, is given by the 
following system of differential equations for the state vector $\left((\Psi,\Phi),\mathcal{U}_{(\text{n})}\right)$:}
\begin{eqnarray}\label{PertSys_single}
\begin{aligned}
 \mathcal{U}^{\prime}_{(\text{n})}&=-\mathcal{U}^{2}_{(\text{n})}-\mathcal{\xi}(\Psi,\Phi)\mathcal{U}_{(\text{n})}-\mathcal{\zeta}(\Psi,\Phi)\\
  \Psi^{\prime}&=2\Psi^{3}-\left(2+\Phi^{2n}\right)\Psi-n\sqrt{6}\Phi^{2n-1}\frac{\partial\Phi}{\partial\phi} \\
  \Phi^{\prime}&=-\frac{\Phi^{2n+1}}{n}+\frac{\left(1+2\Psi^{2}\right)}{n}\Phi+\sqrt{6}\Psi\frac{\partial\Phi}{\partial\phi}
\end{aligned}
\end{eqnarray}
{\it subject to the  background constraint equation}
\begin{equation}
 \Psi^{2}+\Phi^{2n}+K=1
\end{equation}
{\it and with the coefficients given by}
\begin{equation}
 \begin{aligned}
    \mathcal{\xi}(\Phi,\Psi)=&
                    -\left[2+2\Psi^{2}-\Phi^{2n}+\frac{12n}{\sqrt{6}}\frac{\Phi^{2n-1}}{\Psi}\frac{\partial\Phi}{\partial\phi}\right]  \label{pertsyst}\\
  \mathcal{\zeta}(\Phi,\Psi)=&-6\Phi^{2n}-\frac{6n}{\sqrt{6}}\frac{\Phi^{2n-1}}{\Psi}\frac{\partial\Phi}{\partial\phi}\left(10+\frac{12n}{\sqrt{6}}\frac{\Phi^{2n-1}}{\Psi}\frac{\partial\Phi}{\partial\phi}\right) \\
                  &-12n\left((2n-1)\Phi^{2n-2}\left(\frac{\partial\Phi}{\partial\phi}\right)^{2}+\Phi^{2n-1}\frac{\partial^{2}\Phi}{\partial\phi^{2}}\right)
                   +\frac{\text{n}^{2}}{a^{2}H^{2}}.
 \end{aligned}
\end{equation}
Equations (\ref{PertSys_single})-(\ref{pertsyst}) generalise and correct Eqs. (4) of \cite{Alh10,AM11}. 
In fact, a term was missing in the coefficients $\xi(\Phi,\Psi)$ and $\zeta(\Phi,\Psi)$ of the latter equations which, 
as shall see ahead, affects the results quantitatively, but not qualitatively.

As in the qualitative analysis of the background spacetime in Section $2$, we are interested in the dynamics in 
the invariant set of $K=0$ models, corresponding to the background of spatially flat hypersurfaces. 
In particular, since the background evolution equations forms an autonomous subsystem,  the background fixed points, 
given by Lemma \ref{Lemma_1}, are also fixed points of \eqref{PertSys_single} and the following result holds:
\begin{lem}
 For $K=0$, the fixed points of system \eqref{PertSys_single} are given by the conditions:
\begin{equation}\label{PertSys_FP}
 \begin{aligned}
&\mathcal{P}\,:~~ \frac{\partial\Phi}{\partial\phi}=-\frac{\sqrt{6}}{2n}\Phi\Psi \\
&\mathcal{U}^{\pm}_{(\text{n})}(\mathcal{P})=\frac{1}{2}\left(-\mathcal{\xi}(\mathcal{P})\pm\sqrt{\mathcal{\xi}^{2}(\mathcal{P})-4\mathcal{\zeta}(\mathcal{P})}\right) \\
 \end{aligned}
\end{equation}
subject to $\Psi^{2}+\Phi^{2n}=1$, where
\end{lem}
\begin{equation}\label{Coef_PertSys_Single}
 \begin{aligned}
  \mathcal{\xi}(\mathcal{P})&= -\left[4-9\Phi_{\mathcal{P}}^{2n}\right]\\
 \mathcal{\zeta}(\mathcal{P})&=18\left(\frac{1}{n}-\frac{2}{3}\right)\Phi^{2n}_{\mathcal{P}}-18\left(\frac{1}{n}-1\right)\Phi^{4n}_{\mathcal{P}}
                               -12n\Phi^{2n-1}_{\mathcal{P}}\left(\frac{\partial^2\Phi}{\partial\phi^2}\right)_{\mathcal{P}}+\frac{\text{n}^{2}}{H^2a^2}\;.
 \end{aligned}
\end{equation}
From the above considerations, and from the fact that in a flat background we can reduce the linearised matrix at $\mathcal{P}$ given by \eqref{Single_LinMat}, it also follows that the eigenvalues of the linearized system around the fixed points are the ones given by \eqref{EigenValue} together with
\begin{equation}\label{EigenVal_PertSys_Single}
 \omega_{\mathcal{U}^{\pm}(\mathcal{P})}=-2\mathcal{U}^{\pm}(\mathcal{P})-\xi(\mathcal{P})=\mp\sqrt{\mathcal{\xi}^{2}(\mathcal{P})-4\mathcal{\zeta}(\mathcal{P})}\;.
\end{equation}
Thus, from \eqref{PertSys_FP}, the fixed points exist if 
\begin{equation}\label{RealFP_Cond}
\mathcal{\xi}^{2}(\mathcal{P})-4\mathcal{\zeta}(\mathcal{P})\geq0\; 
\end{equation}
and reduce to a single point when the equality is verified. In that case, the eigenvalues coincide and, from \eqref{EigenVal_PertSys_Single}, are identically zero, resulting in a saddle point. If this is not the case, and if $\mathcal{P}$ is an atractor point of the background dynamical system, then it follows that $\mathcal{U}^{+}(\mathcal{P})$ is the late time 
attractor of \eqref{PertSys_single} having the following properties:  
\begin{equation}\label{Stability_Cond}
 \begin{aligned}
&\mathcal{U}^{+}_{(\text{n})}(\mathcal{P})<0\quad\text{if}\quad\zeta(\mathcal{P})>0\quad\text{and}\quad \xi(\mathcal{P})>0, \\
&\mathcal{U}^{+}_{(\text{n})}(\mathcal{P})=0\quad\text{if}\quad\zeta(\mathcal{P})=0, \\
&\mathcal{U}^{+}_{(\text{n})}(\mathcal{P})>0\quad\text{if}\quad\xi(\mathcal{P})<0.
 \end{aligned}
\end{equation}
\subsubsection{Exponential potential: Power-law inflation} 
\begin{figure}[h!]
\centering
\begin{minipage}{\textwidth/2-1pc}
\includegraphics[width=\textwidth]{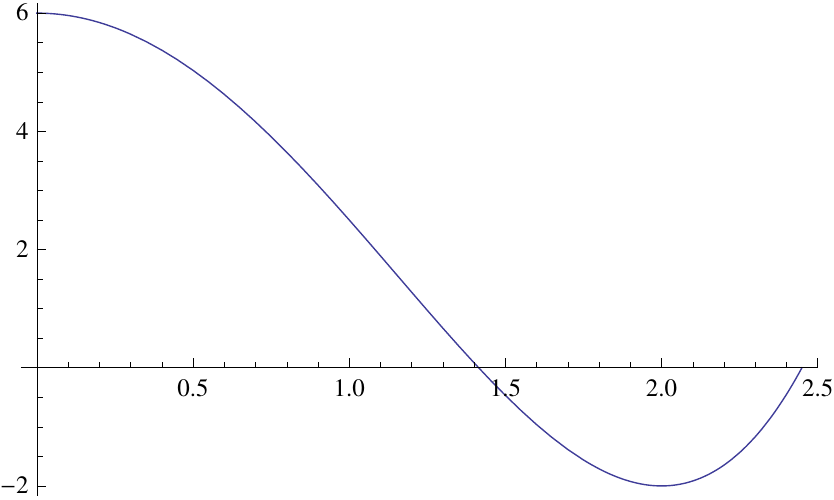}
\caption{\label{Fig5}Plot of $\zeta(\lambda)$ using the long wavelenght limit. The positive region 
gives the values of $\lambda$ for which the perturbations decay, for an exponential potential. 
The zeros are at $\lambda=\sqrt{2}$ and $\lambda=\sqrt{6}$. 
Notice that if $\text{n}^{2}>0$ the graph is shifted upwards along the $\zeta$ axis and, consequently, 
the interval for which $\zeta$ is positive gets bigger.}
\end{minipage}\hspace{2pc}%
\begin{minipage}{\textwidth/2-1pc}
\includegraphics[width=\textwidth]{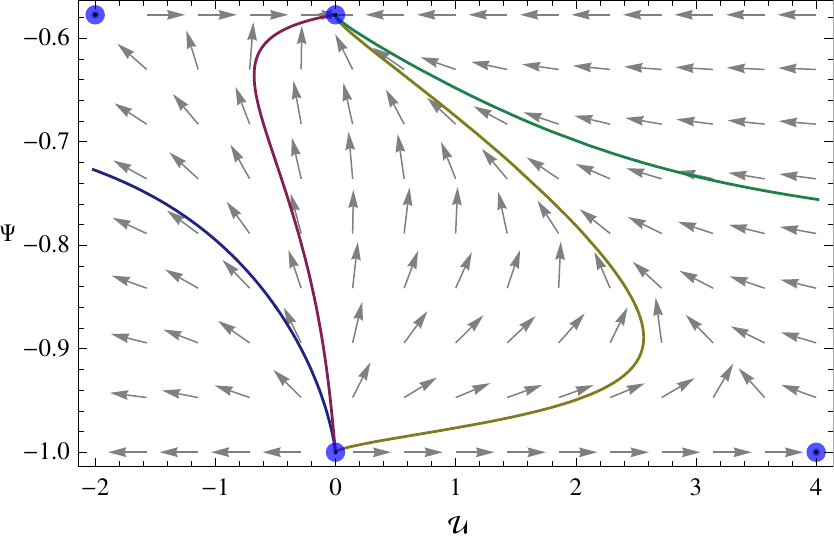}
\caption{\label{Fig6}Density perturbations described by orbits in the phase
plane $(\Psi,\mathcal{U}_{(n)})$ for an exponential potential with
$\lambda=\sqrt{2}$ in the long wavelength limit. The figure
 shows the equilibrium points, one of which is the future attractor 
$(\mathcal{P},\mathcal{U}^{+}_{(n)})$ having $\mathcal{U}_{(n)}=0$.}
\end{minipage} 
\end{figure}

\begin{figure}[h!]
\centering
\begin{minipage}{\textwidth/2-1pc}
\includegraphics[width=\textwidth]{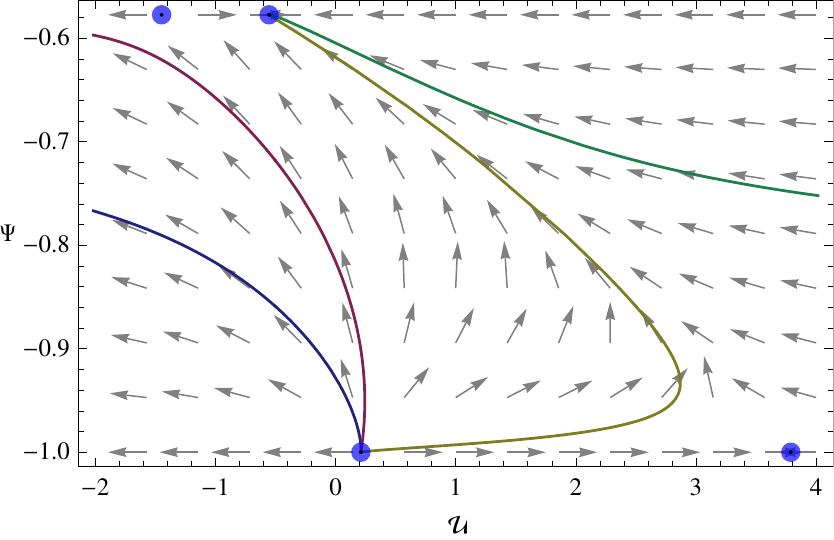}
\caption{\label{Fig7}Density perturbations described by orbits in the phase
plane $(\Psi,\mathcal{U}_{(\text{n})})$ for an exponential potential with
$\lambda=\sqrt{2}$ for $\frac{\text{n}^2}{a^{2}H^{2}}=0.8>\frac{\text{n}^2_{(-)}(\sqrt{2})}{a^{2}H^{2}}=0$. The figure
 shows the equilibrium points, one of which is the future attractor $(\mathcal{P},\mathcal{U}^{+}_{(n)})$ 
having $\mathcal{U}_{(n)}<0$.}
\end{minipage}\hspace{2pc}%
\begin{minipage}{\textwidth/2-1pc}
\includegraphics[width=\textwidth]{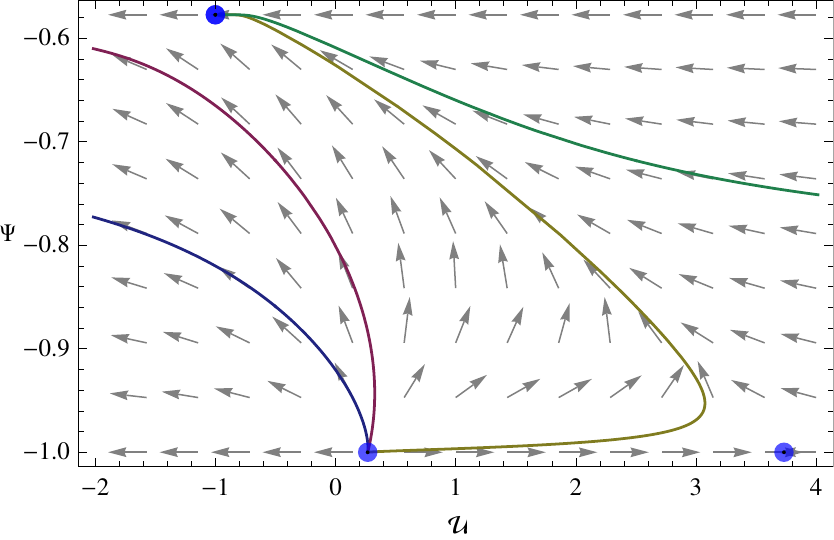}
\caption{\label{Fig8}Density perturbations described by orbits in the phase
plane $(\Psi,\mathcal{U}_{(n)})$ for an exponential potential with
$\lambda=\sqrt{2}$ and $\text{n}^2=\text{n}^{2}_{crit}(\sqrt{2})=1$. The figure
 shows the saddle point in the region $\mathcal{U}_{(n)}<0$.}
\end{minipage} 
\end{figure}

\begin{figure}[h!]
\centering
\begin{minipage}{\textwidth/2-1pc}
\includegraphics[width=\textwidth]{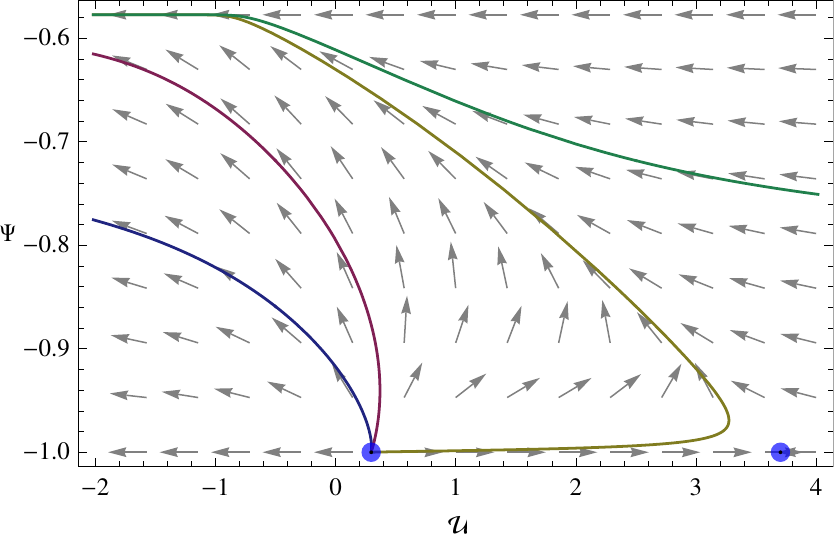}
\caption{\label{Fig9}Density perturbations described by orbits in the phase
plane $(\Psi,\mathcal{U}_{(n)})$ for an exponential potential with
$\lambda=\sqrt{2}$ and $\text{n}^2>\text{n}^{2}_{crit}(\sqrt{2})=1$. 
For this value of the slope parameter the orbits are periodic.}
\end{minipage}\hspace{2pc}%
\begin{minipage}{\textwidth/2-1pc}
\includegraphics[width=\textwidth]{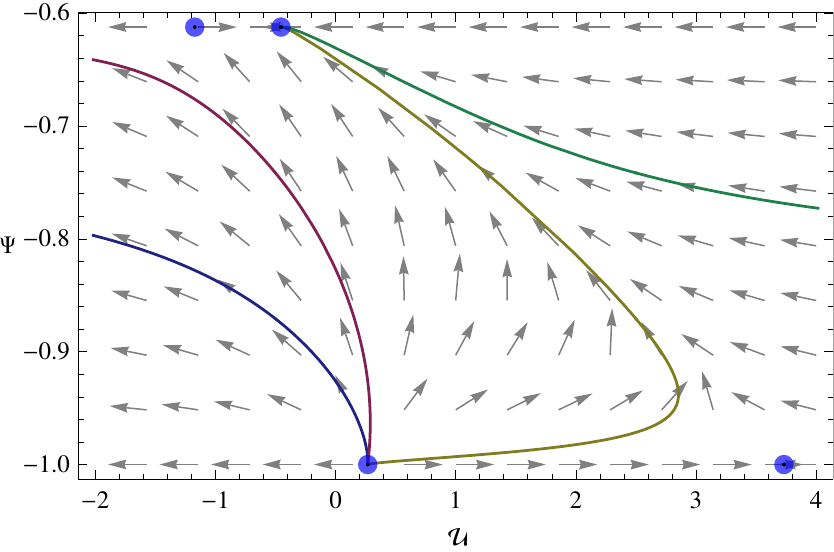}
\caption{\label{Fig10}Density perturbations described by orbits in the phase
plane $(\Psi,\mathcal{U}_{(n)})$ for an exponential potential with
$\lambda=1.5$ and $\frac{\text{n}^2}{a^{2}H^{2}}=1$. The figure
 shows the equilibrium points, one of which is the 
future attractor $(\mathcal{P},\mathcal{U}^{+}_{(n)})$ having $\mathcal{U}_{(n)}<0$.}
\end{minipage} 
\end{figure}

\begin{figure}[h]
\centering
\begin{minipage}{\textwidth/2-1pc}
\includegraphics[width=\textwidth]{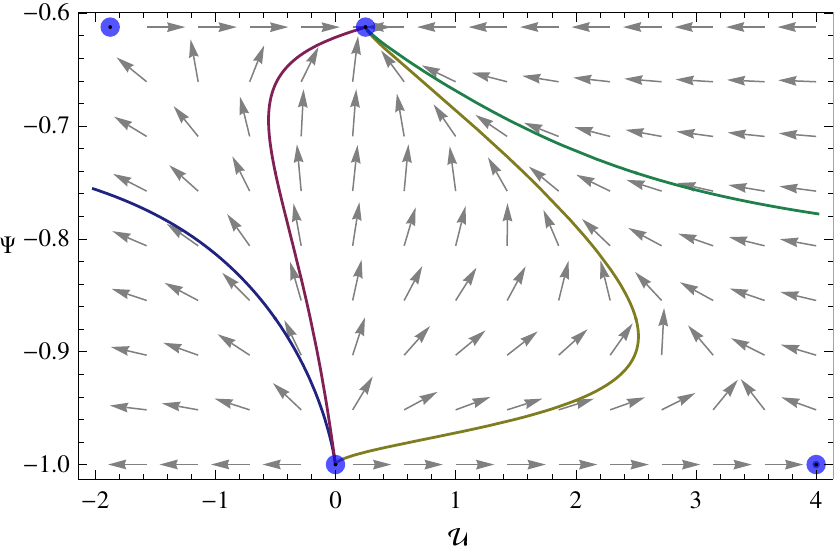}
\caption{\label{Fig11}Density perturbations described by orbits in the phase
plane $(\Psi,\mathcal{U}_{(n)})$ for an exponential potential with
$\lambda=1.5$ taking the long wavelength limit. }
\end{minipage}\hspace{2pc}%
\begin{minipage}{16pc}
\includegraphics[width=\textwidth]{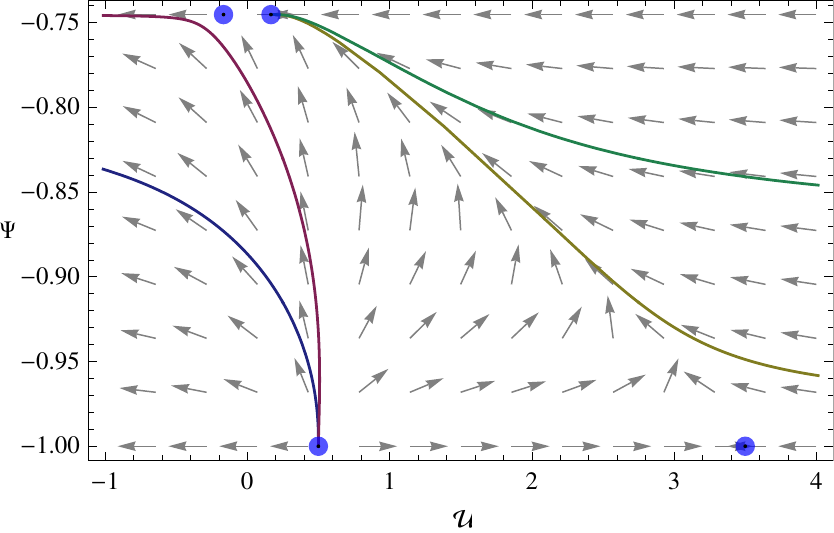}
\caption{\label{Fig12}Density perturbations described by orbits in the phase
plane $(\Psi,\mathcal{U}_{(n)})$ for an exponential potential with
$\lambda=\sqrt{\frac{10}{3}}$ and $\frac{\text{n}^2}{a^{2}H^{2}}=1.75$.}
\end{minipage}
\end{figure}

\begin{figure}[h!]
\centering
\begin{minipage}{\textwidth/2-1pc}
\includegraphics[width=\textwidth]{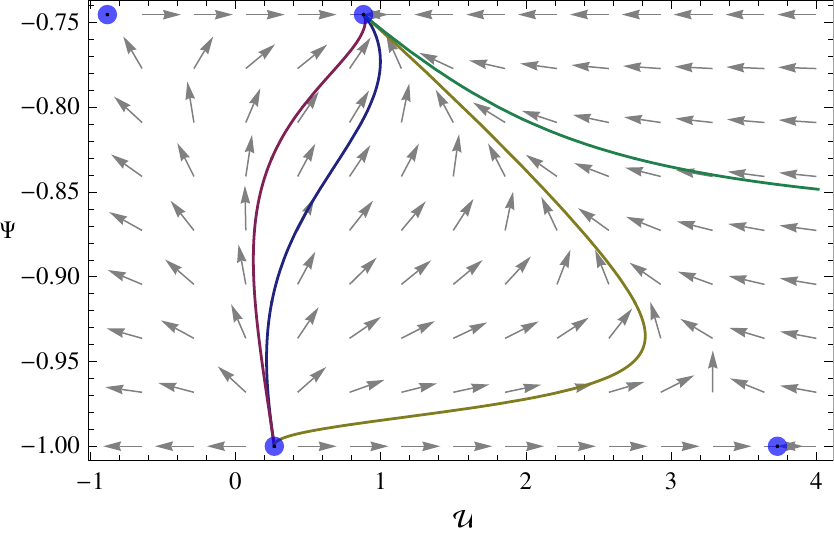}
\caption{\label{Fig13}Density perturbations described by orbits in the phase
plane $(\Psi,\mathcal{U}_{(n)})$ for an exponential potential with
$\lambda=\sqrt{\frac{10}{3}}$ and $\frac{\text{n}^2}{a^{2}H^{2}}=1$. The figure
 shows the equilibrium points, one of which is the 
future attractor $(\mathcal{P},\mathcal{U}^{+}_{(n)})$ having $\mathcal{U}_{(n)}>0$.}
\end{minipage}\hspace{2pc}%
\begin{minipage}{\textwidth/2-1pc}
\includegraphics[width=\textwidth]{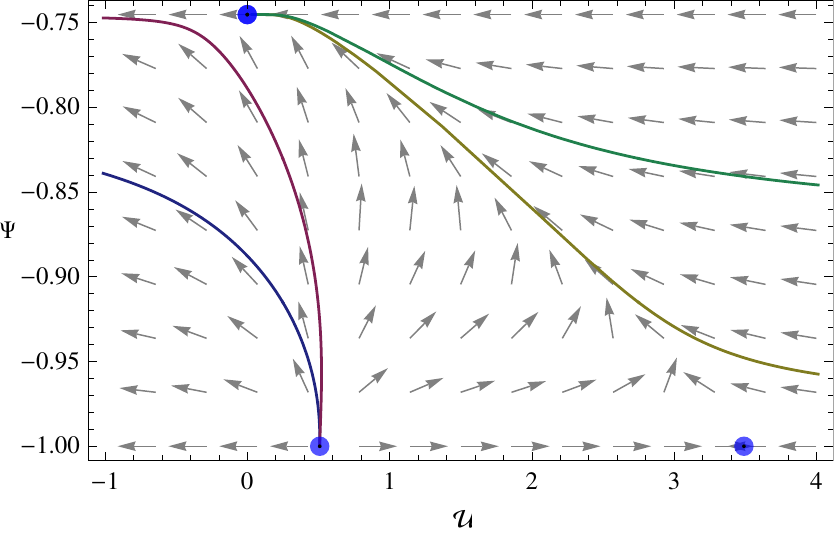}
\caption{\label{Fig14}Density perturbations described by orbits in the phase
plane $(\Psi,\mathcal{U}_{(n)})$ for an exponential potential with
$\lambda=\sqrt{\frac{10}{3}}$ and $\frac{\text{n}^2_{crit}}{a^{2}H^{2}}=\frac{16}{9}$. The figure
 shows the equilibrium points. The saddle point has $\mathcal{U}_{(n)}=0$.}
\end{minipage} 
\end{figure}

We have seen in Section \ref{subsection_4} that, for an exponential potential, the background subsystem has two fixed points, 
$\mathcal{P}_{0}$ and $\mathcal{P}_{1}$ given by \eqref{Single_P0} and \eqref{Single_P1}, respectively. 
Then, for the perturbed system (\ref{PertSys_single}) it follows from Lemma $2$ that there are four fixed points
\begin{equation}
 \left(\mathcal{P}_{0},\mathcal{U}^{\pm}_{(\text{n})}(\mathcal{P}_{0})\right),~~~ \left(\mathcal{P}_{1},\mathcal{U}^{\pm}_{(\text{n})}(\mathcal{P}_{1})\right)
\end{equation}
where, in this case, the coefficients \eqref{Coef_PertSys_Single} are 
\begin{equation}
\begin{aligned}
 \mathcal{\xi}(\mathcal{P}_{0})&=-4~~\text{and}~~~&\mathcal{\xi}(\mathcal{P}_{1})=5-\frac{3}{2}\lambda^{2}
\end{aligned}
\end{equation}
\begin{equation}\label{ExpPot_coeff}
\begin{aligned}
 \mathcal{\zeta}(\mathcal{P}_{0})&=\frac{\text{n}^{2}}{a^{2}H^{2}}~~\text{and}~~~
 &\mathcal{\zeta}(\mathcal{P}_{1})=6-4\lambda^{2}+\frac{\lambda^{4}}{2}+\frac{\text{n}^{2}}{a^{2}H^{2}}.
\end{aligned}
\end{equation}
The fixed point solutions are real  if \eqref{RealFP_Cond} is satisfied which, in turn, implies that the wave number satisfies
\begin{equation}
\mathcal{P}_{0}\,:~~  \text{n}^{2}\leq \text{n}^2_{crit}=\frac{a^{2}H^{2}}{4}, 
\end{equation}
\begin{equation}
\mathcal{P}_{1}\,:~~  \text{n}^{2}\leq \text{n}^2_{crit}=\frac{a^{2}H^{2}}{4}\left(1+\lambda^{2}+\frac{\lambda^{4}}{4}\right).
\end{equation}
When $\text{n}=\text{n}_{crit}$, the points $\mathcal{U}^{\pm}_{(\text{n})}$ merge into a single saddle point. For $\text{n}>\text{n}_{crit}$, the fixed points cease to exist, the orbit is periodic and the perturbations behave as waves. 
We also saw that $\mathcal{P}_{1}$ was the late time attractor corresponding to the flat power-law solution of the background 
dynamical system \eqref{syst-ode}. Thus, the attractor point of the dynamical system \eqref{PertSys_single} is 
$\mathcal{U}^{+}_{\text{n}}(\mathcal{P}_{1})$ and we obtain, from \eqref{Stability_Cond}, that there exists 
$$
\text{n}^{2}_{(-)}=a^{2}H^{2}\left(-6+4\lambda^{2}-\frac{\lambda^{4}}{2}\right)
$$
such that, if $\text{n}^{2}_{(-)}(\lambda)<\text{n}^{2}<\text{n}^{2}_{crit}(\lambda)$, 
then $\mathcal{U}^{+}_{(\text{n})}(\mathcal{P}_{1})<0$ and the perturbations decay. 
When $\text{n}^{2}=\text{n}^{2}_{(-)}(\lambda)$, then $\zeta(\mathcal{P}_{1})=0$ and the perturbations tend to a constant. 
Therefore, when the slope parameter satisfies $0<\lambda<\sqrt{2}$, the density perturbation modes decay for all wavelenghts 
in the range for which there exist fixed points, i.e. for $0\leq\text{n}^{2}<\text{n}^{2}_{crit}(\lambda)$. 
In particular, for $\lambda=\sqrt{2}$ the modes decay, except in the long wavelenght limit, for which the perturbations tend 
to constant since $\text{n}^{2}_{(-)}(\sqrt{2})=0$ (see Figs. \ref{Fig5}, \ref{Fig6}, \ref{Fig7}, \ref{Fig8} and \ref{Fig9}). 
Moreover, if $\sqrt{2}<\lambda<\sqrt{10/3}$, there are perturbation modes which can either decay, tend to constant or grow, depending on the value of $\text{n}^{2}$ (see Figs. \ref{Fig10}, \ref{Fig11}). For $\lambda=\sqrt{10/3}$, all perturbations grow, while $\text{n}$ reaches its critical value. In this case, the saddle point is at $\mathcal{U}_{\text{n}_{crit}}=0$ (see Figs. \ref{Fig12}, \ref{Fig13}, \ref{Fig14}). Finally, for  $\sqrt{\frac{10}{3}}\leq\lambda<\sqrt{6}$, all modes grow for all wavelenghts in the range for which there exist fixed points.

Notice that, by taking the long wavelength limit $\mathcal{U}^{-}(\mathcal{P}_{0})=0$  (see Figs \ref{Fig6} and \ref{Fig11}), which corresponds to the constant mode of \eqref{MSF_LWL}, while $\mathcal{U}^{+}(\mathcal{P}_{0})>0$ corresponds to the growing mode.
Our results are summarised in Table \ref{tb:tablename} which corrects and generalises Table $1$ of \cite{AM11}.
%
\subsubsection{Polynomial potentials: Chaotic inflation}\label{HArm}
\begin{figure}[h!]
\centering
\begin{minipage}{\textwidth/2-1pc}
\includegraphics[width=\textwidth]{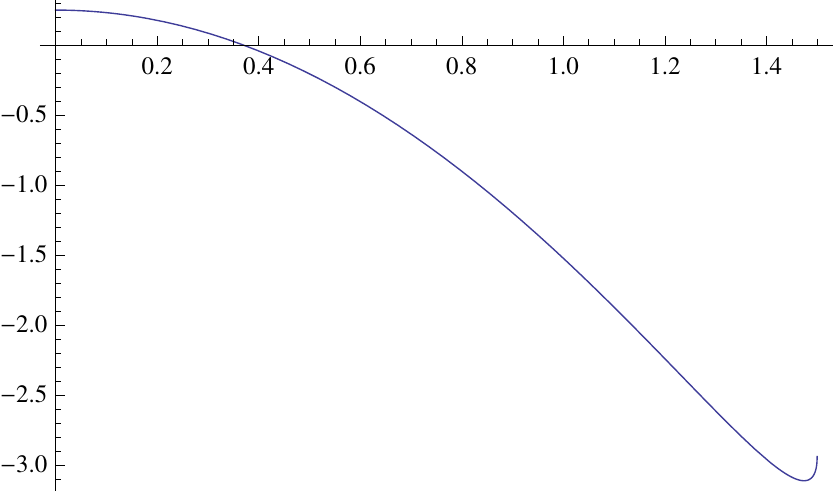}
\caption{\label{fig15}  The non-negative region shows the set of values of $\mathcal{M}$ for which there are fixed points.}
\end{minipage}\hspace{2pc}%
\begin{minipage}{\textwidth/2-1pc}
\includegraphics[width=\textwidth]{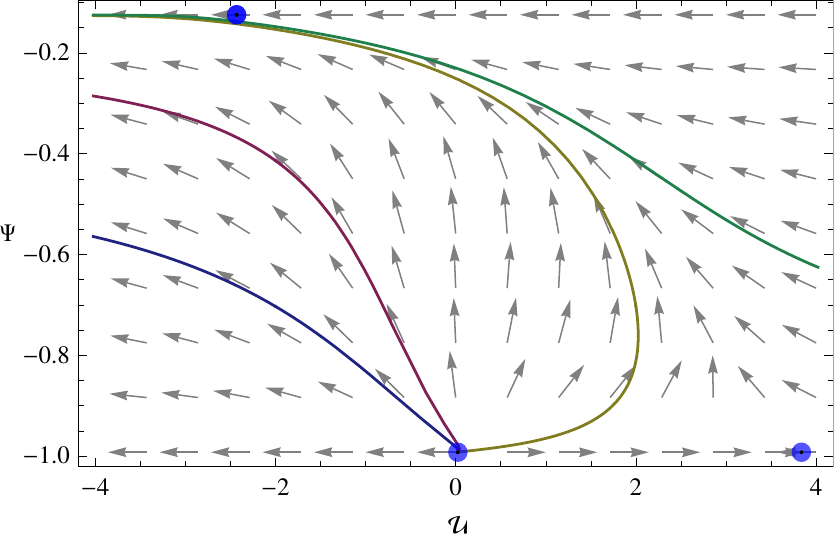}
\caption{\label{fig16} Density perturbations described by orbits in the phase
plane $(\Psi,\mathcal{U}_{(n)})$ for a quadratic potential with $\mathcal{M}=\frac{2}{9}\sqrt{16-5\sqrt{7}}$ 
in the long wavelength limit. The figure
 shows the saddle point in the region $\mathcal{U}_{(n)}<0$.}
\end{minipage}
\end{figure}

\begin{figure}[h!]
\centering
\begin{minipage}{\textwidth/2-1pc}
\includegraphics[width=\textwidth]{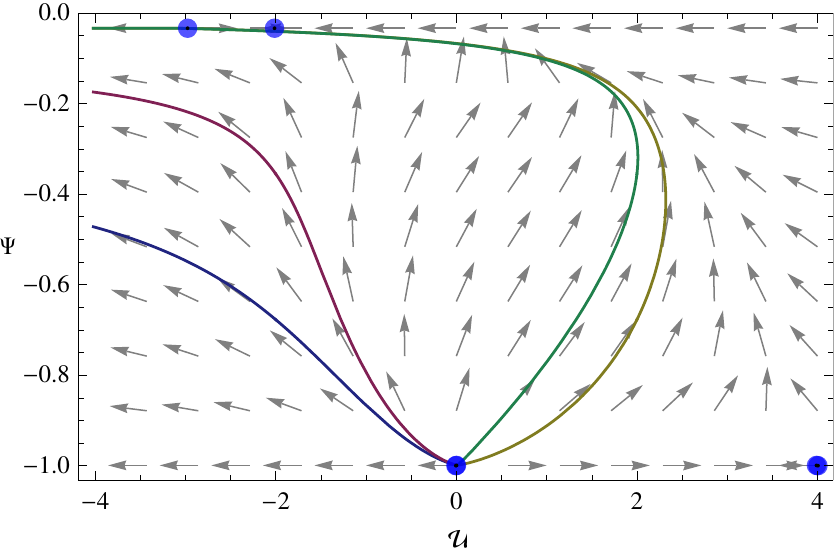}
\caption{\label{fig17}  Density perturbations described by orbits in the phase
plane $(\Psi,\mathcal{U}_{(n)})$ for a quadratic potential with $\mathcal{M}=10^{-1}$ 
in the long wavelength limit. 
The figure shows the attractor point $\mathcal{U}^{+}(\mathcal{P}_{1})$ in the region $\mathcal{U}_{(n)}<0$. See also Fig. $18$.}
\end{minipage}\hspace{2pc}%
\begin{minipage}{\textwidth/2-1pc}
\includegraphics[width=\textwidth]{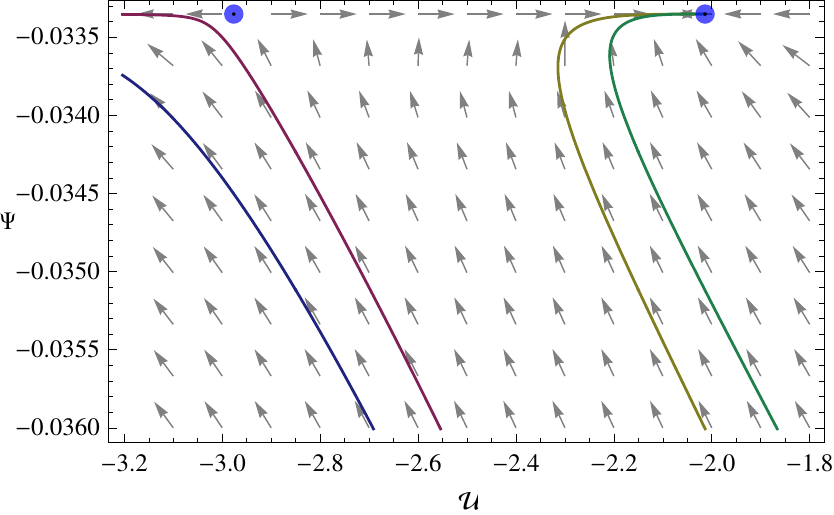}
\caption{\label{fig18} Density perturbations described by orbits in the phase
plane $(\Psi,\mathcal{U}_{(n)})$ for a quadratic potential with $\mathcal{M}=10^{-1}$ 
in the long wavelength limit. The figure  shows in more detail the attractor point in the region $\mathcal{U}_{(n)}<0$ of Fig. $17$.}
\end{minipage}
\end{figure}

\begin{figure}[h!]
\centering
\begin{minipage}{\textwidth/2-1pc}
\includegraphics[width=\textwidth]{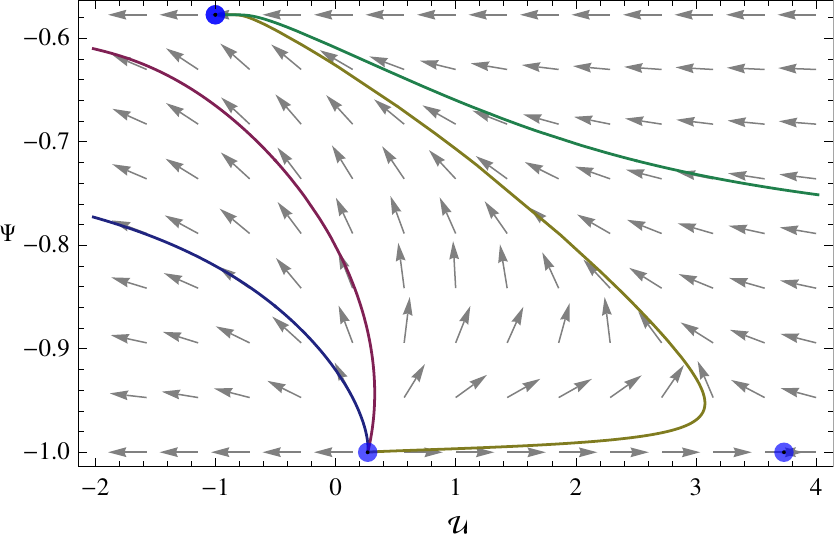}
\caption{\label{fig19}Density perturbations described by orbits in the phase
plane $(\Psi,\mathcal{U}_{(n)})$ for a quadratic potential with
$\mathcal{M}=10^{-1}$ for $n^2=n^{2}_{crit}$. The figure
 shows the saddle point, in the region $\mathcal{U}_{(n)}<0$.}
\end{minipage}\hspace{2pc}%
\begin{minipage}{\textwidth/2-1pc}
\includegraphics[width=\textwidth]{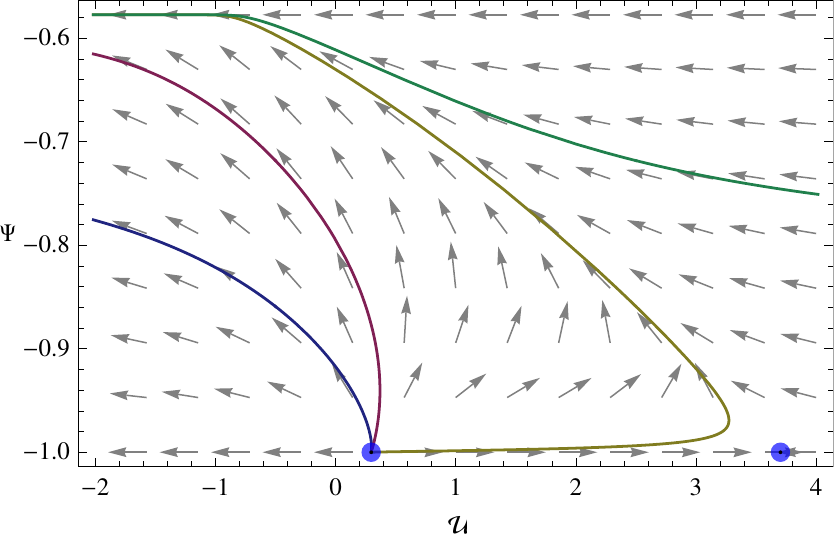}
\caption{\label{fig20}Density perturbations described by orbits in the phase
plane $(\Psi,\mathcal{U}_{(n)})$ for an quadratic potential with
$\mathcal{M}=10^{-1}$ and $n^2>n^{2}_{crit}$. The figure shows the existence of periodic orbits, and the perturbations behave as waves.}
\end{minipage} 
\end{figure}

\begin{figure}[h!]
\centering
\begin{minipage}{\textwidth/2-1pc}
\includegraphics[width=\textwidth]{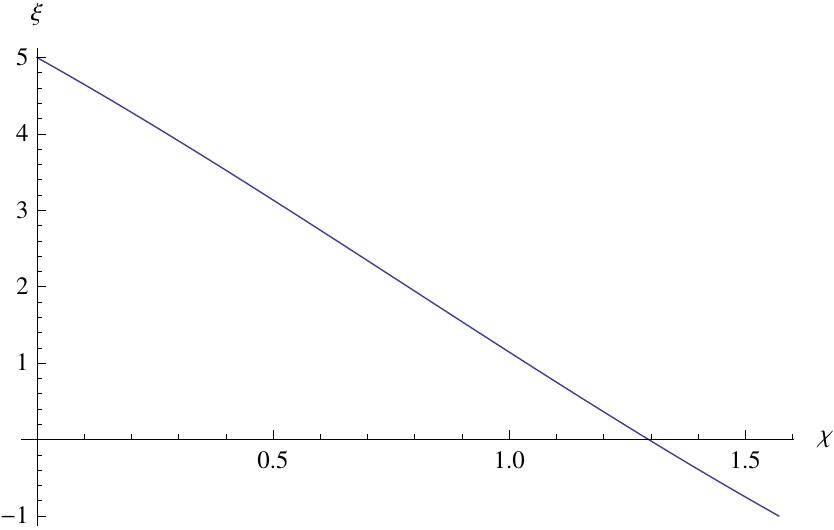}
\caption{\label{fig21} Plot of $\xi(\mathcal{P}_{1})$ in the interval $0\leq\chi\leq\frac{\pi}{2}$. The figure shows a zero at $\chi=\pi-\frac{3}{2}\arccos{\left(\frac{1}{\sqrt{3}}\right)}$. The positive region shows the admitted values for decay as long as there exists fixed points.}
\end{minipage}\hspace{2pc}%
\begin{minipage}{\textwidth/2-1pc}
\includegraphics[width=\textwidth]{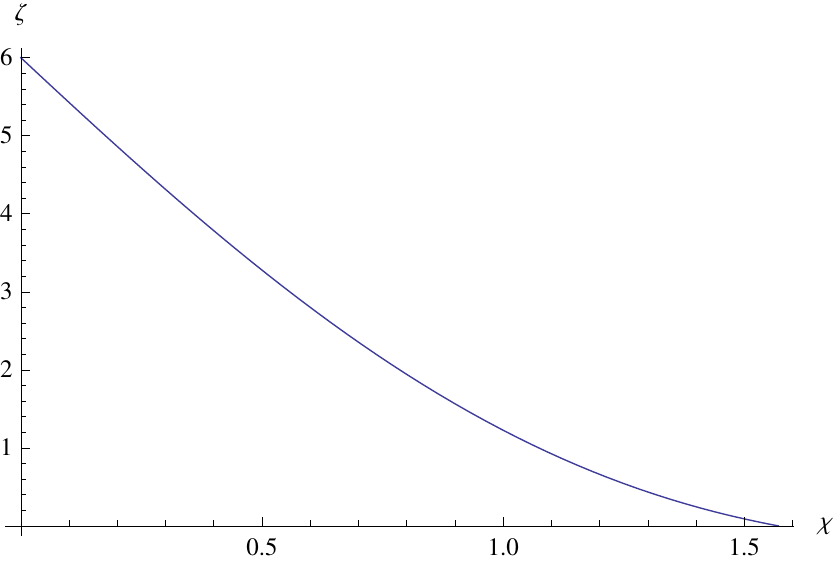}
\caption{\label{fig22} Plot of $\zeta(\mathcal{P}_{1})$ in the interval $0\leq\chi\leq\frac{\pi}{2}$ taking the long-wavelength limit. The figure shows that $\zeta$ is a positive monotone function of $\chi$ achieving the value zero at $\chi=\frac{\pi}{2}$.}
\end{minipage}
\end{figure}

\begin{figure}[h!]
\centering
\begin{minipage}{\textwidth/2-1pc}
\includegraphics[width=\textwidth]{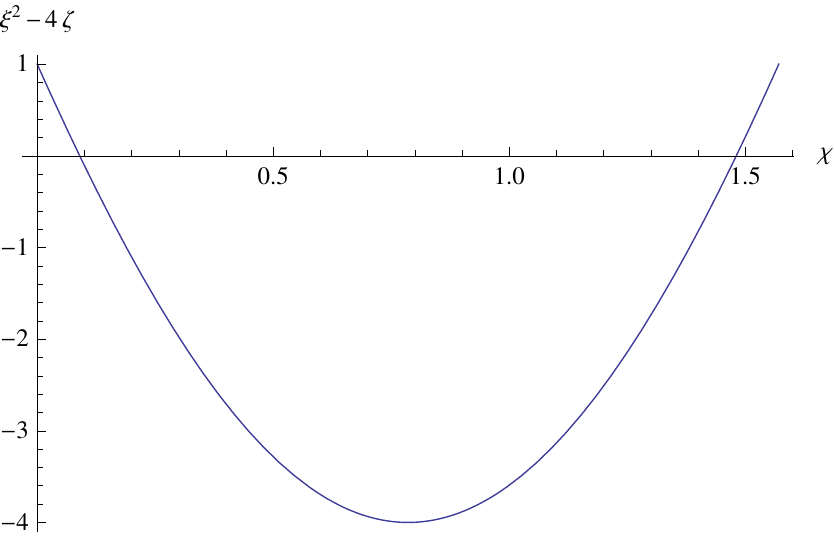}
\caption{\label{fig23} Plot of $\xi^{2}(\mathcal{P}_{1})-4\zeta(\mathcal{P}_{1})$ in the interval $0\leq\chi\leq\frac{\pi}{2}$. The figure shows two zeros at $\chi=\pi-\frac{3}{2}\arccos{\left(\pm\frac{1}{\sqrt{5}}\right)}$. The nonegative region shows the admitted values of $\chi$ for which there exists fixed points.}
\end{minipage}\hspace{2pc}%
\begin{minipage}{\textwidth/2-1pc}
\includegraphics[width=\textwidth]{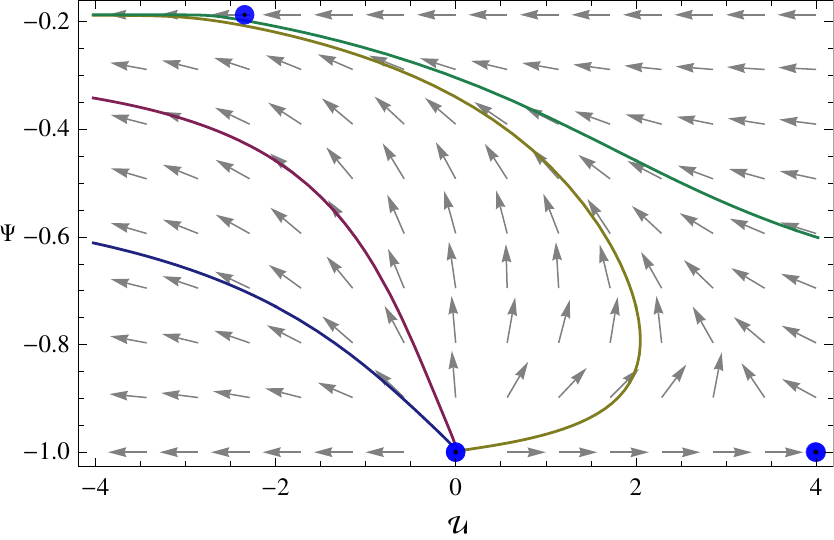}
\caption{\label{fig24} Phase portrait for $\chi=\pi-\frac{3}{2}\arccos{\left(-\frac{1}{\sqrt{5}}\right)}$ taking the long wavelength limit. The figure shows the saddle point in the region $\mathcal{U}<0$.}
\end{minipage}
\end{figure}

\begin{figure}[h!]
\centering
\begin{minipage}{\textwidth/2-1pc}
\includegraphics[width=\textwidth]{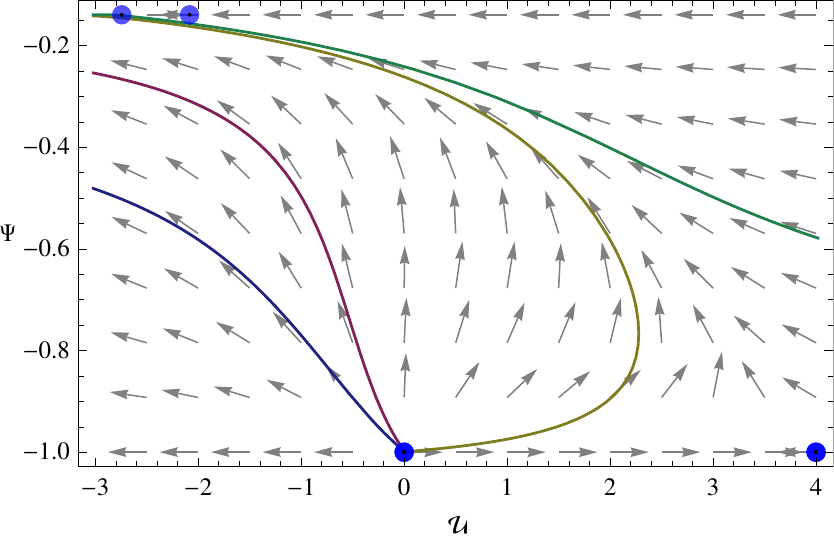}
\caption{\label{fig25} Density perturbations described by orbits in the phase
 plane $(\Psi,\mathcal{U}_{(n)})$ for an quartic potential with $\chi=20^{-1}$ taking the long wavelength limit. The figure shows the late time attractor in the region $\mathcal{U}<0$.}
\end{minipage}\hspace{2pc}%
\begin{minipage}{\textwidth/2-1pc}
\includegraphics[width=\textwidth]{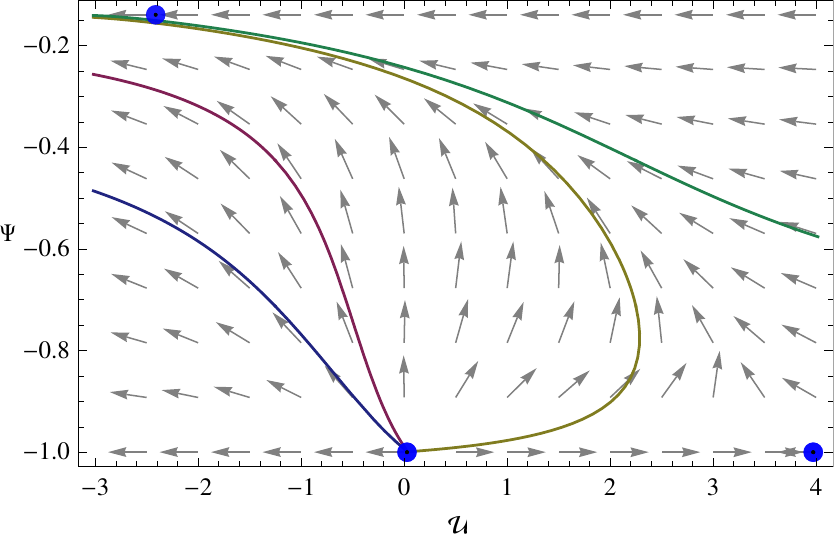}
\caption{\label{fig26} Density perturbations described by orbits in the phase
 plane $(\Psi,\mathcal{U}_{(n)})$ for an quartic potential with $\chi=20^{-1}$ for $n^{2}=n^{2}_{crit}(20^{-1})$. The figure shows the saddle point in the region $\mathcal{U}<0$.}
\end{minipage}
\end{figure}

\begin{figure}[h!]
\centering
\begin{minipage}{\textwidth/2-1pc}
\includegraphics[width=\textwidth]{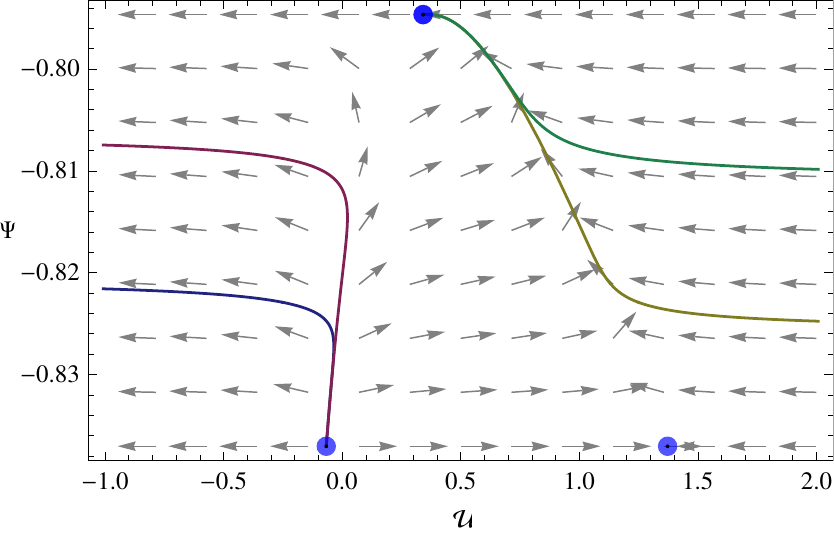}
\caption{\label{fig27} Density perturbations described by orbits in the phase
 plane $(\Psi,\mathcal{U}_{(n)})$ for an quartic potential with $\chi=\pi-\frac{3}{2}\arccos{\left(\frac{1}{\sqrt{5}}\right)}$ taking the long wavelength limit. The figure shows the saddle point in the region $\mathcal{U}>0$.}
\end{minipage}\hspace{2pc}%
\begin{minipage}{\textwidth/2-1pc}
\includegraphics[width=\textwidth]{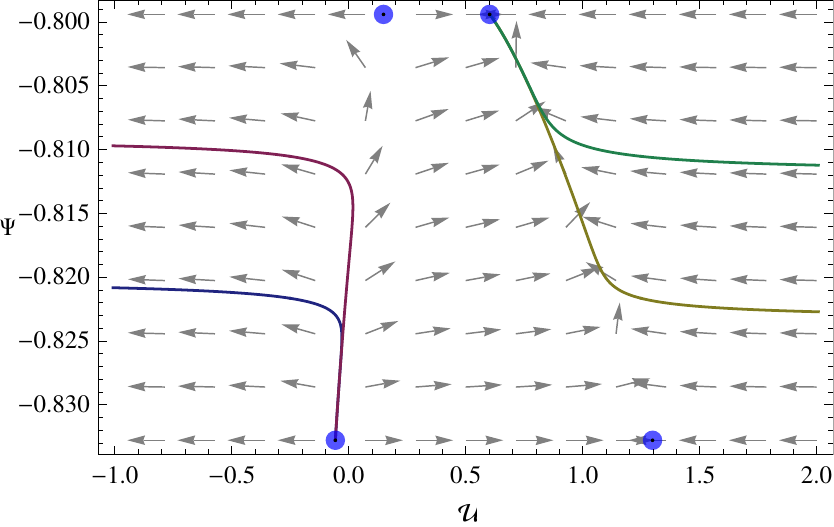}
\caption{\label{fig28} Density perturbations described by orbits in the phase
 plane $(\Psi,\mathcal{U}_{(n)})$ for an quartic potential with $\chi=3/2$ taking the long wavelength limit. The figure shows the late time attractor in the region $\mathcal{U}>0$.}
\end{minipage}
\end{figure}

\begin{figure}[h!]
\centering
\begin{minipage}{\textwidth/2-1pc}
\includegraphics[width=\textwidth]{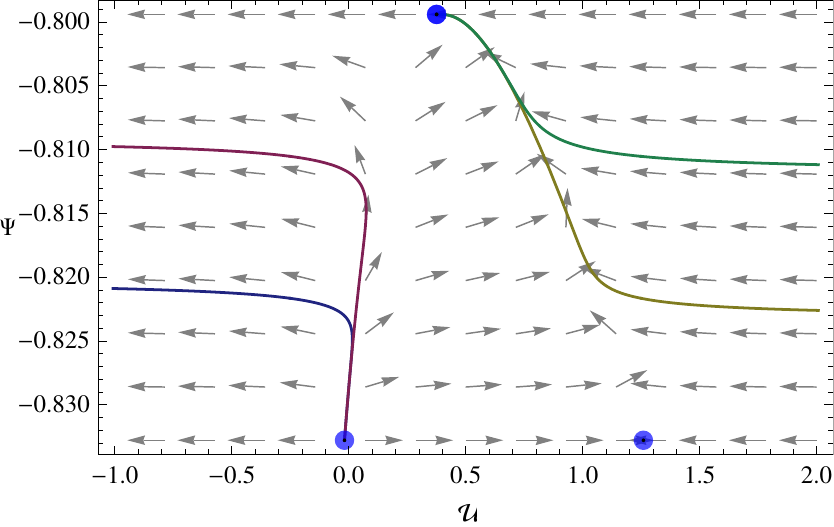}
\caption{\label{fig29} Density perturbations described by orbits in the phase
 plane $(\Psi,\mathcal{U}_{(n)})$ for a quartic potential with $\chi=3/2$ for $n^{2}=n^{2}_{crit}(3/2)$. The figure shows the saddle point in the region $\mathcal{U}>0$.}
\end{minipage}\hspace{2pc}%
\begin{minipage}{\textwidth/2-1pc}
\includegraphics[width=\textwidth]{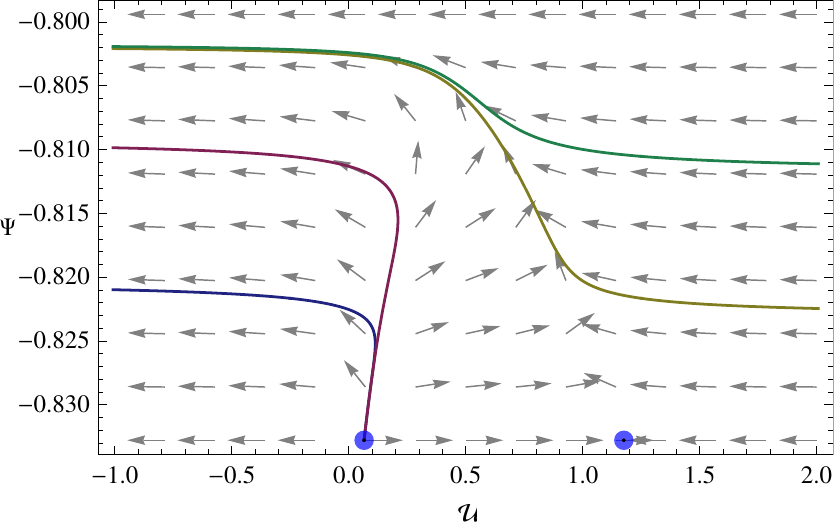}
\caption{\label{fig30} Density perturbations described by orbits in the phase
 plane $(\Psi,\mathcal{U}_{(n)})$ for a quartic potential with $\chi=3/2$, for $n^{2}>n^{2}_{crit}(3/2)$. The figure shows periodic orbits and the perturbations behave as waves.}
\end{minipage}
\end{figure}

\begin{figure}[h!]
\centering
\begin{minipage}{\textwidth/2-1pc}
\includegraphics[width=\textwidth]{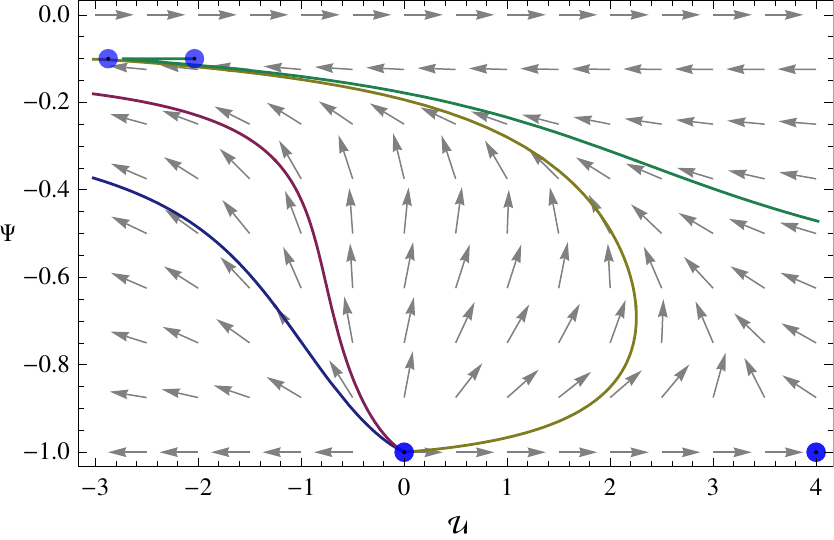}
\caption{\label{Fig31}  Density perturbations described by orbits in the phase
plane $(\Psi,\mathcal{U}_{(n)})$ for new inflation with $\mathcal{M}=0.3$ and $\mathcal{N}=0.1$ in the long wavelength limit.}
\end{minipage}\hspace{2pc}%
\begin{minipage}{\textwidth/2-1pc}
\includegraphics[width=\textwidth]{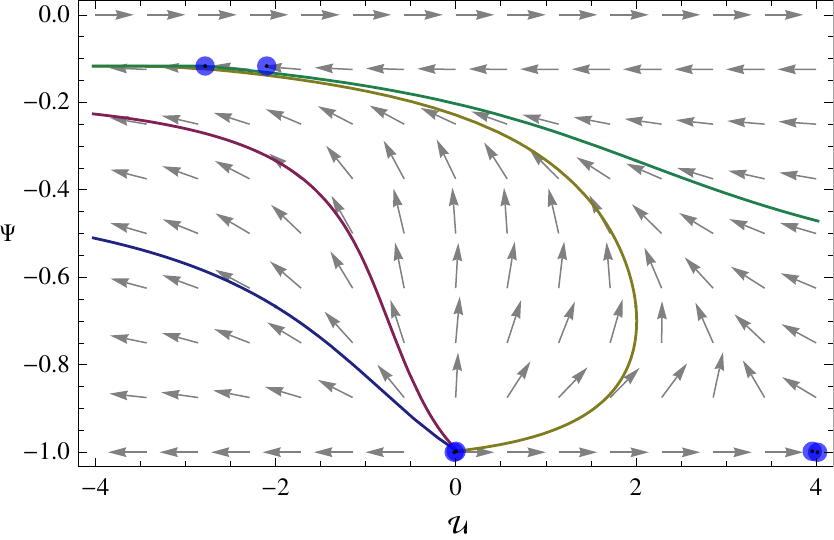}
\caption{\label{Fig32}Density perturbations described by orbits in the phase
plane $(\Psi,\mathcal{U}_{(n)})$ for new inflation with $\mathcal{M}=0.3$ and $\mathcal{N}=2$ in the long wavelength limit.}
\end{minipage}
\end{figure}

%
{\bf (i) Quadratic potential}
\\\\
We have seen, in Section \ref{subsection_5}, that for a quadratic potential the method of \cite{ULRI09} gives two fixed points $\mathcal{P}_{0}$ and $\mathcal{P}_{1}$ given by \eqref{P0_QuadraticPot} and \eqref{P1_QuadraticPot}, respectively. 
Then, for the perturbed system (\ref{PertSys_single}), it follows from Lemma $2$ that there are four fixed points
\begin{equation}
 \left(\mathcal{P}_{0},\mathcal{U}^{\pm}_{(\text{n})}(\mathcal{P}_{0})\right),~~~ \left(\mathcal{P}_{1},\mathcal{U}^{\pm}_{(\text{n})}(\mathcal{P}_{1})\right)
\end{equation}
where the coefficients \eqref{Coef_PertSys_Single} are 
\begin{equation}\label{Coef_QuadraticPot}
\begin{aligned}
 \xi(\mathcal{P}_{0,1})&=\frac{1}{2}\mp\frac{9}{2}\sqrt{1-\frac{4}{9}\mathcal{M}^{2}} \\
 \zeta(\mathcal{P}_{0,1})&=3\mp3\sqrt{1-\frac{4}{9}\mathcal{M}^{2}}+\frac{\text{n}^{2}}{H^{2}a^{2}}\;.
\end{aligned}
\end{equation}
We also saw that $\mathcal{P}_{1}$ is the late time attractor of the background subsystem. 
Therefore, the late time attractor of the perturbed dynamical system \eqref{pert-system} is 
$\left(\mathcal{P}_{1},\mathcal{U}^{+}_{(\text{n})}(\mathcal{P}_{1})\right)$. 
From \eqref{Stability_Cond} and \eqref{Coef_QuadraticPot}, we easily see that this fixed point always lies in the region 
$\mathcal{U}_{(\text{n})}<0$ of the phase-space, and it only exists if \eqref{RealFP_Cond} is satisfied (see Fig. \ref{fig15}), i.e. 
for values of the wave number satisfying
\begin{equation}
 \text{n}^{2}\leq a^{2}H^{2}\left(\frac{17}{8}-\frac{15}{8}\sqrt{1-\frac{4}{9}\mathcal{M}^{2}}-\frac{9}{4}\mathcal{M}^{2}\right),
\end{equation}
which, in turn, implies
\begin{equation}\label{open-set}
 \mathcal{M}\leq \frac{2}{9}\sqrt{16-5\sqrt{7}}.
\end{equation}
Thus, when $\mathcal{M}=\frac{2}{9}\sqrt{16-5\sqrt{7}}$, the fixed point only exists in the long wavelength limit and it is a 
saddle point (see Fig. \ref{fig16}), while for $\mathcal{M}>\frac{2}{9}\sqrt{16-5\sqrt{7}}$, the orbits are periodic. 
See Figs. \ref{fig17}, \ref{fig18}, \ref{fig19} and \ref{fig20} for the case $\mathcal{M}=10^{-1}<\frac{2}{9}\sqrt{16-5\sqrt{7}}$ and 
Table \ref{tb:tablename} for a summary of results.
\\\\
{\bf (ii) Quartic potential}
\\\\
We have seen, in Section \ref{subsection_5}, that for a quartic potential the method of \cite{ULRI09}, used in \cite{KT08}, showed the existence of two fixed points $\mathcal{P}_{0}$ and $\mathcal{P}_{1}$ given by \eqref{P_QuarticPot}. 
Then, for the perturbed system \eqref{PertSys_single} it follows, from \eqref{PertSys_FP}, that there are four fixed points
\begin{equation}
 \left(\mathcal{P}_{0},\mathcal{U}^{\pm}_{(\text{n})}(\mathcal{P}_{0})\right),~~~ \left(\mathcal{P}_{1},\mathcal{U}^{\pm}_{(\text{n})}(\mathcal{P}_{1})\right)
\end{equation}
with the coefficients \eqref{Coef_PertSys_Single} given by 
\begin{equation}\label{QuarticPot_coeff}
\begin{aligned}
\xi(\mathcal{P}_{0})&=2\left(1-3\cos{\left(\frac{2}{3}\chi\right)}\right)~~\text{and}~~~ \mathcal{\xi}(\mathcal{P}_{1})=2\left(1-3\cos{\left(\frac{2}{3}(\chi-\pi)\right)}\right) \\
\zeta(\mathcal{P}_{0})&=\xi(\mathcal{P}_{0})+4\cos^{2}{\left(\frac{2}{3}\chi\right)}+\frac{\text{n}^{2}}{H^{2}a^{2}}~~\text{and}~~~
\zeta(\mathcal{P}_{1})=\xi(\mathcal{P}_{1})+4\cos^{2}{\left(\frac{2}{3}(\chi-\pi)\right)}+\frac{\text{n}^{2}}{H^{2}a^{2}}.
\end{aligned}
\end{equation}
From \eqref{RealFP_Cond}, the fixed points exist if
\begin{equation}
 \text{n}^{2}\leq\text{n}^{2}_{crit}=a^{2}H^{2}\left(5\cos^{2}{\left(\frac{2}{3}(\chi-\pi)\right)}-1\right),
\end{equation}
i.e. for  $\chi$ satisfying (see also Fig. \ref{fig23})
\begin{equation}\label{open-set_Quartic}
 0<\chi\leq\pi-\frac{3}{2}\arccos{\left(-\frac{1}{\sqrt{5}}\right)}\quad\text{or}\quad \pi-\frac{3}{2}\arccos{\left(\frac{1}{\sqrt{5}}\right)}\leq\chi\leq\frac{\pi}{2}.
\end{equation}
Furthermore, we find that $\zeta(\mathcal{P}_{1})>0$ for all values of $0<\chi<\pi/2$ and $\zeta(\mathcal{P}_{1})=0$ for $\chi=\pi/2$, see  Fig. \ref{fig22}. We also find that (see also Fig. \ref{fig21})
\begin{equation}\label{QuarticResults}
\begin{aligned}
 &\text{For}\quad 0<\chi<\pi-\frac{3}{2}\arccos{\left(\frac{1}{\sqrt{3}}\right)}\quad \text{then}\quad\xi(\mathcal{P}_{1})>0 \\
 &\text{For}\quad \chi=\pi-\frac{3}{2}\arccos{\left(\frac{1}{\sqrt{3}}\right)}\quad \text{then}\quad\xi(\mathcal{P}_{1})=0 \\
 &\text{For}\quad \pi-\frac{3}{2}\arccos{\left(\frac{1}{\sqrt{3}}\right)}<\chi\leq\frac{\pi}{2}\quad \text{then}\quad\xi(\mathcal{P}_{1})<0\;.
\end{aligned}
\end{equation}
Therefore, there are perturbation modes which decay for $0<\chi<\pi-\frac{3}{2}\arccos{\left(-\frac{1}{\sqrt{5}}\right)}$, 
see Figs.  \ref{fig24}, \ref{fig25} and \ref{fig26}, and which grow for $\pi-\frac{3}{2}\arccos{\left(\frac{1}{\sqrt{5}}\right)}<\chi<\pi/2$, 
see Figs. \ref{fig27}, \ref{fig28}, \ref{fig29} and \ref{fig30}. 
From \eqref{QuarticPot_CondInf}, the solution is inflationary for $\chi<\pi/4$ and the instabilities 
(in the sense of growing modes) occur when the solution is non-inflationary while, when inflation occurs, 
there are only decaying modes present. 
\\\\
{\bf (iii) New inflation}
\\\\
In the linearly perturbed case, we find that the fixed points of the system (\ref{pert-system}) are
\begin{equation}
  \left(\mathcal{P},\mathcal{U}^{\pm}_{(\text{n})}(\mathcal{P})\right),
\end{equation}
where $\mathcal{U}^{\pm}_{(n)}(\mathcal{P})$ are given by Eqs. \eqref{PertSys_FP}  with
 \begin{equation}
 \begin{aligned}
  \mathcal{\xi}(\mathcal{P})&=-\left[4-9\Phi^{4}_{\mathcal{P}}\right]\\
  \quad\mathcal{\zeta}(\mathcal{P})&=-3\Phi^{4}_{\mathcal{P}}\left(1-3\Phi^{4}_{\mathcal{P}}\right)+\mathcal{M}^{4}\mathcal{N}^{2}+\frac{\text{n}^{2}}{H^{2}a^{2}}\;
  \nonumber
 \end{aligned}
 \end{equation}
and are investigated numerically, see Figs. \ref{Fig31} and \ref{Fig32}. See Table \ref{tb:tablename} for a summary of the results of Section $5.1$.
\begin{small}\begin{center}
\begin{table*}[h!]
{\small
\hfill{}
\begin{tabular}{c |c| c| c| c  }
\hline
\textbf{$\mathcal{V}(\phi)$}&\textbf{Fut. Att.: $(\mathcal{P}_{1},\mathcal{U}^{+}_{\text{n}}(\mathcal{P}_{1}))$}& \textbf{Parameter Space} & \textbf{Phys. Mean.} & \textbf{Inf. Sol.}\\ [1ex]
\hline 
                                            & Pt $\Psi=-\frac{\lambda}{\sqrt{6}}$,\,$\mathcal{U}^+_{(n)}<0$ & $0<\lambda< \sqrt 2 $, $0\leq n^{2}<n^{2}_{crit}(\lambda)$ & decays &     \\ [1ex]
                                            & Pt $\Psi=-\frac{\lambda}{\sqrt{6}}$,\,$\mathcal{U}^+_{(n)}<0$ & $ \lambda=\sqrt 2 $, $0<n^{2}<n^{2}_{crit}(\lambda)$ & decays &     \\ [1ex]
                                            & Pt $\Psi=-\frac{\lambda}{\sqrt{6}}$,\,$\mathcal{U}^+_{(n)}=0$ & $ \lambda=\sqrt 2 $, $n^{2}=0$ & tend to a const. &     \\ [1ex]    
 $\Lambda e^{\lambda\phi}$& Pt $\Psi=-\frac{\lambda}{\sqrt{6}}$,\,$\mathcal{U}^+_{(n)}<0$ & $ \sqrt 2<\lambda<\sqrt \frac{10}{3}$, $0<n^{2}_{-}(\lambda)<n^{2}<n^{2}_{crit}(\lambda)$ & decays & $0<\lambda<\sqrt{2}$ \\ [1ex]
                                            & Pt $\Psi=-\frac{\lambda}{\sqrt{6}}$,\,$\mathcal{U}^+_{(n)}=0$ & $ \sqrt 2<\lambda<\sqrt \frac{10}{3}$, $n^{2}=n^{2}_{-}(\lambda)$ & tend to a const. &  \\ [1ex]
                                            & Pt $\Psi=-\frac{\lambda}{\sqrt{6}}$,\,$\mathcal{U}^+_{(n)}>0$ & $ \sqrt 2<\lambda<\sqrt \frac{10}{3}$, $0\leq n^{2}<n^{2}_{-}(\lambda)$ & grows &  \\ [1ex]
                                            & Pt $\Psi=-\frac{\lambda}{\sqrt{6}}$,\,$\mathcal{U}^+_{(n)}>0$ & $ \frac{10}{3}\leq\lambda<\sqrt{6}$, $0\leq n^{2}<n^{2}_{crit}(\lambda)$ & grows &  \\ [1ex]
                                            & Per. orb. $\Psi=-\frac{\lambda}{\sqrt{6}}$ & for all $\lambda$, $n^{2}>n^{2}_{crit}(\lambda)$ & Pert. is a wave \\ [1ex]
                           
\hline 
\multirow{2}{*}{$\frac{m^{2}}{2}\phi^{2}$}   & Point $\Psi=\Psi_{{\cal P}_1}$,\,$\mathcal{U}^+_{(n)}<0$ & $0<\mathcal{M}<\frac{2}{9}\sqrt{16-5\sqrt{7}}$, $n^{2}<n^{2}_{crit}$ & Pert. decays     & \multirow{2}{*}{$0<\mathcal{M}<\sqrt{2}$}    \\ [1ex]
                                              & Periodic orbit $\Psi=\Psi_{{\cal P}_1}$              &       $\mathcal{M}>\frac{2}{9}\sqrt{16-5\sqrt{7}}$                             & Pert. is a wave  & \\  
\hline
\multirow{2}{*}{$\frac{\lambda^{4}}{4}\phi^{4}$}   & Point $\Psi=\Psi_{\mathcal{P}_{1}}$,\,$\mathcal{U}^+_{(n)}<0$ & & Pert. decays     & \multirow{2}{*}{$0<\mathcal{M}<6^{\frac{1}{4}}$}    \\ [1ex]
                                              & Periodic orbit $\Psi=\Psi_{\mathcal{P}_{1}}$  &  See \eqref{QuarticResults} & Pert. is a wave  & \\  
                                              & Point $\Psi=\Psi_{\mathcal{P}_{1}},\mathcal{U}^+_{(n)}>0$ &  & Pert. grows & \\
\hline
\end{tabular}}
\hfill{}
\caption{Summary of the results of Section $5.1$ about the behaviour of density perturbations in flat FLRW scalar field backgrounds with exponential, quadratic and quartic potentials.}
\label{tb:tablename}
\end{table*}
\end{center}            
\end{small}
\subsection{Two scalar fields}
In this section, we consider the simplified case where two scalar fields $\phi_{1}$ and $\phi_2$ do not interact with each 
other, i.e. with $\mathcal{W}(\phi_1,\phi_2)=0$. In this case, we find from Proposition \ref{PERTSYSTEM} the following result:
\begin{cor}
The evolution for the phase of first order scalar perturbations on a FLRW background, with two nonlinear smooth scalar fields, is given by the following system of differential equations for the state vector $\left((\Psi_{1},\Psi_{2},\Phi_{1},\Phi_{1}),\mathcal{U},\mathcal{X}_{[12]},\mathcal{Y}_{[12]}\right)$:
\begin{eqnarray}\label{2SFs_pert-system}
\begin{aligned}
  \mathcal{U}^{\prime}&=-\mathcal{U}^{2}-\xi\mathcal{U}-\zeta+\gamma_{12}\mathcal{X}_{[12]}+\eta_{12}\mathcal{Y}_{[12]} \\
 \mathcal{X}^{\prime}_{[12]}&=\left(-\mathcal{U}+\varsigma_{12}\right)\mathcal{X}_{[12]}+\varpi_{12}\mathcal{Y}_{[12]}  \\
 \mathcal{Y}^{\prime}_{[12]}&=\left(-\mathcal{U}+\iota_{12}\right)\mathcal{Y}_{[12]}-\mathcal{X}_{[12]}\\
  \Psi^{\prime}_{1}&=(q-2)\Psi_{1}-n\sqrt{6}\Phi^{2n-1}_{1}\frac{\partial\Phi_{1}}{\partial\phi_{1}} \\
  \Psi^{\prime}_{2}&=(q-2)\Psi_{2}-n\sqrt{6}\Phi^{2n-1}_{2}\frac{\partial\Phi_{2}}{\partial\phi_{2}} \\
  \Phi^{\prime}_{1}&=\frac{1}{n}(q+1)\Phi_{1}+\sqrt{6}\Psi_{1}\frac{\partial\Phi_{1}}{\partial\phi_{1}} \\
  \Phi^{\prime}_{2}&=\frac{1}{n}(q+1)\Phi_{2}+\sqrt{6}\Psi_{2}\frac{\partial\Phi_{2}}{\partial\phi_{2}}\;, \\
\end{aligned}
  \end{eqnarray}
subject to the  background constraint equation
\begin{equation}
 \Psi^{2}_{1}+\Psi^{2}_{2}+\Phi^{2n}_{1}+\Phi^{2n}_{2}=1-K,
\end{equation}
{\it where}
\begin{equation}
 q=2\left(\Psi^{2}_{1}+\Psi^{2}_{2}\right)-\left(\Phi^{2n}_{1}+\Phi^{2n}_{2}\right)
\end{equation}
and the coefficients are given in Section $6.2$ of the appendix.
\end{cor}
We also prove from \eqref{2SFs_pert-system} and Lemma \ref{Lemma_1}:
\begin{lem}
\label{Lemma4}
For $K=0$, the fixed points of system (\ref{pert-system}) are given by Lemma \ref{Lemma_1} together with:
\begin{equation*}
-\mathcal{U}^{2}-\xi(\mathcal{P})\mathcal{U}-\zeta(\mathcal{P})=0\quad\text{and}\quad \gamma_{12}(\mathcal{P})\mathcal{X}_{[12]}+\eta_{12}(\mathcal{P})\mathcal{Y}_{[12]}=0 \\
\end{equation*}
or
\begin{equation*}
-\mathcal{U}^{2}-\left(\varsigma_{12}(\mathcal{P})+\iota_{12}(\mathcal{P})\right)\mathcal{U}+\left(\varsigma_{12}(\mathcal{P})\iota_{12}(\mathcal{P})+\varpi_{12}(\mathcal{P})\right)=0\quad\text{and}\quad \gamma_{12}(\mathcal{P})\mathcal{X}_{[12]}+\eta_{12}(\mathcal{P})\mathcal{Y}_{[12]}\neq0,
\end{equation*}
where the coefficients $\xi$, $\zeta$, $\gamma_{12}$, $\eta_{12}$, $\varsigma_{12}$, $\iota_{12}$ and $\varpi_{12}$ at $\mathcal{P}$ are given in appendix.
\end{lem}
In particular, if $$\gamma_{12}(\mathcal{P})=\eta_{12}(\mathcal{P})=0$$ 
then 
\begin{equation}\label{Specialcase}
\mathcal{Y}_{[12]}(\mathcal{P})=\mathcal{X}_{[12]}(\mathcal{P})=0\quad \text{and}\quad \mathcal{U}^{\pm}_{\text{n}}(\mathcal{P})=\frac{1}{2}\left(-\xi(\mathcal{P})\pm\sqrt{\xi(\mathcal{P})^2-4\zeta(\mathcal{P})}\right)\;. 
\end{equation} 
and the linearised matrix of the system at the fixed points has
eigenvalues given by the background eigenvalues together with
\begin{equation}
 -2\mathcal{U}^{\pm}(\mathcal{P})-\xi(\mathcal{P})=\mp\sqrt{\xi^{2}(\mathcal{P})-4\zeta(\mathcal{P})}\;
\end{equation}
and
\begin{small}\begin{equation}\label{2NDEigen}
 \frac{1}{2}\left[-2\mathcal{U}^{\pm}+\iota_{12}+\varsigma_{12}\pm\sqrt{\left(2\mathcal{U}^{\pm}-\iota_{12}-\varsigma_{12}\right)^{2}-4\left(\left(\mathcal{U}^{\pm}\right)^{2}-\left(\iota_{12}+\varsigma_{12}\right)\mathcal{U}^{\pm}+\varpi_{12}+\iota_{12}\varsigma_{12}\right)}\right].
\end{equation}              
\end{small}
In the next section, we shall apply Lemma~\ref{Lemma4} to the case of two scalar fields with independent exponential potentials.
%
\subsubsection{Assisted power-law inflation}
In this case, we recall from Section $2.12$ that the fixed points are characterised by
\begin{equation}
 \mathcal{P}:~~ \frac{\partial^{2}\Phi_{A}}{\partial\phi^{2}_{A}}=-\frac{\sqrt{6}}{4}\lambda_{A}\frac{\Phi_{A}}{\Psi_{A}}\Psi^{2}\quad,\quad \frac{\partial^{3}\Phi_{A}}{\partial\phi^{3}_{A}}=-\frac{\sqrt{6}}{8}\lambda^{2}_{A}\frac{\Phi_{A}}{\Psi_{A}}\Psi^{2}
\end{equation}
so that, using equations \eqref{P0_APLI}-\eqref{P3_APLI}, the coefficients of the system \eqref{2SFs_pert-system} become
\begin{equation*}
 \xi(\mathcal{P}_0)=-4  \quad,\quad \xi(\mathcal{P}_{1,2})=5-\frac{3}{2}\lambda^{2}_{1,2} \quad,\quad \xi(\mathcal{P}_3)=5-\frac{3}{2}\lambda^{2}
\end{equation*}
\begin{equation*}
\begin{aligned}
 \zeta(\mathcal{P}_0)=\frac{\text{n}^{2}}{a^{2}H^{2}}\quad,\quad \zeta(\mathcal{P}_{1,2})= 6-4\lambda^{2}_{1,2}+\frac{\lambda^{4}_{1,2}}{2}+\frac{\text{n}^{2}}{a^{2}H^{2}}\quad,\quad \zeta(\mathcal{P}_{3})= 6-4\lambda^{2}+\frac{\lambda^{4}}{2}+\frac{\text{n}^{2}}{a^{2}H^{2}}  
\end{aligned}
\end{equation*}
\begin{equation*}
 \gamma_{12}(\mathcal{P}_0)=0 \quad,\quad \gamma_{12}(\mathcal{P}_{1,2})=0 \quad,\quad \gamma_{12}(\mathcal{P}_3)=0
\end{equation*}
\begin{equation*}
 \eta_{12}(\mathcal{P}_0)=0 \quad,\quad \eta_{12}(\mathcal{P}_{1,2})=0 \quad,\quad \eta_{12}(\mathcal{P}_3)=0
\end{equation*}
\begin{equation*}
 \varsigma_{12}(\mathcal{P}_0)=0 \quad,\quad \varsigma_{12}(\mathcal{P}_{1,2})=0 \quad,\quad \varsigma_{12}(\mathcal{P}_3)=\frac{6-\lambda^{2}}{2}
\end{equation*}
\begin{equation*}
 \iota_{12}(\mathcal{P}_0)=0 \quad,\quad \iota_{12}(\mathcal{P}_{1,2})=\frac{6-\lambda^{2}_{1,2}}{2} \quad,\quad \iota_{12}(\mathcal{P}_3)=0
\end{equation*}
\begin{equation*}
 \begin{aligned}
  \varpi_{12}(\mathcal{P}_{0})=\frac{\text{q}^{2}}{a^{2}H^{2}}\quad,\quad   \varpi_{12}(\mathcal{P}_{1,2})=\frac{\text{q}^{2}}{a^{2}H^{2}}\quad,\quad   \varpi_{12}(\mathcal{P}_{3})=\frac{\lambda^{2}}{2}\left(6-\lambda^{2}\right)+\frac{\text{q}^{2}}{a^{2}H^{2}}\\
 \end{aligned}
\end{equation*}
where $\text{q}$ is the wave number associated with the velocity scalar perturbations. 
Then, from equations \eqref{Specialcase}-\eqref{2NDEigen}, the fixed points of the system reduce to
\begin{equation}
 \left(\left(\Psi_{A},\Phi_{A}\right),\mathcal{U},\mathcal{X}_{[12]},\mathcal{Y}_{[12]}\right)=\left(\mathcal{P}_{i},\mathcal{U}^{\pm}_{\text{n}}(\mathcal{P}_{i}),0,0\right)\quad,\quad i=0,1,2,3,
\end{equation}
with eigenvalues given by the background solution \eqref{P0_Eigen}-\eqref{P3_Eigen} together with
\begin{equation}
\omega^{\pm}(\mathcal{P}_{0})=\mp2\sqrt{4-\frac{\text{n}^{2}}{a^{2}H^{2}}}\;,\;  -2\mp\sqrt{4-\frac{\text{n}^{2}}{a^{2}H^{2}}}\pm\sqrt{-\frac{\text{q}^{2}}{a^{2}H^{2}}}\;,
\end{equation}
\begin{equation}
\begin{aligned}
 \omega^{\pm}(\mathcal{P}_{1,2})=&\mp2\sqrt{\frac{1}{4}\left(1+\lambda^{2}_{1,2}+\frac{\lambda^{4}_{1,2}}{4}\right)-\frac{\text{n}^{2}}{a^{2}H^{2}}}\;, \\
                                 &\left(4-\lambda^{2}_{1,2}\right)\mp\sqrt{\frac{1}{4}\left(1+\lambda^{2}_{1,2}+\frac{\lambda^{4}_{1,2}}{4}\right)-\frac{\text{n}^{2}}{a^{2}H^{2}}}\pm\sqrt{\frac{1}{4}\left(\frac{6-\lambda^{2}_{1,2}}{2}\right)^{2}-\frac{\text{q}^{2}}{a^{2}H^{2}}}\;,
\end{aligned}
\end{equation}
\begin{small}\begin{equation}
\begin{aligned}
 \omega^{\pm}(\mathcal{P}_{3})=&\mp2\sqrt{\frac{1}{4}\left(1+\lambda^{2}+\frac{\lambda^{4}}{4}\right)-\frac{\text{n}^{2}}{a^{2}H^{2}}}\;, \\
                                 &\left(4-\lambda^{2}\right)\mp\sqrt{\frac{1}{4}\left(1+\lambda^{2}+\frac{\lambda^{4}}{4}\right)-\frac{\text{n}^{2}}{a^{2}H^{2}}}\pm\sqrt{\frac{1}{4}\left(\frac{6-\lambda^{2}}{2}\right)^{2}-\frac{\lambda^{2}}{2}(6-\lambda^{2})-\frac{\text{q}^{2}}{a^{2}H^{2}}}\;.
\end{aligned}
\end{equation}              \end{small}
Thus, the behaviour of $\mathcal{U}^{+}$, being the future attractor, is similar to the case of a single scalar field solution with more restrictions due to the new eigenvalue \eqref{2NDEigen}. 
In particular, the parameter space constraints of Table 1 also apply to this case for $\lambda$ defined by 
$1/\lambda^2=1/\lambda_1^2+1/\lambda_2^2$. However, there are further constraints to the parameters due to the velocity perturbations 
(which were not present in the single scalar field case) that can be inferred, in each case, from the sign of the second eigenvalue of $\omega^{+}(\mathcal{P}_{3})$.
For instance, the fixed points for the perturbations of the massless scalar field solution only exist if $\text{q}/(aH)\ll1$.

The results of this section appear to be easily extendible to the case of an arbitrary number of scalar fields 
with this type of potentials.
\section*{Acknowledgments}
 We thank Claes Uggla for very helpful comments  and Ure\~na-L\'opez for pointing out useful references. 
 We acknowledge support from CMAT, Univ. Minho (through the FEDER Funds COMPETE and FCT Project Est-C/MAT/UI0013/2011), 
 as well as from FCT projects PTDC/MAT/108921/2008 and CERN/FP/123609/2011.
\section{Appendix} 
In this appendix, we present the explicit expressions for the derivatives of the coefficients in equations \eqref{ddotDelta} and for the coefficients of \eqref{pert-system}.
\subsection{Evolution equations for $A, B_{AB}$ and $C_{AB}$}
From
\begin{equation}
\begin{aligned}
\dot{\alpha}_{A}\beta^{2}_{A}&+\dot{\alpha}_{B}\beta^{2}_{B}+\alpha_{A}\dot{(\beta^{2}_{A})}+\alpha_{B}\dot{(\beta^{2}_{B})}= \\
                            =&3H\left(\alpha_{A}\beta^{2}_{A}+\alpha_{B}\beta^{2}_{B}\right)+\left(\alpha^{2}_{A}\beta^{2}_{A}+\alpha^{2}_{B}\beta^{2}_{B}\right)-2\left(\alpha_{A}-\alpha_{B}\right)B_{AB}  \\
              &+\left(\frac{d^{2}\mathcal{V}}{d\phi^{2}_{A}}+\frac{\partial^{2}\mathcal{W}}{\partial\phi^{2}_{A}}+\frac{\psi_{B}}{\psi_{A}}\frac{\partial^{2}\mathcal{W}}{\partial\phi_{B}\partial\phi_{A}}\right)\beta^{2}_{A}+\left(\frac{d^{2}\mathcal{V}}{d\phi^{2}_{B}}+\frac{\partial^{2}\mathcal{W}}{\partial\phi^{2}_{B}}+\frac{\psi_{A}}{\psi_{B}}\frac{\partial^{2}\mathcal{W}}{\partial\phi_{A}\partial\phi_{B}}\right)\beta^{2}_{B}\\
              &+\sum^{N}_{C\neq A,B}\psi_{C}\left[\left(\frac{\beta^{2}_{A}}{\psi_{A}}\frac{\partial^{2}\mathcal{W}}{\partial\phi_{C}\partial\phi_{A}}+\frac{\beta^{2}_{B}}{\psi_{B}}\frac{\partial^{2}\mathcal{W}}{\partial\phi_{C}\partial\phi_{B}}\right)-2\left(\alpha_{A}B_{AC}+\alpha_{B}B_{BC}\right)\right]
\end{aligned}
\end{equation}
and equations \eqref{Der_alpha}-\eqref{Der_beta} we get
\begin{small}\begin{equation}
\begin{aligned}
\label{Aevol}
  \dot{A}=&3\dot{H}+6H\left(\alpha_{A}\beta^{2}_{A}+\alpha_{B}\beta^{2}_{B}\right)+2\left(\alpha^{2}_{A}\beta^{2}_{A}+\alpha^{2}_{B}\beta^{2}_{B}\right) -4\left(\alpha_{A}-\alpha_{B}\right)B_{AB} \\
          &+2\left(\frac{d^{2}\mathcal{V}}{d\phi^{2}_{A}}+\frac{\partial^{2}\mathcal{W}}{\partial\phi^{2}_{A}}+\frac{\psi_{B}}{\psi_{A}}\frac{\partial^{2}\mathcal{W}}{\partial\phi_{B}\partial\phi_{A}}\right)\beta^{2}_{A}+2\left(\frac{d^{2}\mathcal{V}}{d\phi^{2}_{B}}+\frac{\partial^{2}\mathcal{W}}{\partial\phi^{2}_{B}}+\frac{\psi_{A}}{\psi_{B}}\frac{\partial^{2}\mathcal{W}}{\partial\phi_{A}\partial\phi_{B}}\right)\beta^{2}_{B}\\
          &+2\sum^{N}_{C\neq A,B}\left[-2\left(\alpha_{C}-\alpha_{A}\right)B_{CA}-2\left(\alpha_ {C}-\alpha_{B}\right)B_{CB}\right] \\
          &+2\sum^{N}_{C\neq A,B}\left[\frac{\psi_{A}}{\psi_{C}}\beta^{2}_{C}\frac{\partial^{2}\mathcal{W}}{\partial\phi_{A}\partial\phi_{C}}+\frac{\psi_{B}}{\psi_{C}}\beta^{2}_{C}\frac{\partial^{2}\mathcal{W}}{\partial\phi_{B}\partial\phi_{C}}+\psi_{C}\left(\frac{\beta^{2}_{A}}{\psi_{A}}\frac{\partial^{2}\mathcal{W}}{\partial\phi_{C}\partial\phi_{A}}+\frac{\beta^{2}_{B}}{\psi_{B}}\frac{\partial^{2}\mathcal{W}}{\partial\phi_{C}\partial\phi_{B}}\right)\right] \\
          &+2\sum^{N}_{C\neq A,B}\sum^{N}_{D\neq A,B, C}\left[\left(3H+\alpha_{C}\right)\alpha_{C}+\frac{d^{2}\mathcal{V}}{d\phi^{2}_{C}}+\frac{\partial^{2}\mathcal{W}}{\partial\phi^{2}_{C}}+\frac{\psi_{D}}{\psi_{C}}\frac{\partial^{2}\mathcal{W}}{\partial\phi_{D}\partial\phi_{C}}-2\alpha_{C}\left(\alpha_{C}-\alpha_{D}\right)\beta^{2}_{D}\right]\beta^{2}_{C} \\
\end{aligned}
\end{equation}\end{small}
and
\begin{equation}
 \begin{aligned}
\dot{B}_{AB}=
              &\left[3H+\left(\alpha_{A}+\alpha_{B}\right)+2\left(\alpha_{A}-\alpha_{B}\right)\left(\beta^{2}_{A}-\beta^{2}_{B}\right)\right]B_{AB} \\   
              &+\left[\left(\frac{d^{2}\mathcal{V}}{d\phi^{2}_{A}}-\frac{d^{2}\mathcal{V}}{d\phi^{2}_{B}}\right)+\left(\frac{\partial^{2}\mathcal{W}}{\partial\phi^{2}_{A}}-\frac{\partial^{2}\mathcal{W}}{\partial\phi^{2}_{B}}\right)+\left(\frac{\psi_{B}}{\psi_{A}}\frac{\partial^{2}\mathcal{W}}{\partial\phi_{B}\partial\phi_{A}}-\frac{\psi_{A}}{\psi_{B}}\frac{\partial^{2}\mathcal{W}}{\partial\phi_{A}\partial\phi_{B}}\right)\right]\beta^{2}_{A}\beta^{2}_{B} \\
              &+\sum^{N}_{C\neq A,B}\left[-2B_{AB}\left(\alpha_{A}+\alpha_{B}-2\alpha_{C}\right)\beta^{2}_{C}+\left(\frac{\psi_{C}}{\psi_{A}}\frac{\partial^{2}\mathcal{W}}{\partial\phi_{C}\partial\phi_{A}}-\frac{\psi_{C}}{\psi_{B}}\frac{\partial^{2}\mathcal{W}}{\partial\phi_{C}\partial\phi_{B}}\right)\beta^{2}_{A}\beta^{2}_{B}\right],
              \label{Bevol}
 \end{aligned}
\end{equation}

\begin{equation}
\begin{aligned}
\label{Cevol}
 \dot{C}_{AB}=&\left[3\dot{H}+3H\left(\alpha_{A}\beta^{2}_{A}+\alpha_{B}\beta^{2}_{B}\right)+\left(\alpha^{2}_{A}\beta^{2}_{A}+\alpha^{2}_{B}\beta^{2}_{B}\right)-2\left(\alpha_{A}-\alpha_{B}\right)B_{AB}\right]B_{AB} \\
              &+\left[\left(\frac{d^{2}\mathcal{V}}{d\phi^{2}_{A}}+\frac{\partial^{2}\mathcal{W}}{\partial\phi^{2}_{A}}+\frac{\psi_{B}}{\psi_{A}}\frac{\partial^{2}\mathcal{W}}{\partial\phi_{B}\partial\phi_{A}}\right)\beta^{2}_{A}+\left(\frac{d^{2}\mathcal{V}}{d\phi^{2}_{B}}+\frac{\partial^{2}\mathcal{W}}{\partial\phi^{2}_{B}}+\frac{\psi_{A}}{\psi_{B}}\frac{\partial^{2}\mathcal{W}}{\partial\phi_{A}\partial\phi_{B}}\right)\beta^{2}_{B}\right]B_{AB} \\
              &+\left(\frac{\psi_{A}}{\psi_{B}}\frac{\partial^{2}\mathcal{W}}{\partial\phi_{B}\partial\phi_{A}}-\frac{\psi_{B}}{\psi_{A}}\frac{\partial^{2}\mathcal{W}}{\partial\phi_{A}\partial\phi_{B}}\right)B_{AB}+\left(3H+\left(\alpha_{A}\beta^{2}_{A}+\alpha_{B}\beta^{2}_{B}\right)\right)\dot{B}_{AB} \\
              &+2\left[\left(\frac{d^{2}\mathcal{V}}{d\phi^{2}_{A}}-\frac{d^{2}\mathcal{V}}{d\phi^{2}_{B}}\right)+\left(\frac{\partial^{2}\mathcal{W}}{\partial\phi^{2}_{A}}-\frac{\partial^{2}\mathcal{W}}{\partial\phi^{2}_{B}}\right)-\left(\frac{\psi_{A}}{\psi_{B}}-\frac{\psi_{B}}{\psi_{A}}\right)\frac{\partial^{2}\mathcal{W}}{\partial\phi_{A}\partial\phi_{B}}\right]\left(\beta^{2}_{A}-\beta^{2}_{B}\right)B_{AB} \\
              &+\left[\left(\frac{d^{3}\mathcal{V}}{d\phi^{3}_{A}}+\frac{\partial^{3}\mathcal{W}}{\partial\phi^{3}_{A}}-\frac{\partial^{3}\mathcal{W}}{\partial\phi_{A}\partial\phi^{2}_{B}}-\frac{\partial^{3}\mathcal{W}}{\partial\phi^{2}_{B}\partial\phi_{A}}-\frac{\psi_{A}}{\psi_{B}}\frac{\partial^{3}\mathcal{W}}{\partial\phi_{A}\partial\phi_{B}\partial\phi_{A}}\right)\psi_{A}\right]\beta^{2}_{A}\beta^{2}_{B} \\
              &-\left[\left(\frac{d^{3}\mathcal{V}}{d\phi^{3}_{B}}+\frac{\partial^{3}\mathcal{W}}{\partial\phi^{3}_{B}}-\frac{\partial^{3}\mathcal{W}}{\partial\phi_{B}\partial\phi^{2}_{A}}-\frac{\partial^{3}\mathcal{W}}{\partial\phi^{2}_{A}\partial\phi_{B}}-\frac{\psi_{B}}{\psi_{A}}\frac{\partial^{3}\mathcal{W}}{\partial\phi_{B}\partial\phi_{A}\partial\phi_{B}}\right)\psi_{B}\right]\beta^{2}_{A}\beta^{2}_{B} \\
              &+2\sum^{N}_{C\neq A,B}\left[\left(\frac{d^{2}\mathcal{V}}{d\phi^{2}_{A}}-\frac{d^{2}\mathcal{V}}{d\phi^{2}_{B}}\right)+\left(\frac{\partial^{2}\mathcal{W}}{\partial\phi^{2}_{A}}-\frac{\partial^{2}\mathcal{W}}{\partial\phi^{2}_{B}}\right)\right]\left(B_{CA}\beta^{2}_{B}+B_{CB}\beta^{2}_{A}\right) \\
              &-2\sum^{N}_{C\neq A,B}\left[\left(\frac{\psi_{A}}{\psi_{B}}-\frac{\psi_{B}}{\psi_{A}}\right)\frac{\partial^{2}\mathcal{W}}{\partial\phi_{B}\partial\phi_{A}}\right]\left(B_{CA}\beta^{2}_{B}+B_{CB}\beta^{2}_{A}\right) \\
              &+\sum^{N}_{C\neq A,B}\left[(3H+\alpha_{C})\alpha_{C}+\frac{d^{2}\mathcal{V}}{d\phi^{2}_{C}}+\frac{\partial^{2}\mathcal{W}}{\partial\phi^{2}_{C}}+\frac{\psi_{A}}{\psi_{C}}\frac{\partial^{2}\mathcal{W}}{\partial\phi_{A}\partial\phi_{C}}+\frac{\psi_{B}}{\psi_{C}}\frac{\partial^{2}\mathcal{W}}{\partial\phi_{B}\partial\phi_{C}}\right]\beta^{2}_{A}\beta^{2}_{B} \\
              &+\sum^{N}_{C\neq A,B}\left[-2\left(\alpha_{A}B_{AC}+\alpha_{B}B_{BC}\right)B_{AB}-2\alpha_{C}B_{AB}\left(B_{CA}+B_{CB}\right)+\alpha_{C}\beta^{2}_{C}\dot{B}_{AB}\right] \\
              &+\sum^{N}_{C\neq A,B}\left[\psi_{C}\left(\frac{\beta^{2}_{A}}{\psi_{A}}\frac{\partial^{2}\mathcal{W}}{\partial\phi_{C}\partial\phi_{A}}+\frac{\beta^{2}_{B}}{\psi_{B}}\frac{\partial^{2}\mathcal{W}}{\partial\phi_{C}\partial\phi_{B}}\right)B_{AB}\right] \\
              &+\sum^{N}_{C\neq A,B}\left[\left(\alpha_{A}-\alpha_{C}\right)\frac{\psi_{C}}{\psi_{A}}\frac{\partial^{2}\mathcal{W}}{\partial\phi_{A}\partial\phi_{C}}-\left(\alpha_{B}-\alpha_{C}\right)\frac{\psi_{C}}{\psi_{B}}\frac{\partial^{2}\mathcal{W}}{\partial\phi_{B}\partial\phi_{C}}\right] \\
              &+\sum^{N}_{C\neq A,B}\psi_{C}\left[\frac{\partial^{3}\mathcal{W}}{\partial\phi_{C}\partial\phi_{A}^{2}}-\frac{\partial^{3}\mathcal{W}}{\partial\phi_{C}\partial\phi_{B}^{2}}-\left(\frac{\psi_{A}}{\psi_{B}}\frac{\partial^{3}\mathcal{W}}{\partial\phi_{A}\partial\phi_{B}\partial\phi_{C}}-\frac{\psi_{B}}{\psi_{A}}\frac{\partial^{3}\mathcal{W}}{\partial\phi_{B}\partial\phi_{A}\partial\phi_{C}}\right)\right]\beta^{2}_{A}\beta^{2}_{B} \\
              &+\sum^{N}_{C\neq A,B}\left[\frac{\psi_{C}}{\psi_{A}}\left(\psi_{A}\frac{\partial^{3}\mathcal{W}}{\partial\phi_{C}\partial\phi_{A}^{2}}+\psi_{B}\frac{\partial^{3}\mathcal{W}}{\partial\phi_{B}\partial\phi_{A}\partial\phi_{C}}+\psi_{C}\frac{\partial^{3}\mathcal{W}}{\partial\phi^{2}_{C}\partial\phi_{A}}\right)\right] \\
              &-\sum^{N}_{C\neq A,B}\left[\frac{\psi_{C}}{\psi_{B}}\left(\psi_{A}\frac{\partial^{3}\mathcal{W}}{\partial\phi_{A}\partial\phi_{B}\partial\phi_{C}}+\psi_{B}\frac{\partial^{3}\mathcal{W}}{\partial\phi_{C}\partial\phi_{B}^{2}}+\psi_{C}\frac{\partial^{3}\mathcal{W}}{\partial\phi_{C}\partial\phi_{B}\partial\phi_{C}}\right)\right] \\
              &+\sum^{N}_{C\neq A,B}\sum^{N}_{D\neq A,B,C}\left[-2\alpha_{C}B_{AB}B_{CD}+B_{AB}\beta^{2}_{C}\frac{\psi_{D}}{\psi_{C}}\frac{\partial^{2}\mathcal{W}}{\partial\phi_{D}\partial\phi_{C}}\right] \\
              &+\sum^{N}_{C\neq A,B}\sum^{N}_{D\neq A,B,C}\left[\psi_{D}\left(\frac{\psi_{C}}{\psi_{A}}\frac{\partial^{3}\mathcal{W}}{\partial\phi_{D}\partial\phi_{A}\partial\phi_{C}}-\frac{\psi_{C}}{\psi_{B}}\frac{\partial^{3}\mathcal{W}}{\partial\phi_{D}\partial\phi_{B}\partial\phi_{C}}\right)\right].
\end{aligned}
\end{equation}
\subsection{Coefficients $\xi, \zeta, \gamma_{AB}, \eta_{AB}, \varsigma_{AB}, \varpi_{AB}$ and $\iota_{AB}$ }
The coefficients $\xi, \zeta, \gamma_{AB}, \eta_{AB}, \varsigma_{AB}, \varpi_{AB}$ and $\iota_{AB}$ are given by
\begin{equation}
\begin{aligned}
  \xi=&-\left[q+2+2\left(\frac{\alpha_{A}}{H}\beta^{2}_{A}+\frac{\alpha_{B}}{H}\beta^{2}_{B}\right)+2\sum^{N}_{C\neq A,B}\beta^{2}_{C}\frac{\alpha_{C}}{H}\right] \\
\end{aligned}
\end{equation}

\begin{small}\begin{equation}
\begin{aligned}
\zeta=&-\left[\frac{2HA+3\dot{H}+2\psi^2+\dot{A}}{H^{2}}-\frac{\text{n}^{2}}{a^{2}H^{2}}\right] \\
     =&-\left[6\left(1+\frac{\dot{H}}{H^{2}}\right)+2\frac{\psi^{2}}{H^{2}}+2\left[\frac{\alpha_{A}}{H}\beta^{2}_{A}\left(5+\frac{\alpha_{A}}{H}\right)+\frac{\alpha_{B}}{H}\beta^{2}_{B}\left(5+\frac{\alpha_{B}}{H}\right)\right]-4\frac{\left(\alpha_{A}-\alpha_{B}\right)}{H}B_{AB}\right] \\
     &-\frac{2}{H^{2}}\left[\left(\frac{d^{2}\mathcal{V}}{d\phi^{2}_{A}}+\frac{\partial^{2}\mathcal{W}}{\partial\phi^{2}_{A}}+\frac{\psi_{B}}{\psi_{A}}\frac{\partial^{2}\mathcal{W}}{\partial\phi_{B}\partial\phi_{A}}\right)\beta^{2}_{A}+\left(\frac{d^{2}\mathcal{V}}{d\phi^{2}_{B}}+\frac{\partial^{2}\mathcal{W}}{\partial\phi^{2}_{B}}+\frac{\psi_{A}}{\psi_{B}}\frac{\partial^{2}\mathcal{W}}{\partial\phi_{A}\partial\phi_{B}}\right)\beta^{2}_{B}\right] \\
     &-2\sum^{N}_{C\neq A,B}\left[-2\left(\frac{\alpha_{C}}{H}-\frac{\alpha_{A}}{H}\right)\frac{B_{CA}}{H}-2\left(\frac{\alpha_ {C}}{H}-\frac{\alpha_{B}}{H}\right)\frac{B_{CB}}{H}\right] \\
     &-\frac{2}{H^{2}}\sum^{N}_{C\neq A,B}\left[\frac{\Psi_{A}}{\Psi_{C}}\beta^{2}_{C}\frac{\partial^{2}\mathcal{W}}{\partial\phi_{A}\partial\phi_{C}}+\frac{\Psi_{B}}{\Psi_{C}}\beta^{2}_{C}\frac{\partial^{2}\mathcal{W}}{\partial\phi_{B}\partial\phi_{C}}+\Psi_{C}\left(\frac{\beta^{2}_{A}}{\Psi_{A}}\frac{\partial^{2}\mathcal{W}}{\partial\phi_{C}\partial\phi_{A}}+\frac{\beta^{2}_{B}}{\Psi_{B}}\frac{\partial^{2}\mathcal{W}}{\partial\phi_{C}\partial\phi_{B}}\right)\right] \\
          &-2\sum^{N}_{C\neq A,B}\sum^{N}_{D\neq A,B, C}\left[\left(3+\frac{\alpha_{C}}{H}\right)\frac{\alpha_{C}}{H}+\frac{1}{H^{2}}\frac{d^{2}\mathcal{V}}{d\phi^{2}_{C}}+\frac{1}{H^{2}}\frac{\partial^{2}\mathcal{W}}{\partial\phi^{2}_{C}}+\frac{1}{H^{2}}\frac{\Psi_{D}}{\Psi_{C}}\frac{\partial^{2}\mathcal{W}}{\partial\phi_{D}\partial\phi_{C}}\right]\beta^{2}_{C}
 \end{aligned}
\end{equation}              \end{small}

\begin{small}
\begin{equation*}
\begin{aligned}
 \gamma_{AB}=&-\left[\frac{\dot{B}_{AB}+\left(2H-D_{AB}\right)B_{AB}+C_{AB}}{H^2}\right] \\
            =&-\left[8+\frac{\left(\alpha_{A}+\alpha_{B}\right)}{H}+3\frac{\left(\alpha_{A}-\alpha_{B}\right)}{H}\left(\beta^{2}_{A}-\beta^{2}_{B}\right)\right]\frac{B_{AB}}{H} \\  
             &-\frac{1}{H^{2}}\left[2\left(\frac{d^{2}\mathcal{V}}{d\phi^{2}_{A}}-\frac{d^{2}\mathcal{V}}{d\phi^{2}_{B}}\right)+2\left(\frac{\partial^{2}\mathcal{W}}{\partial\phi^{2}_{A}}-\frac{\partial^{2}\mathcal{W}}{\partial\phi^{2}_{B}}\right)+\left(\frac{\partial^{2}\mathcal{W}}{\partial\phi_{B}\partial\phi_{A}}+\frac{\partial^{2}\mathcal{W}}{\partial\phi_{A}\partial\phi_{B}}\right)\left(\frac{\psi_{B}}{\psi_{A}}-\frac{\psi_{A}}{\psi_{B}}\right)\right]\beta^{2}_{A}\beta^{2}_{B} \\
             &+\frac{B_{AB}}{H}\sum^{N}_{C\neq A,B}\left[\left(2\frac{\alpha_{A}}{H}+2\frac{\alpha_{B}}{H}-3\frac{\alpha_{C}}{H}\right)\beta^{2}_{C}\right] \\
             &-\frac{1}{H^{2}}\sum^{N}_{C\neq A,B}\left[\frac{\psi_{C}}{\psi_{A}}\left(\frac{\partial^{2}\mathcal{W}}{\partial\phi_{A}\partial\phi_{C}}+\frac{\partial^{2}\mathcal{W}}{\partial\phi_{C}\partial\phi_{A}}\right)-\frac{\psi_{C}}{\psi_{B}}\left(\frac{\partial^{2}\mathcal{W}}{\partial\phi_{B}\partial\phi_{C}}+\frac{\partial^{2}\mathcal{W}}{\partial\phi_{C}\partial\phi_{B}}\right)\right] \\
\end{aligned}
\end{equation*} 
\end{small}
\begin{small}
\begin{equation*}
\begin{aligned}
 \eta_{AB}=&\left[\frac{\dot{C}_{AB}+\left(5H+D_{AB}\right)C_{AB}}{H^{3}}+\frac{B_{AB}}{H}\left(\frac{3\dot{H}+\psi^2-E_{AB}}{H^{2}}-2\frac{\text{q}^{2}}{a^{2}H^{2}}\right)\right] \\
                               =&\left\{6\frac{\dot{H}}{H^{2}}+24+\frac{\psi^{2}}{H^{2}}+\frac{\left(\alpha^{2}_{A}\beta^{2}_{A}+\alpha^{2}_{B}\beta^{2}_{B}\right)}{H^{2}}+3\frac{D_{AB}}{H}-\frac{E_{AB}}{H^{2}}-\frac{q^{2}}{a^{2}H^{2}}\right\}\frac{B_{AB}}{H}  \\  
                                &+\left\{2\frac{\left(\alpha_{A}-\alpha_{B}\right)}{H}\left[\left[3+\left(\frac{\alpha_{A}}{H}\beta^{2}_{A}+\frac{\alpha_{B}}{H}\beta^{2}_{B}\right)\right]\left(\beta^{2}_{A}-\beta^{2}_{B}\right)-\frac{B_{AB}}{H}\right]\right\}\frac{B_{AB}}{H}  \\  
                                &+\left\{\frac{\left(\alpha_{A}+\alpha_{B}\right)}{H}\left[3+\left(\frac{\alpha_{A}}{H}\beta^{2}_{A}+\frac{\alpha_{B}}{H}\beta^{2}_{B}\right)\right]+\left[11+\frac{D_{AB}}{H}\right]\left(\frac{\alpha_{A}}{H}\beta^{2}_{A}+\frac{\alpha_{B}}{H}\beta^{2}_{B}\right)\right\}\frac{B_{AB}}{H}  \\
                                &+\frac{1}{H^{2}}\left[\left(\frac{d^{2}\mathcal{V}}{d\phi^{2}_{A}}-\frac{d^{2}\mathcal{V}}{d\phi^{2}_{B}}\right)+\left(\frac{\partial^{2}\mathcal{W}}{\partial\phi^{2}_{A}}-\frac{\partial^{2}\mathcal{W}}{\partial\phi^{2}_{B}}\right)-\left(\frac{\Psi_{A}}{\Psi_{B}}-\frac{\Psi_{B}}{\Psi_{A}}\right)\frac{\partial^{2}\mathcal{W}}{\partial\phi_{A}\partial\phi_{B}}\right]\left[\left(8+\left(\frac{\alpha_{A}}{H}\beta^{2}_{A}+\frac{\alpha_{B}}{H}\beta^{2}_{B}\right)\frac{D_{AB}}{H}\right)\beta^{2}_{A}\beta^{2}_{B}\right] \\
                                &+\frac{1}{H^{2}}\left[\left(\frac{d^{2}\mathcal{V}}{d\phi^{2}_{A}}-\frac{d^{2}\mathcal{V}}{d\phi^{2}_{B}}\right)+\left(\frac{\partial^{2}\mathcal{W}}{\partial\phi^{2}_{A}}-\frac{\partial^{2}\mathcal{W}}{\partial\phi^{2}_{B}}\right)-\left(\frac{\Psi_{A}}{\Psi_{B}}-\frac{\Psi_{B}}{\Psi_{A}}\right)\frac{\partial^{2}\mathcal{W}}{\partial\phi_{A}\partial\phi_{B}}\right]\left[2\left(\beta^{2}_{A}-\beta^{2}_{B}\right)\frac{B_{AB}}{H}\right] \\
                               &\frac{1}{H^{2}}\left[\left(\frac{d^{2}\mathcal{V}}{d\phi^{2}_{A}}+\frac{\partial^{2}\mathcal{W}}{\partial\phi^{2}_{A}}+\frac{\Psi_{B}}{\Psi_{A}}\frac{\partial^2 \mathcal{W}_{AB}}{\partial \phi_A\partial\phi_B}\right)\beta^{2}_{A}+\left(\frac{d^{2}\mathcal{V}}{d\phi^{2}_{B}}+\frac{\partial^{2}\mathcal{W}}{\partial\phi^{2}_{B}}+\frac{\Psi_{A}}{\Psi_{B}}\frac{\partial^2 \mathcal{W}_{AB}}{\partial \phi_A\partial\phi_B}\right)\beta^{2}_{B}+\left(\frac{\Psi_{A}}{\Psi_{B}}-\frac{\Psi_{B}}{\Psi_{A}}\right)\frac{\partial^{2}\mathcal{W}}{\partial\phi_{A}\partial\phi_{B}}\right]\frac{B_{AB}}{H} \\
                               &+\frac{1}{H^{3}}\left[\left(\frac{d^{3}\mathcal{V}}{d\phi^{3}_{A}}+\frac{\partial^{3}\mathcal{W}}{\partial\phi^{3}_{A}}-\frac{\partial^{3}\mathcal{W}}{\partial\phi_{A}\partial\phi^{2}_{B}}-\frac{\partial^{3}\mathcal{W}}{\partial\phi^{2}_{B}\partial\phi_{A}}-\frac{\psi_{A}}{\psi_{B}}\frac{\partial^{3}\mathcal{W}}{\partial\phi_{A}\partial\phi_{B}\partial\phi_{A}}\right)\psi_{A}\right]\beta^{2}_{A}\beta^{2}_{B} \\
              &-\frac{1}{H^{3}}\left[\left(\frac{d^{3}\mathcal{V}}{d\phi^{3}_{B}}+\frac{\partial^{3}\mathcal{W}}{\partial\phi^{3}_{B}}-\frac{\partial^{3}\mathcal{W}}{\partial\phi_{B}\partial\phi^{2}_{A}}-\frac{\partial^{3}\mathcal{W}}{\partial\phi^{2}_{A}\partial\phi_{B}}-\frac{\psi_{B}}{\psi_{A}}\frac{\partial^{3}\mathcal{W}}{\partial\phi_{B}\partial\phi_{A}\partial\phi_{B}}\right)\psi_{B}\right]\beta^{2}_{A}\beta^{2}_{B} \\ 
              &+\left(\frac{5H+D_{AB}}{H}\right)\sum^{N}_{C\neq A,B}\left[\alpha_{C}\beta^{2}_{C}B_{AB}+\left(\frac{\psi_{C}}{\psi_{A}}\frac{\partial^{2}\mathcal{W}}{\partial\phi_{A}\partial\phi_{C}}-\frac{\psi_{C}}{\psi_{B}}\frac{\partial^{2}\mathcal{W}}{\partial\phi_{B}\partial\phi_{C}}\right)\right] \\
              &+\left(\frac{3H+\alpha_{A}\beta^{2}_{A}+\alpha_{B}\beta^{2}_{B}}{H}\right)\sum^{N}_{C\neq A,B}\left[-2B_{AB}\left(\alpha_{A}+\alpha_{B}-2\alpha_{C}\right)\beta^{2}_{C}+\left(\frac{\psi_{C}}{\psi_{A}}\frac{\partial^{2}\mathcal{W}}{\partial\phi_{C}\partial\phi_{A}}-\frac{\psi_{C}}{\psi_{B}}\frac{\partial^{2}\mathcal{W}}{\partial\phi_{C}\partial\phi_{B}}\right)\beta^{2}_{A}\beta^{2}_{B}\right] \\
              &+2\sum^{N}_{C\neq A,B}\left[\left(\frac{d^{2}\mathcal{V}}{d\phi^{2}_{A}}-\frac{d^{2}\mathcal{V}}{d\phi^{2}_{B}}\right)+\left(\frac{\partial^{2}\mathcal{W}}{\partial\phi^{2}_{A}}-\frac{\partial^{2}\mathcal{W}}{\partial\phi^{2}_{B}}\right)-\left(\frac{\psi_{A}}{\psi_{B}}\frac{\partial^{2}\mathcal{W}}{\partial\phi_{B}\partial\phi_{A}}-\frac{\psi_{B}}{\psi_{A}}\frac{\partial^{2}\mathcal{W}}{\partial\phi_{A}\partial\phi_{B}}\right)\right]\left(B_{CA}\beta^{2}_{B}+B_{CB}\beta^{2}_{A}\right) \\
              &+\sum^{N}_{C\neq A,B}\left[(3H+\alpha_{C})\alpha_{C}+\frac{d^{2}\mathcal{V}}{d\phi^{2}_{C}}+\frac{\partial^{2}\mathcal{W}}{\partial\phi^{2}_{C}}+\frac{\psi_{A}}{\psi_{C}}\frac{\partial^{2}\mathcal{W}}{\partial\phi_{A}\partial\phi_{C}}+\frac{\psi_{B}}{\psi_{C}}\frac{\partial^{2}\mathcal{W}}{\partial\phi_{B}\partial\phi_{C}}\right]\beta^{2}_{A}\beta^{2}_{B} \\
              &+\sum^{N}_{C\neq A,B}\left[-2\left(\alpha_{A}B_{AC}+\alpha_{B}B_{BC}\right)B_{AB}-2\alpha_{C}B_{AB}\left(B_{CA}+B_{CB}\right)+\alpha_{C}\beta^{2}_{C}\dot{B}_{AB}\right] \\
              &+\sum^{N}_{C\neq A,B}\left[\psi_{C}\left(\frac{\beta^{2}_{A}}{\psi_{A}}\frac{\partial^{2}\mathcal{W}}{\partial\phi_{C}\partial\phi_{A}}+\frac{\beta^{2}_{B}}{\psi_{B}}\frac{\partial^{2}\mathcal{W}}{\partial\phi_{C}\partial\phi_{B}}\right)B_{AB}+\left(\alpha_{A}-\alpha_{C}\right)\frac{\psi_{C}}{\psi_{A}}\frac{\partial^{2}\mathcal{W}}{\partial\phi_{A}\partial\phi_{C}}-\left(\alpha_{B}-\alpha_{C}\right)\frac{\psi_{C}}{\psi_{B}}\frac{\partial^{2}\mathcal{W}}{\partial\phi_{B}\partial\phi_{C}}\right] \\
              &+\sum^{N}_{C\neq A,B}\psi_{C}\left[\frac{\partial^{3}\mathcal{W}}{\partial\phi_{C}\partial\phi_{A}^{2}}-\frac{\partial^{3}\mathcal{W}}{\partial\phi_{C}\partial\phi_{B}^{2}}-\left(\frac{\psi_{A}}{\psi_{B}}\frac{\partial^{3}\mathcal{W}}{\partial\phi_{A}\partial\phi_{B}\partial\phi_{C}}-\frac{\psi_{B}}{\psi_{A}}\frac{\partial^{3}\mathcal{W}}{\partial\phi_{B}\partial\phi_{A}\partial\phi_{C}}\right)\right]\beta^{2}_{A}\beta^{2}_{B} \\
              &+\sum^{N}_{C\neq A,B}\left[\frac{\psi_{C}}{\psi_{A}}\left(\psi_{A}\frac{\partial^{3}\mathcal{W}}{\partial\phi_{C}\partial\phi_{A}^{2}}+\psi_{B}\frac{\partial^{3}\mathcal{W}}{\partial\phi_{B}\partial\phi_{A}\partial\phi_{C}}+\psi_{C}\frac{\partial^{3}\mathcal{W}}{\partial\phi_{C}\partial\phi_{A}\partial\phi_{C}}\right)\right] \\
              &-\sum^{N}_{C\neq A,B}\left[\frac{\psi_{C}}{\psi_{B}}\left(\psi_{A}\frac{\partial^{3}\mathcal{W}}{\partial\phi_{A}\partial\phi_{B}\partial\phi_{C}}+\psi_{B}\frac{\partial^{3}\mathcal{W}}{\partial\phi_{C}\partial\phi_{B}^{2}}+\psi_{C}\frac{\partial^{3}\mathcal{W}}{\partial\phi_{C}\partial\phi_{B}\partial\phi_{C}}\right)\right] \\
              &+\sum^{N}_{C\neq A,B}\sum^{N}_{D\neq A,B,C}\left[-2\alpha_{C}B_{AB}B_{CD}+B_{AB}\beta^{2}_{C}\frac{\psi_{D}}{\psi_{C}}\frac{\partial^{2}\mathcal{W}}{\partial\phi_{D}\partial\phi_{C}}+\psi_{D}\left(\frac{\psi_{C}}{\psi_{A}}\frac{\partial^{3}\mathcal{W}}{\partial\phi_{D}\partial\phi_{A}\partial\phi_{C}}-\frac{\psi_{C}}{\psi_{B}}\frac{\partial^{3}\mathcal{W}}{\partial\phi_{D}\partial\phi_{B}\partial\phi_{C}}\right)\right].
\end{aligned}
\end{equation*} 
\end{small}

\begin{equation*}
\begin{aligned}
 \varsigma_{AB}&=-\frac{D_{AB}}{H}=-\frac{1}{H}\left(\alpha_A\beta^{2}_{B}+\alpha_B\beta^{2}_{A}\right) \\
\end{aligned}
\end{equation*}

\begin{equation*}
\begin{aligned}
 \varpi_{AB}&=\frac{E_{AB}}{H^{2}}+\frac{\text{q}^{2}}{a^{2}H^{2}}
\end{aligned}
\end{equation*}

\begin{equation*}
\begin{aligned}
 \iota_{AB}&=\left[1-(1+q)+2+\frac{D_{AB}}{H}\right]
\end{aligned}
\end{equation*}
%
%
In the case of two scalar fields which do not interact with each other we have $\mathcal{W}(\phi_{1},\phi_{2})=0$ and, 
using expansion normalised variables \eqref{ENV}, the above coefficient read
\begin{equation}
\begin{aligned}
  \xi
     =&-\left[2+2\left(\Psi^{2}_{1}+\Psi^{2}_{2}\right)-\left(\Phi^{2n}_{1}+\Phi^{2n}_{2}\right)+2\sqrt{6}n\left[\frac{\Phi^{2n-1}_{1}}{\Psi_{1}}\left(\frac{\Psi^{2}_{1}}{\Psi^{2}}\frac{\partial\Phi_{1}}{\partial\phi_{1}}\right)+\frac{\Phi^{2n-1}_{2}}{\Psi_{2}}\left(\frac{\Psi^{2}_{2}}{\Psi^{2}}\frac{\partial\Phi_{2}}{\partial\phi_{2}}\right)\right]\right] \\
\end{aligned}
\end{equation}

\begin{small}\begin{equation}
\begin{aligned}
\zeta=&-\left[6\Phi^{2n}+2\sqrt{6}n\left[\frac{\Phi^{2n-1}_{1}}{\Psi_{1}}\frac{\partial\Phi_{1}}{\partial\phi_{1}}\frac{\Psi^{2}_{1}}{\Psi^{2}}\left(5+\sqrt{6}n\frac{\Phi^{2n-1}_{1}}{\Psi_{1}}\frac{\partial\Phi_{1}}{\partial\phi_{1}}\right)+\frac{\Phi^{2n-1}_{2}}{\Psi_{2}}\frac{\partial\Phi_{2}}{\partial\phi_{2}}\frac{\Psi^{2}_{2}}{\Psi^{2}}\left(5+\sqrt{6}n\frac{\Phi^{2n-1}_{2}}{\Psi_{2}}\frac{\partial\Phi_{2}}{\partial\phi_{2}}\right)\right]\right] \\
      &+24n^{2}\left[\frac{\Phi^{2n-1}_{1}}{\Psi_{1}}\frac{\partial\Phi_{1}}{\partial\phi_{1}}-\frac{\Phi^{2n-1}_{2}}{\Psi_{2}}\frac{\partial\Phi_{2}}{\partial\phi_{2}}\right]^{2}\frac{\Psi^{2}_{1}\Psi^{2}_{2}}{\Psi^{4}} \\
      &-12n(2n-1)\left[\Phi^{2(n-1)}_{1}\left(\frac{\partial\Phi_{1}}{\partial\phi_{1}}\right)^{2}\frac{\Psi^{2}_{1}}{\Psi^{2}}+\Phi^{2(n-1)}_{2}\left(\frac{\partial\Phi_{2}}{\partial\phi_{2}}\right)^{2}\frac{\Psi^{2}_{2}}{\Psi^{2}}\right] \\
      &-12n\left[\Phi^{2n-1}_{1}\frac{\partial^{2}\Phi_{1}}{\partial\phi^{2}_{1}}\frac{\Psi^{2}_{1}}{\Psi^{2}}+\Phi^{2n-1}_{2}\frac{\partial^{2}\Phi_{2}}{\partial\phi^{2}_{2}}\frac{\Psi^{2}_{2}}{\Psi^{2}}\right]+\frac{\text{n}^{2}}{a^{2}H^{2}} \\
 \end{aligned}
\end{equation}\end{small}
\begin{equation*}
\begin{aligned}
  \gamma_{12}
            =&-8\sqrt{6}n\left(\frac{\Phi^{2n-1}_{1}}{\Psi_{1}}\frac{\partial\Phi_{1}}{\partial\phi_{1}}-\frac{\Phi^{2n-1}_{2}}{\Psi_{2}}\frac{\partial\Phi_{2}}{\partial\phi_{2}}\right)\frac{\Psi^{2}_{1}}{\Psi^{2}}\frac{\Psi^{2}_{2}}{\Psi^{2}} \\
            &-18n^{2}\left(\frac{\Phi^{2n-1}_{1}}{\Psi_{1}}\frac{\partial\Phi_{1}}{\partial\phi_{1}}-\frac{\Phi^{2n-1}_{2}}{\Psi_{2}}\frac{\partial\Phi_{2}}{\partial\phi_{2}}\right)^{2}\frac{\left(\Psi^{2}_{1}-\Psi^{2}_{2}\right)}{\Psi^{2}}\frac{\Psi^{2}_{1}}{\Psi^{2}}\frac{\Psi^{2}_{2}}{\Psi^{2}} \\
             &-6n^{2}\left[\left(\frac{\Phi^{2n-1}_{1}}{\Psi_{1}}\frac{\partial\Phi_{1}}{\partial\phi_{1}}\right)^{2}-\left(\frac{\Phi^{2n-1}_{2}}{\Psi_{2}}\frac{\partial\Phi_{2}}{\partial\phi_{2}}\right)^{2}\right]\frac{\Psi^{2}_{1}}{\Psi^{2}}\frac{\Psi^{2}_{2}}{\Psi^{2}} \\ 
             &-12n(2n-1)\left(\Phi^{2(n-1)}_{1}\left(\frac{\partial\Phi_{1}}{\partial\phi_{1}}\right)^{2}-\Phi^{2(n-1)}_{2}\left(\frac{\partial\Phi_{2}}{\partial\phi_{2}}\right)^{2}\right)\frac{\Psi^{2}_{1}}{\Psi^{2}}\frac{\Psi^{2}_{2}}{\Psi^{2}} \\
             &-12n\left(\Phi^{2n-1}_{1}\frac{\partial^{2}\Phi_{1}}{\partial\phi_{1}^{2}}-\Phi^{2n-1}_{2}\frac{\partial^{2}\Phi_{2}}{\partial\phi_{2}^{2}}\right)\frac{\Psi^{2}_{1}}{\Psi^{2}}\frac{\Psi^{2}_{2}}{\Psi^{2}} \\
\end{aligned}
\end{equation*}

\begin{footnotesize}\begin{equation*}
\begin{aligned}
  \eta_{12}
            =&\left[18\sqrt{6}n-6\sqrt{6}n\left(\Psi^{2}-\Phi^{2n}\right)-2\sqrt{6}n\frac{\text{q}^{2}}{a^{2}H^{2}}\right]\left(\frac{\Phi^{2n-1}_{1}}{\Psi_{1}}\frac{\partial\Phi_{1}}{\partial\phi_{1}}-\frac{\Phi^{2n-1}_{2}}{\Psi_{2}}\frac{\partial\Phi_{2}}{\partial\phi_{2}}\right)\frac{\Psi^{2}_{1}\Psi^{2}_{2}}{\Psi^{4}} \\
             &+48n^{2}\left[\frac{\Phi^{2n-1}_{1}}{\Psi_{1}}\frac{\Psi^{2}_{1}}{\Psi^{2}}\frac{\partial\Phi_{1}}{\partial\phi_{1}}+\frac{\Phi^{2n-1}_{2}}{\Psi_{2}}\frac{\Psi^{2}_{2}}{\Psi^{2}}\frac{\partial\Phi_{2}}{\partial\phi_{2}}\right]\left(\frac{\Phi^{2n-1}_{1}}{\Psi_{1}}\frac{\partial\Phi_{1}}{\partial\phi_{1}}-\frac{\Phi^{2n-1}_{2}}{\Psi_{2}}\frac{\partial\Phi_{2}}{\partial\phi_{2}}\right)\frac{\Psi^{2}_{1}\Psi^{2}_{2}}{\Psi^{4}} \\             
             &+18\sqrt{6}n^{3}\left[\frac{\Phi^{2n-1}_{1}}{\Psi_{1}}\frac{\Psi^{2}_{1}}{\Psi^{2}}\frac{\partial\Phi_{1}}{\partial\phi_{1}}+\frac{\Phi^{2n-1}_{2}}{\Psi_{2}}\frac{\Psi^{2}_{2}}{\Psi^{2}}\frac{\partial\Phi_{2}}{\partial\phi_{2}}\right]^{2}\left(\frac{\Phi^{2n-1}_{1}}{\Psi_{1}}\frac{\partial\Phi_{1}}{\partial\phi_{1}}-\frac{\Phi^{2n-1}_{2}}{\Psi_{2}}\frac{\partial\Phi_{2}}{\partial\phi_{2}}\right)\frac{\Psi^{2}_{1}\Psi^{2}_{2}}{\Psi^{4}} \\             
             &+6\sqrt{6}n^{3}\left[\left(\frac{\Phi^{2n-1}_{1}}{\Psi_{1}}\frac{\partial\Phi_{1}}{\partial\phi_{1}}\right)^{2}\frac{\Psi^{2}_{1}}{\Psi^{2}}+\left(\frac{\Phi^{2n-1}_{2}}{\Psi_{2}}\frac{\partial\Phi_{2}}{\partial\phi_{2}}\right)^{2}\frac{\Psi^{2}_{2}}{\Psi^{2}}\right]\left(\frac{\Phi^{2n-1}_{1}}{\Psi_{1}}\frac{\partial\Phi_{1}}{\partial\phi_{1}}-\frac{\Phi^{2n-1}_{2}}{\Psi_{2}}\frac{\partial\Phi_{2}}{\partial\phi_{2}}\right)\frac{\Psi^{2}_{1}\Psi^{2}_{2}}{\Psi^{4}} \\             
             &+36n^{2}\left(\frac{\Phi^{2n-1}_{1}}{\Psi_{1}}\frac{\partial\Phi_{1}}{\partial\phi_{1}}-\frac{\Phi^{2n-1}_{2}}{\Psi_{2}}\frac{\partial\Phi_{2}}{\partial\phi_{2}}\right)^{2}\frac{\Psi^{2}_{1}-\Psi^{2}_{2}}{\Psi^{2}}\frac{\Psi^{2}_{1}\Psi^{2}_{2}}{\Psi^{4}} \\
             &+24n^{2}\left[\left(\frac{\Phi^{2n-1}_{1}}{\Psi_{1}}\frac{\partial\Phi_{1}}{\partial\phi_{1}}\right)^{2}-\left(\frac{\Phi^{2n-1}_{2}}{\Psi_{2}}\frac{\partial\Phi_{2}}{\partial\phi_{2}}\right)^{2}\right]\frac{\Psi^{2}_{1}\Psi^{2}_{2}}{\Psi^{4}} \\
             &-18\sqrt{6}n^{3}\left(\frac{\Phi^{2n-1}_{1}}{\Psi_{1}}\frac{\partial\Phi_{1}}{\partial\phi_{1}}-\frac{\Phi^{2n-1}_{2}}{\Psi_{2}}\frac{\partial\Phi_{2}}{\partial\phi_{2}}\right)^{3}\frac{\Psi^{4}_{1}\Psi^{4}_{2}}{\Psi^{8}} \\
             &+6n(2n-1)\left[8+\sqrt{6}n\left(\frac{\Phi^{2n-1}_{1}}{\Psi_{1}}\frac{\partial\Phi_{1}}{\partial\phi_{1}}+\frac{\Phi^{2n-1}_{2}}{\Psi_{2}}\frac{\partial\Phi_{2}}{\partial\phi_{2}}\right)\right]\left(\Phi^{2(n-1)}_{1}\left(\frac{\partial\Phi_{1}}{\partial\phi_{1}}\right)^{2}-\Phi^{2(n-1)}_{2}\left(\frac{\partial\Phi_{2}}{\partial\phi_{2}}\right)^{2}\right)\frac{\Psi^{2}_{1}\Psi^{2}_{2}}{\Psi^{4}} \\
             &+6n\left[8+\sqrt{6}n\left(\frac{\Phi^{2n-1}_{1}}{\Psi_{1}}\frac{\partial\Phi_{1}}{\partial\phi_{1}}+\frac{\Phi^{2n-1}_{2}}{\Psi_{2}}\frac{\partial\Phi_{2}}{\partial\phi_{2}}\right)\right]\left(\Phi^{2n-1}_{1}\frac{\partial^{2}\Phi_{1}}{\partial\phi_{1}^{2}}-\Phi^{2n-1}_{2}\frac{\partial^{2}\Phi_{2}}{\partial\phi_{2}^{2}}\right)\frac{\Psi^{2}_{1}\Psi^{2}_{2}}{\Psi^{4}} \\
             &+18\sqrt{6}n(2n-1)\left[\Phi^{2(n-1)}_{1}\Psi_{1}\frac{\partial\Phi_{1}}{\partial\phi_{1}}\frac{\partial^{2}\Phi_{1}}{\partial\phi^{2}_{1}}-\Phi^{2(n-1)}_{2}\Psi_{2}\frac{\partial\Phi_{2}}{\partial\phi_{2}}\frac{\partial^{2}\Phi_{2}}{\partial\phi^{2}_{2}}\right]\frac{\Psi^{2}_{1}\Psi^{2}_{2}}{\Psi^{4}} \\
             &+12\sqrt{6}n(2n-1)(n-1)\left[\Phi^{2n-3}_{1}\Psi_{1}\left(\frac{\partial\Phi_{1}}{\partial\phi_{1}}\right)^{3}-\Phi^{2n-3}_{2}\Psi_{2}\left(\frac{\partial\Phi_{2}}{\partial\phi_{2}}\right)^{3}\right]\frac{\Psi^{2}_{1}\Psi^{2}_{2}}{\Psi^{4}} \\
             &+6\sqrt{6}n\left[\Phi^{2n-1}_{1}\Psi_{1}\frac{\partial^{3}\Phi_{1}}{\partial\phi^{3}_{1}}-\Phi^{2n-1}_{2}\Psi_{2}\frac{\partial^{3}\Phi_{2}}{\partial\phi^{3}_{2}}\right]\frac{\Psi^{2}_{1}\Psi^{2}_{2}}{\Psi^{4}}
\end{aligned}
\end{equation*}\end{footnotesize}

\begin{equation*}
\begin{aligned}
\varsigma_{12}
              =&-\sqrt{6}n\left[\frac{\Phi^{2n-1}_{1}}{\Psi_{1}}\frac{\partial\Phi_{1}}{\partial\phi_{1}}\frac{\Psi^{2}_{2}}{\Psi^{2}}+\frac{\Phi^{2n-1}_{2}}{\Psi_{2}}\frac{\partial\Phi_{2}}{\partial\phi_{2}}\frac{\Psi^{2}_{1}}{\Psi^{2}}\right]  \\ 
\end{aligned}
\end{equation*}

\begin{equation*}
\begin{aligned}
\varpi_{12}=
            &6n^{2}\left[\frac{\Phi^{2n-1}_{1}}{\Psi_{1}}\frac{\partial\Phi_{1}}{\partial\phi_{1}}-\frac{\Phi^{2n-1}_{2}}{\Psi_{2}}\frac{\partial\Phi_{2}}{\partial\phi_{2}}\right]^{2}\frac{\Psi^{2}_{1}\Psi^{2}_{2}}{\Psi^{4}} \\
            &+6n\left[(2n-1)\Phi^{2(n-1)}_{1}\left(\frac{\partial\Phi_{1}}{\partial\phi_{1}}\right)^{2}+\Phi^{2n-1}_{1}\frac{\partial^{2}\Phi_{1}}{\partial\phi^{2}_{1}}\right]\frac{\Psi^{2}_{2}}{\Psi^{2}} \\
            &+6n\left[(2n-1)\Phi^{2(n-1)}_{2}\left(\frac{\partial\Phi_{2}}{\partial\phi_{2}}\right)^{2}+\Phi^{2n-1}_{2}\frac{\partial^{2}\Phi_{2}}{\partial\phi^{2}_{2}}\right]\frac{\Psi^{2}_{1}}{\Psi^{2}}+\frac{\text{q}^{2}}{a^{2}H^{2}}
\end{aligned}
\end{equation*}

\begin{equation*}
\begin{aligned}
\iota_{12}
          =2-2\left(\Psi^{2}_{1}+\Psi^{2}_{2}\right)+\left(\Phi^{2n}_{1}+\Phi^{2n}_{2}\right)-\varsigma_{12}\;. 
\end{aligned}
\end{equation*}
%
%
At $\mathcal{P}$, it follows from Lemma \ref{Lemma_1} that
\begin{equation*}
 \begin{aligned}
  \xi(\mathcal{P})=&-\left[4-9\Phi_{1}^{2n}(\mathcal{P})-9\Phi_{2}^{2n}(\mathcal{P})\right]\\
 \end{aligned}
\end{equation*}

\begin{equation*}
 \begin{aligned}
   \zeta(\mathcal{P})=&\left[24-9\left(4-\frac{2}{n}\right)\Psi^{2}\right]\left(\Phi^{2n}_{1}+\Phi^{2n}_{2}\right)                                              -18\left[\frac{\Phi^{4n}_{1}}{\Psi^{2}_{1}}+\frac{\Phi^{4n}_{2}}{\Psi^{2}_{2}}\right]\Psi^{2} \\
                      &+36\left[\Phi^{2n}_{1}\frac{\Psi_{2}}{\Psi_{1}}-\Phi^{2n}_{2}\frac{\Psi_{1}}{\Psi_{2}}\right]^{2}  -12n\left[\Phi^{2n-1}_{1}\frac{\Psi^{2}_{1}}{\Psi^{2}}\frac{\partial^{2}\Phi_{1}}{\partial\phi^{2}_{1}}+\Phi^{2n-1}_{2}\frac{\Psi^{2}_{2}}{\Psi^{2}}\frac{\partial^{2}\Phi_{2}}{\partial\phi^{2}_{2}}\right]+\frac{\text{n}^{2}}{a^{2}H^{2}} \\
 \end{aligned}
\end{equation*}

\begin{equation*}
 \begin{aligned}
  \gamma_{12}(\mathcal{P})
                                       =&24\left[\Phi^{2n}_{1}\frac{\Psi^{2}_{2}}{\Psi^{2}_{1}+\Psi^{2}_{2}}-\Phi^{2n}_{2}\frac{\Psi^{2}_{1}}{\Psi^{2}_{1}+\Psi^{2}_{2}}\right]-9\left[\Phi^{4n}_{1}\frac{\Psi^{2}_{2}}{\Psi^{2}_{1}}-\Phi^{4n}_{2}\frac{\Psi^{2}_{1}}{\Psi^{2}_{2}}\right] \\
                                        &-24\left[\Phi^{2n}_{1}\frac{\Psi_{2}}{\Psi_{1}}-\Phi^{2n}_{2}\frac{\Psi_{1}}{\Psi_{2}}\right]^{2}\frac{\Psi^{2}_{1}-\Psi^{2}_{2}}{\Psi^{2}}-9\left(4-\frac{2}{n}\right)\left[\Phi^{2n}_{1}\Psi^{2}_{2}-\Phi^{2n}_{2}\Psi^{2}_{1}\right] \\
                                        &-12n\left[\Phi^{2n-1}_{1}\frac{\partial^{2}\Phi_{1}}{\partial\phi^{2}_{1}}-\Phi^{2n-1}_{2}\frac{\partial^{2}\Phi_{2}}{\partial\phi^{2}_{2}}\right]\frac{\Psi^{2}_{1}\Psi^{2}_{2}}{\left(\Psi^{2}_{1}+\Psi^{2}_{2}\right)^{2}} \\
 \end{aligned}
\end{equation*}

\begin{equation*}
 \begin{aligned}
   \eta_{12}(\mathcal{P})
                                =&-\left[36\Psi^{2}+81\Phi^{4n}+24\left(\frac{\Phi^{4n}_{1}}{\Psi^{2}_{1}}+\frac{\Phi^{4n}_{2}}{\Psi^{2}_{2}}\right)\Psi^{2}-6\frac{\text{q}^{2}}{a^{2}H^{2}}\right]\left[\Phi^{2n}_{1}\frac{\Psi^{2}_{2}}{\Psi^{2}}-\Phi^{2n}_{2}\frac{\Psi^{2}_{1}}{\Psi^{2}}\right] \\
                                 &+54\left[\Phi^{2n}_{1}\frac{\Psi_{2}}{\Psi_{1}}-\Phi^{2n}_{2}\frac{\Psi_{1}}{\Psi_{2}}\right]^{2}\frac{\Psi^{2}_{1}-\Psi^{2}_{2}}{\Psi^{2}}+36\left[\Phi^{4n}_{1}\frac{\Psi^{2}_{2}}{\Psi^{2}_{1}}-\Phi^{4n}_{2}\frac{\Psi^{2}_{1}}{\Psi^{2}_{2}}\right] \\
                                 &+81\left[\Phi^{2n}_{1}\frac{\Psi_{2}}{\Psi_{1}}-\Phi^{2n}_{2}\frac{\Psi_{1}}{\Psi_{2}}\right]^{3}\frac{\Psi_{1}\Psi_{2}}{\Psi^{4}} \\
                                 &+9\left(2-\frac{1}{n}\right)\left[8-3\left(\frac{\Phi^{2n}_{1}}{\Psi^{2}_{1}}+\frac{\Phi^{2n}_{2}}{\Psi^{2}_{2}}\right)\Psi^{2}\right]\left(\Phi^{2n}_{1}\Psi^{2}_{2}-\Phi^{2n}_{2}\Psi^{2}_{1}\right) \\
                                 &-\left(\frac{\sqrt{6}}{2n}\right)^{3}12\sqrt{6}n(2n-1)(n-1)\left[\Phi^{2n}_{1}\Psi^{2}_{2}-\Phi^{2n}_{2}\Psi^{2}_{1}\right]\Psi^{2} \\
                                 &+6n\left[8-3\left(\frac{\Phi^{2n}_{1}}{\Psi^{2}_{1}}+\frac{\Phi^{2n}_{2}}{\Psi^{2}_{2}}\right)\Psi^{2}\right]\left[\Phi^{2n-1}_{1}\frac{\partial^{2}\Phi_{1}}{\partial\phi^{2}_{1}}-\Phi^{2n-1}_{2}\frac{\partial^{2}\Phi_{2}}{\partial\phi^{2}_{2}}\right]\frac{\Psi^{2}_{1}\Psi^{2}_{2}}{\left(\Psi^{2}_{1}+\Psi^{2}_{2}\right)^{2}}  \\
                                 &-18\left(\frac{\sqrt{6}}{2n}\right)\sqrt{6}n(2n-1)\left[\Phi^{2n-1}_{1}\frac{\partial^{2}\Phi_{1}}{\partial\phi^{2}_{1}}-\Phi^{2n-1}_{2}\frac{\partial^{2}\Phi_{2}}{\partial\phi^{2}_{2}}\right]\frac{\Psi^{2}_{1}\Psi^{2}_{2}}{\Psi^{4}} \\
                                 &+6\sqrt{6}n\left[\Phi^{2n-1}_{1}\Psi_{1}\frac{\partial^{3}\Phi_{1}}{\partial\phi^{3}_{1}}-\Phi^{2n-1}_{2}\Psi_{2}\frac{\partial^{3}\Phi_{2}}{\partial\phi^{3}_{2}}\right]\frac{\Psi^{2}_{1}\Psi^{2}_{2}}{\Psi^{4}}
                                 \end{aligned}
\end{equation*}

\begin{equation*}
\begin{aligned}
     \varsigma_{12}(\mathcal{P})=&\frac{3}{\Psi^{2}_{1}\Psi^{2}_{2}}\left[\Phi^{2n}_{1}\Psi^{4}_{2}+\Phi^{2n}_{2}\Psi^{4}_{1}\right] \\
   \varpi_{12}(\mathcal{P})=&9\frac{\left(\Phi^{2n}_{1}\Psi^{2}_{2}-\Phi^{2n}_{2}\Psi^{2}_{1}\right)^{2}}{\Psi^{2}_{1}\Psi^{2}_{2}}+\frac{9}{n}(2n-1)\Psi^{2}\left(\Phi^{2n}_{1}\left(\frac{\Psi_2}{\Psi_1}\right)^{2}+\Phi^{2n}_{2}\left(\frac{\Psi_1}{\Psi_2}\right)^{2}\right) \\ 
                                  &+\frac{6n}{\Psi^{2}}\left(\Phi^{2n-1}_{1}\Psi^{2}_{2}\frac{\partial^{2}\Phi_1}{\partial\phi^{2}_{1}}+\Phi^{2n-1}_{2}\Psi^{2}_{1}\frac{\partial^{2}\Phi_{2}}{\partial\phi^{2}_{2}}\right)+\frac{\text{q}^{2}}{a^{2}H^{2}} \\
\iota_{12}(\mathcal{P})=&
                        3\left(\Phi^{2}_{1}+\Phi^{2}_{2}\right)-\varsigma_{12}(\mathcal{P}) \\
 \end{aligned}
\end{equation*} 


\end{document}